\documentclass[twocolumn]{aastex631}

\usepackage{amsthm,latexsym,amssymb,amsmath, amsfonts}
\usepackage{xcolor}
\usepackage{subfloat}
\usepackage{graphicx}
\newcommand{\rutgers}{Rutgers University, Department of Physics and Astronomy, 136 Frelinghuysen Road, Piscataway, NJ 08854, USA}

\newcommand{\Dolphot}{\texttt{DOLPHOT}}
\newcommand{\Beast}{\texttt{BEAST}}

\newcommand\jwst{\textit{JWST}}
\newcommand\hst{\textit{HST}}
\newcommand{\cmd}[3]{\ensuremath{#1-#2 \textrm{ vs. } #3}}
\newcommand{\trgbcalfull}[6]{TRGB\,(#1)\ensuremath{=} #2-(#3)\ensuremath{\times\left[\left(#4-#5\right)-#6\right]}~mag}
\received{June 11th, 2024}
\revised{September 2nd, 2024}
\accepted{September 8th, 2024}

\begin{document}

\title{An Empirical Calibration of the Tip of the Red Giant Branch Distance Method in the Near Infrared. II. \jwst\  NIRCam Wide Filters}

\author[0000-0002-8092-2077]{Max J. B. Newman}
\affiliation{\rutgers}

\author[0000-0001-5538-2614]{Kristen B. W. McQuinn}
\affiliation{Space Telescope Science Institute, 3700 San Martin Drive, Baltimore, MD 21218, USA}
\affiliation{\rutgers}

\author[0000-0003-0605-8732]{Evan D. Skillman}
\affiliation{University of Minnesota, Minnesota Institute for Astrophysics, School of Physics and Astronomy, 116 Church Street, S.E., Minneapolis,\\
MN 55455, USA}

\author[0000-0003-4850-9589]{Martha L. Boyer}
\affiliation{Space Telescope Science Institute, 3700 San Martin Drive, Baltimore, MD 21218, USA}

\author[0000-0002-2970-7435]{Roger E. Cohen}
\affiliation{\rutgers}

\author[0000-0001-8416-4093]{Andrew E. Dolphin}
\affiliation{Raytheon, 1151 E. Hermans Road, Tucson, AZ 85756, USA}
\affiliation{Steward Observatory, University of Arizona, 933 North Cherry Avenue, Tucson, AZ 85721, USA}

\newcommand{\princeton}{Department of Astrophysical Sciences, Princeton University, 4 Ivy Lane, Princeton, NJ 08544, USA}
\newcommand{\carnegie}{The Observatories of the Carnegie Institution for Science, 813 Santa Barbara Street, Pasadena, CA 91101, USA}

\author[0000-0003-4122-7749]{O. Grace Telford}
\affiliation{\princeton}
\affiliation{\carnegie}
\affiliation{\rutgers}

\correspondingauthor{Max J. B. Newman}
\email{mjn125@physics.rutgers.edu}

\begin{abstract}
The tip of the red giant (TRGB) is a standardizable candle, identifiable as the discontinuity at the bright extreme of the red giant branch (RGB) stars in color-magnitude diagram (CMD) space. The TRGB-based distance method has been used to measure distances to galaxies out to $D\leq20$~Mpc with the Hubble Space Telescope (\hst) $F814W$ filter, and is an important rung in the distance ladder to measure the Hubble constant, $H_0$. In the infrared (IR), the TRGB apparent magnitude ranges from $1-2$~magnitudes brighter than in the optical. Now with the James Webb Space Telescope (\jwst) the feasible distance range of the TRGB method can reach $\sim50$~Mpc. However, the IR TRGB luminosity depends to varying degrees on stellar metallicity/age. Here we standardize the TRGB luminosity using stellar colors as a proxy for metallicity/age to derive color-based corrections for the \jwst\ Near-Infrared Camera (NIRCam) short wavelength (SW) filters $F090W$, $F115W$, $F150W$ and the long wavelength (LW) filters $F277W$, $F356W,$ and $F444W$. We provide recommended filters for distance measurements depending on the requisite precision. For science requiring high precision ($\leq1\%$ in distance) we recommend measuring the TRGB in $F090W$ vs $F090W-F150W$ or $F115W$ vs. $F115W-F277W$ with the caveat that even with \jwst\ long integration times will be necessary at further distances. If lower precision ($>1.5\%$ in distance) can be tolerated, or if shorter integration times are desirable, we recommend measuring the TRGB in either $F115W$ or $F150W$. We do not recommend $F444W$ for precision TRGB measurements due to its lower angular resolution. \end{abstract}


\section{Introduction}
Distance is one of the most important and fundamental measurements required to place myriad galaxy properties of astronomical systems on absolute scales in the near and far Universe. Distances to galaxies have also been used to produce detailed maps of visible (baryonic) structure and the motion of galaxies on supercluster scales \citep[e.g.,][]{Tully2014}, and are used to construct models of orbital histories \citep[e.g.,][]{vanderMarel2002, Kallivayalil2013, Patel2020}. On cosmological scales, extragalactic distances are a key component to measuring the local expansion rate of the Universe \citep[the Hubble constant; $H_0$; e.g.,][]{Freedman2021, Riess2022}. 

Significant progress has been made in measuring precise distances to nearby galaxies from both ground- and space-based observatories using a number of standard candles. These methods include  RR-Lyrae stars \citep[e.g.,][]{Carretta2000}, Horizontal Branch stars \citep[e.g.,][]{Pietrzynksi2008}, pulsating Cepheids stars via the period-luminosity(-color) relation \citep[e.g.,][]{Sandage2006,Riess2022},  Mira variable starts \citep[e.g.,][]{Soszynski2013,Huang2018,Huang2020}, the J-region Asymptotic Giant Branch stars \cite[e.g.,][]{Lee2023}, and the tip of the red giant branch (TRGB) stars \citep[e.g.,][]{Mould1986, Freedman1988, DaCosta1990}. Of these, the TRGB-based distance method is regarded as one of the more robust techniques \citep[e.g.,][]{Lee1993, Beaton2018}. Of equal import, TRGB stars are found in \textit{every} galaxy regardless of morphology and therefore provide a method that can be applied to any galaxy as long as there are a sufficient number of RGB stars and the TRGB stars can be individually resolved with high-fidelity. 

While the bolometric luminosity of the TRGB stars is only weakly dependent on stellar mass and composition, in practice we observe the stars within specific wavelengths ranges (i.e., bandpasses of filters) where the TRGB luminosity can depend on stellar content \citep{Serenelli2017}. Fortuitously, the luminosities of TRGB stars as measured in ground based I-band filters and in the Hubble Space Telescope's (\hst 's) $F814W$ filter are nearly constant as a function of the age and metallicity of the stars, with only a modest decreasing brightness at younger ages/higher metallicites. In practice, we use color as a proxy for metallicity and age. The change in age and metallicity manifests as a change in color which is well understood, has been empirically characterized in several studies \citep[e.g.,][]{Rizzi2007,Jang2017}, and makes the TRGB standardizable. At the absolute brightness of the TRGB in $I$ band equivalent filters \citep[$M^{\text{TRGB}}_{\text{F814W}}=-4.049$~mag;][]{Freedman2021}, the TRGB method can be used to measure robust distances to $D\approx15$~Mpc with high-quality \hst\ data. 

When observed at wavelengths $>0.9\mu$m (i.e., the near-infrared (NIR)) the TRGB stars appear brighter due to bolometric corrections (BCs) by $\sim1-2$ magnitudes and can therefore expand the feasible range over which distances can be measured up to a factor 5 \citep[e.g.,][]{Serenelli2017}. However, the color dependence of the TRGB luminosity in the NIR is more significant than in the $I$-band. For the TRGB method to be realized as robust and accurate in the NIR, its color-dependence must be characterized. 

Recognizing the utility of the TRGB in the NIR for distance work, previous studies have presented calibrations of the NIR TRGB from theoretical stellar evolutionary models and from observational data. \citet{Valenti2004a, Valenti2004b} derived a TRGB luminosity calibration based on ESO-MPI 2.2-m telescope and Two-Micron Sky Survey (2MASS) $JHK$ observations of 24 Galactic Globular Clusters (GGCs). Their TRGB slope and zero-point calibrations were derived as a function of GCC metallicity ([Fe/H]). \citet{Dalcanton2012a} \citep[and later reanalyzed by ][]{Durbin2020} used \hst\  Wide Field Camera 3 Infrared (WFC3/IR) observations of galaxies in the $F110W$ and $F160W$ filters to derive an $F160W$ ($H$-band) TRGB luminosity calibration based on the color of the TRGB stars.  \citet{Dalcanton2012a} found that the \citet{Valenti2004a,Valenti2004b} TRGB calibration was offset to brighter magnitudes compared to their own $F160W$ calibration and to the \texttt{PARSEC} models. A potential source of the offset is that many GCCs are impacted by large median foreground-extinctions ($A_H\sim0.5$). 

\citet{Wu2014} derived empirical TRGB calibrations in the WFC3/IR $F110W$ and $F160W$ filters based largely on the sample in \citet{Dalcanton2012a}. At the blue end of their calibration ($F110W-F160W<0.95$) they found agreement with \citet{Dalcanton2012a} within $0.07$~mag. However, at $F110W-F160W=1.05$, the calibrations were offset by up to $0.15$~mag. \citet{Wu2014} include M~31 in their study which anchors the red end of the calibration and can be a potential source of the difference between two studies. More recently, \citet{Durbin2020} reanalyzed the data from \citet{Dalcanton2012a} and derived a new TRGB calibration. They found modest difference from the \citet{Dalcanton2012a} calibration of $0.02$~mag in the $F160W$ zero-point and $0.04$ in the slope. 

Separately, \citet{Serenelli2017} theoretically calibrated the metallicity/age dependence of the TRGB via the stellar evolutionary packages \texttt{BaSTI} \citep{Pietrinferni2004,Pietrinferni2006} and \texttt{GARSTEC} \citep{Weiss2008}. They found broad agreement with \citet{Wu2014} at bluer colors, but above $F110W-F160W=0.95$ find the results between theoretical libraries and the empirical calibration show large dispersion. 

\citet{Madore2018} reported empirical $JHK$ TRGB calibrations based on observations of the dwarf galaxy IC 1613 obtained from the FourStar NIR camera on the 6.5 m Baade-Magellan telescope. When compared to \citet{Wu2014} they found that both their $J$-band and $H$-band magnitudes were $+0.29$~mag brighter while the slopes agreed at the $12\%$ level over the color range $0.7<F110W-F160W<0.95$. In a companion study, \citet{Hoyt2018} adopted the slope presented in \citet{Madore2018} to derive $JHK$ zero-points tied to the Large Magellanic Cloud using Near-Infrared Synoptic Survey data \citep[NISS; ][]{Macri2015}. While the slope reported in \citet{Hoyt2018} was shallower and the zero-point was fainter in the $K$-band compared to \citet{Valenti2004a} and \cite{Serenelli2017}, the values agreed with each other within the uncertainties.

Recently, we reported an empirical calibration of the NIR TRGB in the \hst\ WFC3/IR $F110W$ and $F160W$ filters \citep[][hereafter Paper I]{Newman2024}. We found broad agreement with \citet{Dalcanton2012a} and \citet{Durbin2020} in both the slope and zero-point in $F160W$. We also found agreement with \citet{Wu2014} over a narrow color range. 

We emphasize that \hst\ is still a powerful observatory for NIR TRGB measurements. While \hst\ is now operating in a reduced gyroscope mode\footnote{see\,\url{https://science.nasa.gov/missions/hubble/nasa-to-change-how-it-points-hubble-space-telescope/}} making new proposals for precision TRGB distance work more challennig, the Barbara A. Mikulski Archive for Space Telescopes (MAST) hosts an incredible volume of archival NIR \hst\ observations of galaxies that can be used to measure TRGB distances. In addition we use the  \hst\ TRGB NIR calibrations as stepping stones to the \jwst\ TRGB. The \hst\ calibration in Paper I directly informs our \jwst\ TRGB calibration.

However, \hst\ can only efficiently and reliably measure the TRGB out to a maximum of$D\leq20$~Mpc, but reaches the limits of its angular resolution beyond this distance. \jwst\ will provide the means to advance and extend the TRGB method to distances beyond 20~Mpc. Here, we use the \hst\ TRGB NIR calibrations as stepping stones to the \jwst\ TRGB. The HST calibration in Paper I directly informs our \jwst\ TRGB calibration.

\subsection{The Tip of the Red Giant Branch in the Era of \jwst\ }
The \jwst\ is designed to observe at infrared (IR) wavelengths and, combined with its significantly higher angular resolution over \hst, enables TRGB-based distance measurements over larger volumes than were previously accessible. 

Recently, \citet{McQuinn2019} explored the behavior of the TRGB as a function of metallicity, stellar age, and from optical to IR wavelengths from synthetic photometry. Their overarching goal was to determine whether the TRGB can be standardized at wavelengths redder than the $I$-band over a range of stellar properties. The synthetic photometry was generated using the solar-scaled PARSEC stellar evolution library \citep{Bressan2012} in several ground-based filters, in \hst\ ($F475W$, $F606W$, and $F814W$, $F110W$, and $F160W$ filters), in \jwst\ ($F090W$, $F115W$, $F150W$, $F200W$, and $F277W$), and in \textit{Spitzer Space Telescope} IRAC ($3.6\mu$m and $4.5\mu$m filters; similar to NIRCam $F356W$ and $F444W$ filters, respectively), and over a range of stellar ages and metallicites \citep[see Table 1;][]{McQuinn2019}. 

\citet{McQuinn2019} reported several findings based on their analysis of the TRGB in synthetic CMDs. First, the TRGB magnitude increased by up to 2~magnitudes as a function of increasing wavelength from the \hst\ $F814W$ filter to the $F444W$-similar \textit{Spitzer} IRAC $4.5\mu$m filter, with the implication that the NIR TRGB enables observational gains if carefully calibrated. Second, the TRGB luminosity in the \hst\ F814W filter was nearly constant as a function of both age and metallicity. However, in the \hst\ F110W filter, \jwst\ $F115W$ filter, and $J$-band, the TRGB magnitude varied by up to 0.3~mag as a function of stellar metallicity and 0.09~mag as a function of stellar age. In filters redder than F115W the magnitude varied by up to 0.6~mag or 0.1~mag in metallicity or age, respectively. Third, after applying a color-based slope correction (measured from their fiducial synthetic photometry) to the TRGB stars, the TRGB magnitudes measured in the \jwst\ bands exhibit a spread of only $0.02$ to $0.05$~mag $\left(0.9\%-2.0\%\right)$.

From the observational side there exists only one empirical calibration of the TRGB in the \jwst\ Near Infrared Camera (NIRCam) SW $F090W$ filter \citep{Anand2024}. Their calibration is derived from one target, NGC~4258 and in the NIRCam $F090W$ and $F150W$ filters. To date, there exist no calibrations of the TRGB at longer \jwst\ wavelengths in the literature. In addition, the Near Infrared Imager and Slitless Spectroscopy (NIRISS) will be a viable tool for TRGB measurements in addition to NIRCam. While there is not currently a TRGB calibration available for NIRISS filters, we obtained new NIRISS data as part of our program and a TRGB calibration for the filters $F090W$, $F115W$, and $F150W$ is forthcoming.

In this work we calibrate the TRGB in 6 \jwst\ NIRCam filters: the SW filters $F090W$, $F115W$, and $F150W$, and the long wavelength (LW) filters $F277W$, $F356W$, and $F444W$. As in Paper I, our calibration is data driven and the methodology is independent of stellar evolution models (i.e., empirical). The absolute magnitude scale for these data is set by uniformly measured distances in the \hst\  ACS $F814W$ filter \citep[see Paper I for details; see also][for WLM]{Albers2019}, with the exception of Sextans A, for which the TRGB was measured in the \hst\ Wide Field Planetary Camera 2 (WFPC2) $F814W$ filter \citep{Dolphin2003}. 

The structure of this paper is as follows: we present new and archival \jwst\ observations, \jwst\ pipeline processing of the exposures, photometry, and artificial star tests (ASTs) in \S~\ref{sec:obs_and_data}. We show foreground-extinction-corrected color-magnitude diagrams (CMDs) in \S~\ref{sec:CMDs} for all possible mutations of \jwst\ filters available in these data. We describe our methodology for characterizing the slope and zero-point of the TGRB in \S~\ref{sec:calibration_method}. Next, we present the final TRGB calibrations, discuss the advantages and disadvantages of using each filter for TRGB-based distance measurements, and verify the precision of our calibrations in \S~\ref{sec:results}. We then compare our $F090W$ vs. $F090W-F150W$ calibration to the results presented in \citet{Anand2024} in \S~\ref{sec:sh0es_comp}. Finally we provide concluding remarks and an outlook for future TRGB calibrations in \S~\ref{sec:conclusion}.

\section{Observations and Data}\label{sec:obs_and_data}
\subsection{The Galaxy Sample}
In Paper I, we selected the 4 galaxies M~81, NGC~253, NGC~300, and NGC~2403 to bracket a range of metallicity and age in order to sample a wide range of stellar properties that impact the brightness of the TRGB. Note that there is observational evidence in the literature that suggests stellar metallicity can vary across the outer stellar disk of a galaxy \citep[e.g.,][]{Ibata2014, Gilbert2014, Cohen2020, Conroy2019, Monachesi2016}. Thus, we designed our observing strategy to account for these potential changes in the stellar content of our calibration sample. In particular, we observed 2 fields per target with \hst\, each with overlapping ACS and WFC3/IR imaging (see Paper I for more details). In Paper I, in order to have a more robust fit to the TRGB slope we found it beneficial to extend the metallicity baseline further and added archival data on more metal-poor galaxies. 

Here, the core sample for our study includes the original 4 galaxies targeted with \hst. Similar to Paper I, we augment the original sample with \jwst\ archival data on metal-poor dwarfs to extend the color baseline. However, the 4 metal-poor dwarf galaxies used to augment the \hst\ calibration did not have archival \jwst\ data available at the time of this study. We instead include archival \jwst\ data of two different metal-poor dwarf galaxies: Sextans A (JWST-GO-01619) and WLM \citep[JWST-ERS-01334]{Weisz2023}. Sextans A and WLM are both metal-poor, low-mass dwarf irregular galaxies that are actively forming stars \citep{Weisz2014, McQuinn2024}. These data are available for different subsets of the full filter selection in our original sample. \jwst\ imaging is available in the NIRCam filters $F090W$, $F150W$, $F277W$, and $F444W$, for Sextans A, and in the NIRCam filters $F090W$ and $F150W$ for WLM. 

\subsection{Observational Setup}
In Paper I, we described the guiding principles for optimal field placement to make robust TRGB measurements. Briefly, the observing strategy targeted the outer stellar fields of galaxies in order to (1) reduce the impact from crowding effects where the apparent brightness of a star can be biased to higher or lower magnitudes in regions of relatively high stellar densities (e.g., in stellar disks); (2) minimize the presence of younger stellar populations (AGB stars) which can act to blur the location of the TRGB; and (3) reduce the internal extinction relative to the disk. The \hst\ program also observed two fields in each target in order to account for variations in the metal content and/or age of stellar populations. 

\begin{figure*}[!th]
\centering
    \includegraphics[width=0.3\textwidth]{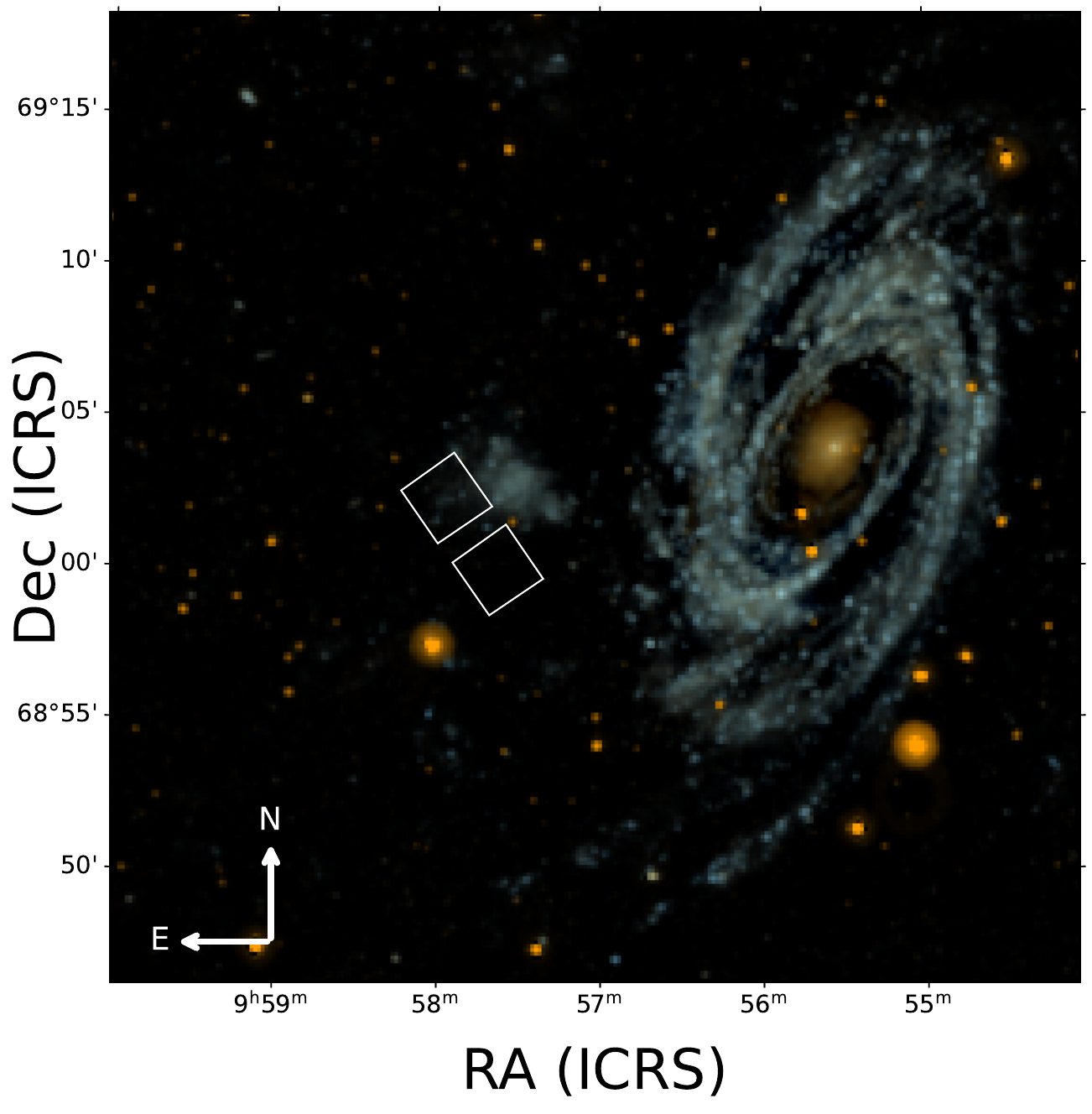}
    \includegraphics[width=0.3\textwidth]{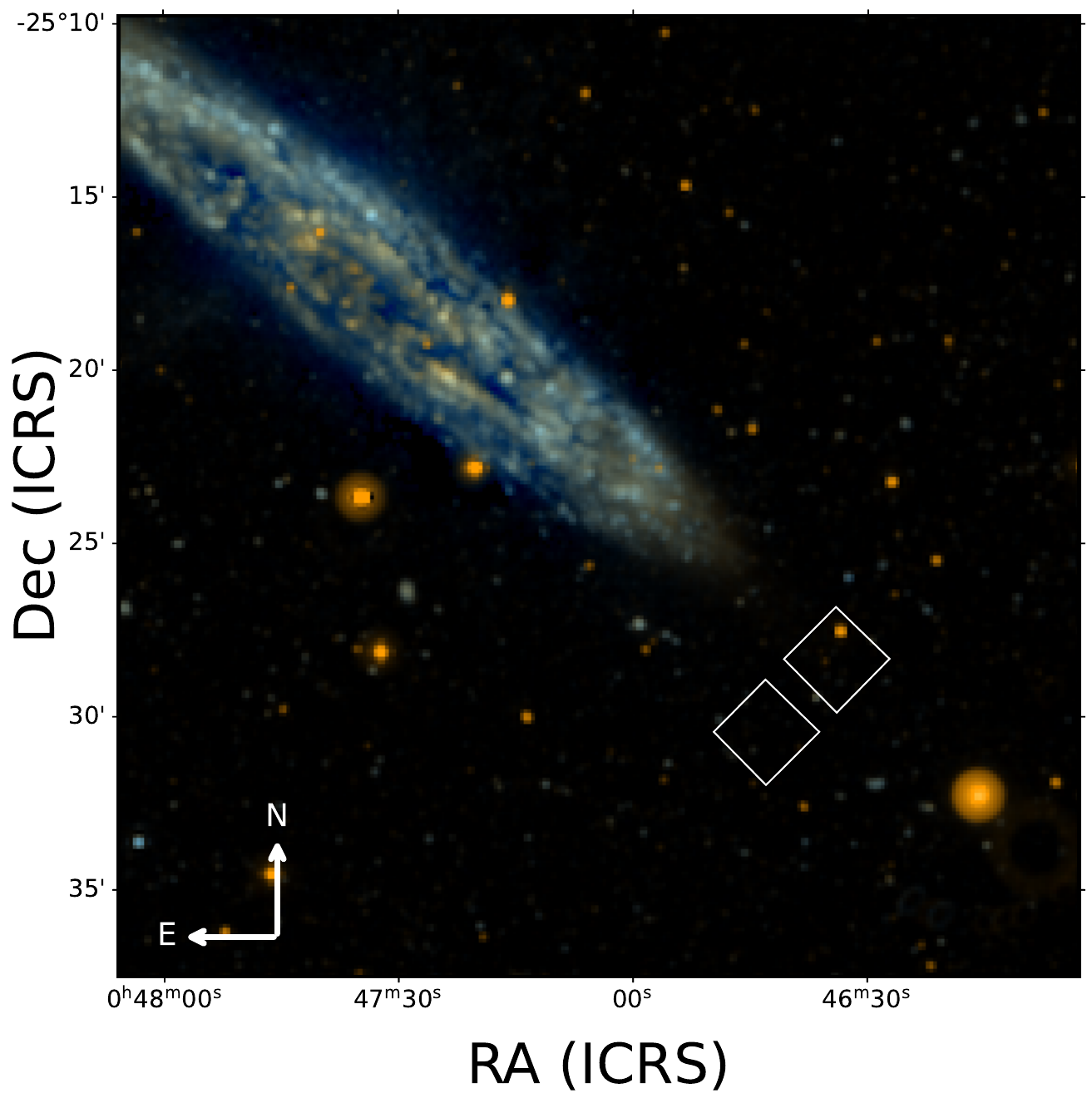}
    \includegraphics[width=0.3\textwidth]{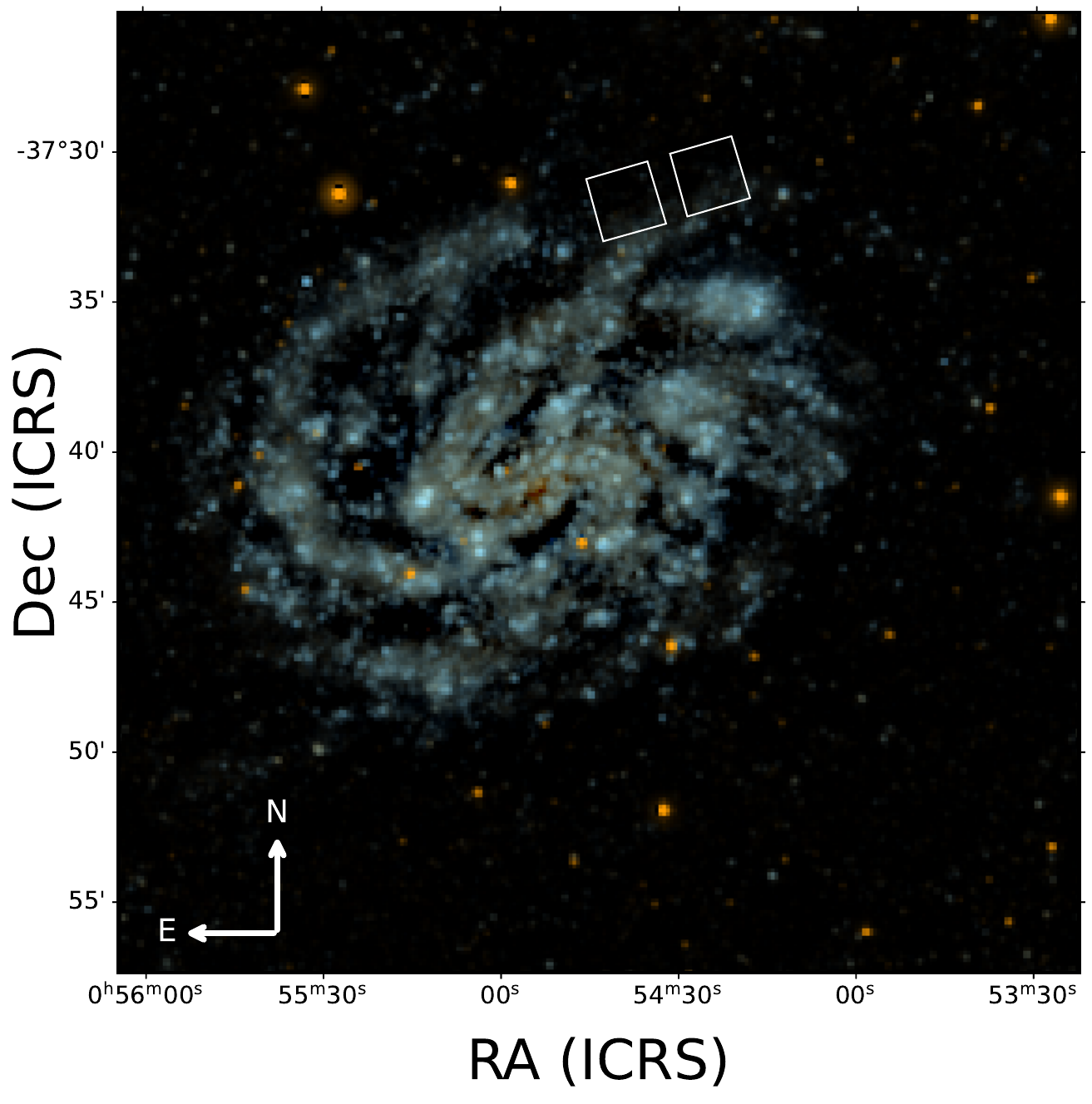}
    \includegraphics[width=0.3\textwidth]{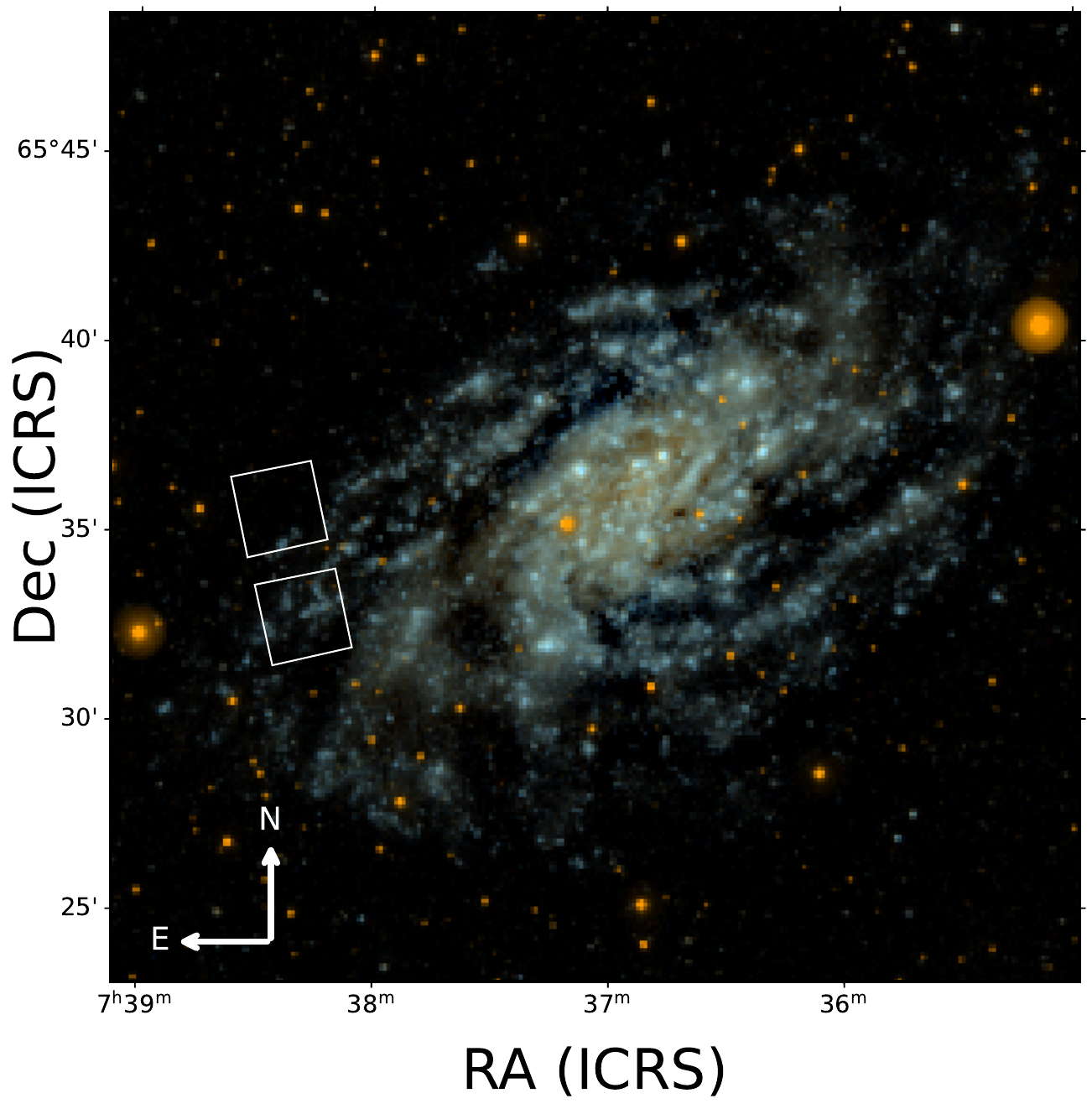}
    \includegraphics[width=0.3\textwidth]{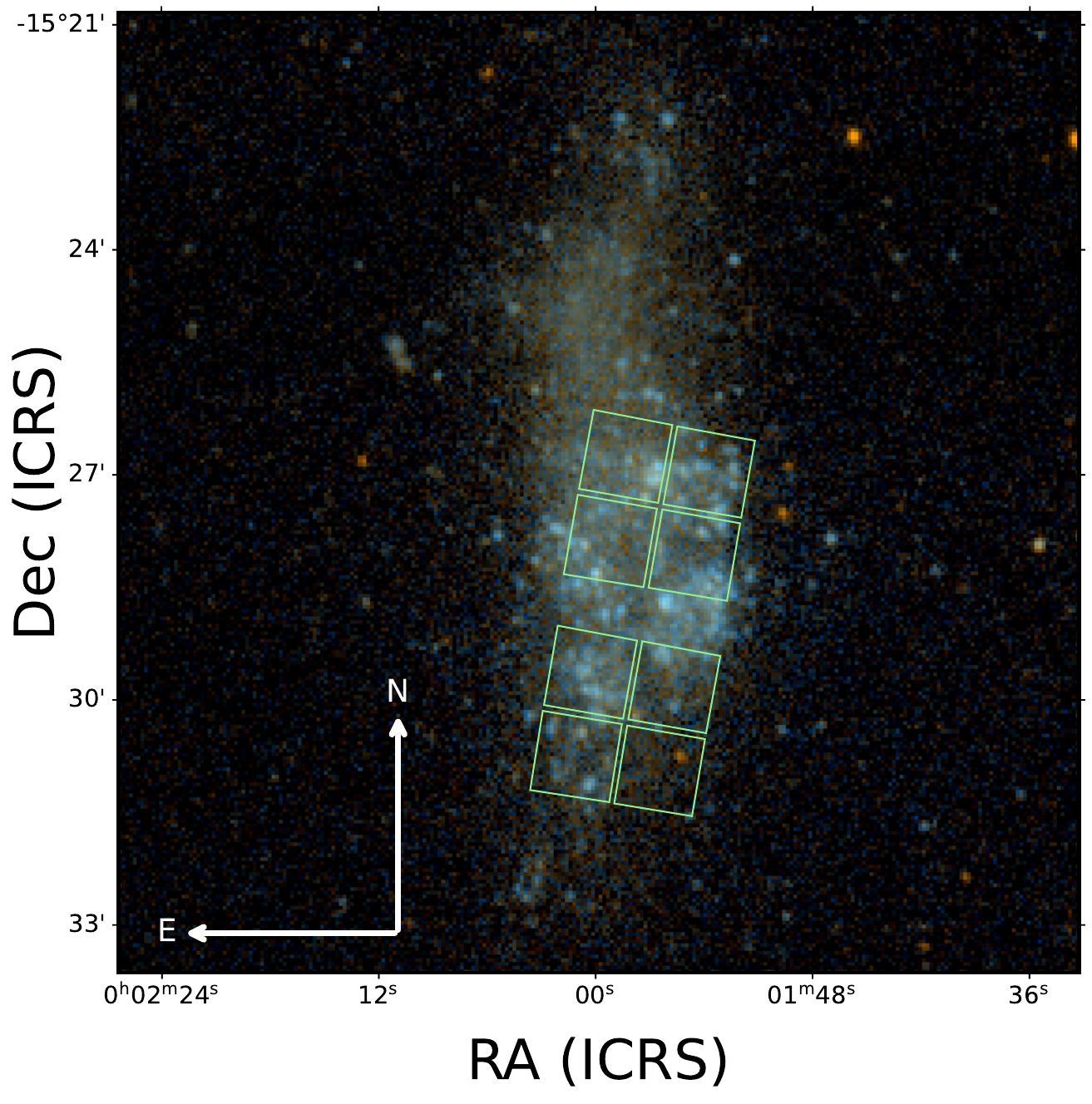}
    \includegraphics[width=0.3\textwidth]{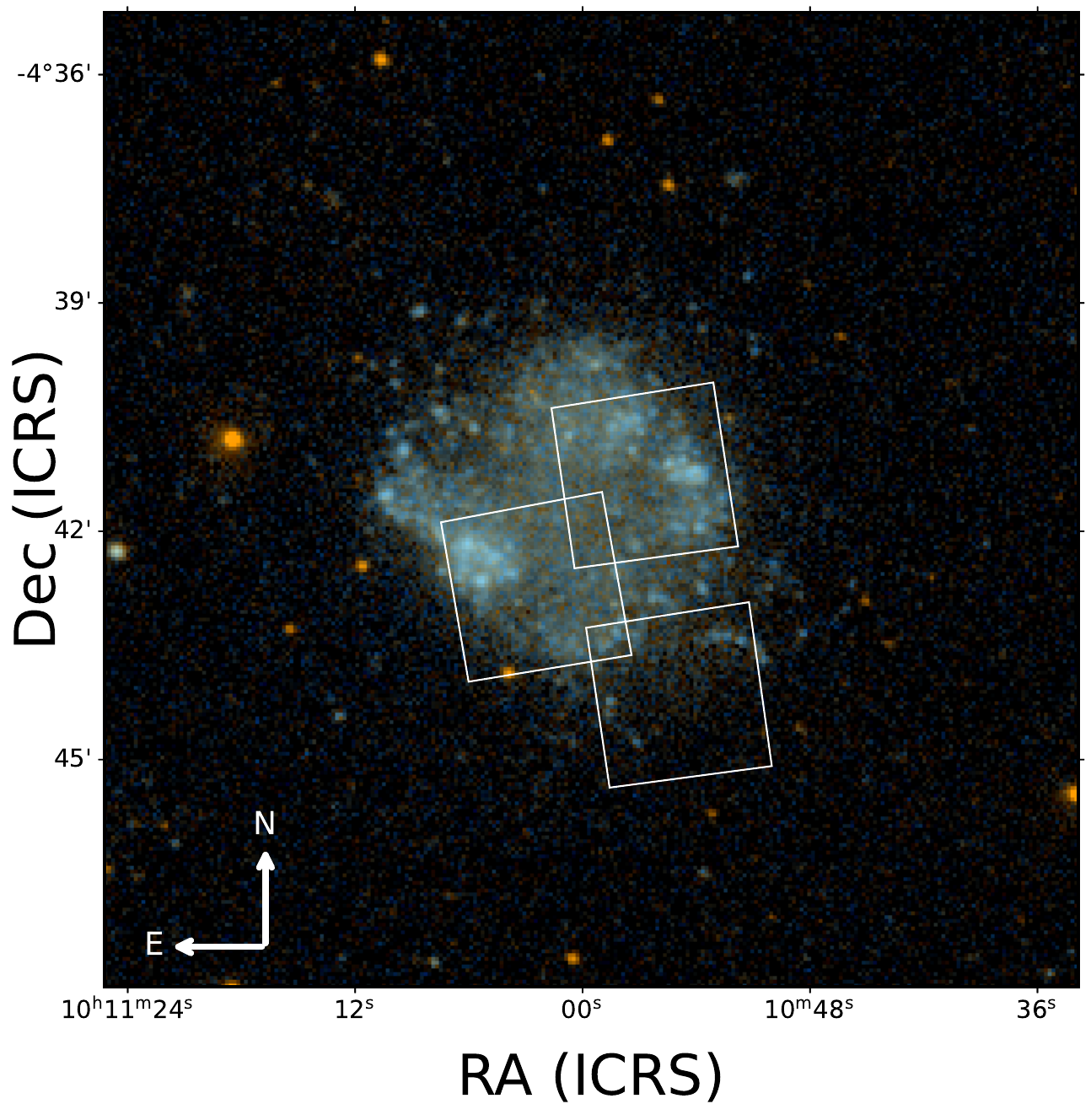}
    \caption{\jwst\ TRGB calibration NIRCam field placements. Top row: M~81, NGC~253, and NGC~300. Bottom row: NGC2403, WLM, and Sextans A. Background 3-color images were created from GALEX NUV (red), NUV+FUV (green), and FUV (blue). The NIRCam LW footprint is show as white regions with the exception of WLM for which only SW imaging was available in the filters appropriate to our calibration. Arrows in the lower left corner of the panels indicate north up east left orientations.}
    \label{fig:FOV}
\end{figure*}

We designed our \jwst\ program to target one \hst\ field with the Near Infrared Camera \citep[NIRCam;][]{Rieke2005, Rieke2003, Beichman2012} and the other field with the Near-Infrared Imager and Slitless Spectrograph \citep[NIRISS;][]{Doyon2023}. We selected the NIRCam observations to overlap with the \hst\ field with the highest stellar density as determined in the target\hst\ FOVs, and we used the relative orientation of NIRISS to NIRCam to place NIRISS on the other \hst\ field. We found that one field per galaxy has sufficient star counts for our TRGB calibration, allowing us to reduce observing times. The exceptions were NGC~300 and M~81 NIRCam module B (see \S~\ref{sec:excluded_data} for details). In Figure~\ref{fig:FOV} we show the position of the NIRCam FOV on each target as regions (excluding WLM for which only SW were available in the filters applicable to this study). Background 3-color Galaxy Evolution Explorer (GALEX) images are created from a combination of far-ultraviolet (FUV; blue filter), near-ultraviolet (NUV; red filter) and an average of the FUV and NUV images (green filter).

Our program includes 6 NIRCam filters that span a wavelength range from $0.9\mu$m -- $4.44\mu$m: NIRCam short wavelength (SW) F090W, F115W, F150W filters; and NIRCam long wavelength (LW) F277W, F356W, and F444W filters.  The new observations analyzed can be accessed via MAST at \dataset[DOI: 10.17909/vknr-3562]{https://doi.org/10.17909/vknr-3562}. The NIRCam SW and LW channels are conveniently set up to allow for simultaneous observing which we take advantage of to observe the same field simultaneously in both channels, significantly reducing overhead. In addition the program includes NIRISS SW observations in the $F090W$, $F115W$, and $F150W$ filters which will be presented in a separate paper.

Our data reach to between $2-4$ mag below the TRGB, depending on the filter. The readout-patterns and dithers to reach this depth were selected for NIRCam as follows: for targets at a distance $D>3$~Mpc we used a readout-pattern of SHALLOW2 or SHALLOW4 depending on the filter. For NGC~300, the closest galaxy, we used SHALLOW2 for all filters.  We also summarize our program and the archival program's observational parameters including total integrated exposure time, groups per integration, dither patterns, and number of dithers in Table~\ref{tab:SampleObservations}.

\begin{table*}
    \footnotesize
    \caption{Galaxy Sample and Summary of the \jwst\ NIRCam Observations}
    \centering
    \label{tab:SampleObservations}
    \begin{tabular}{|lcclccccc|}
    \hline
    Target & R.A. & Decl. & Camera & NIRCam Filter & $t_{exp}$ & Groups/Int. & No. Dithers & Dither Pattern\\
    
           & (J2000) & (J2000) & & &  (seconds) & & & \\
    (1) & (2) & (3) & (4) & (5) & (6) & & & \\
    
    \hline 
    \hline
     M81& 149.4467 &  69.0178  & NIRCam & All Filters & 1031 & 6 & 4  & 4-pt-small-with-niriss\\
     \hline
     NGC~253& 11.66020875 &  -25.4901361  & NIRCam & All Filters & 1031 & 6 & 4 & 4-pt-small-with-niriss\\
     \hline
     NGC~300 & 13.632835 &  -37.520927  & NIRCam & All Filters & 1031  & 6 & 3 & 3-pt-small-with-niriss\\
     \hline
     NGC~2403 & 1114.587072 &  65.5693361 & NIRCam & All Filters & 1031 & 6 & 4 & 4-pt-small-with-niriss \\
     \hline
     WLM & 0.48928958 &  -15.48124722 & NIRCam &  Both Filters & 23449 & 8 & 4 & 4-pt-medium-with-niriss \\
     \hline 
     Sextans A & 152.73610708 &  -4.68750556 & NIRCam &  All Filters & 1246 & 7 & 4 & 4-point-with-miri-2550w \\
     \hline
    \end{tabular}
    \tablecomments{\raggedleft SW filters are $F090W$, $F115W$, and $F150W$; LW filters are $F277W$, $F356W$, and $F444W$. Exceptions are Sextans A which was observed in only $F090W$, $F150W$, $F277W$, and $F444W$, and WLM which was observed in $F090W$ and $F150W$ only.}
\end{table*}

\subsection{Data Processing:}
\subsubsection{\jwst\ NIRCam Image Pipeline}
Individual science exposures obtained from \jwst\ are first processed through a calibration pipeline before these data are ready for photometry.  There are two methods to obtain these data. These data can be retrieved from the MAST directly with the package \texttt{jwst\_mast\_query}\footnote{\url{https://github.com/spacetelescope/jwst_mast_query}} or from the web page. Alternatively, the uncalibrated (uncal.fits) images can be retrieved from MAST or with \texttt{jwst\_mast\_query} and the calibration pipeline can be run on a local machine. We generated our science exposures via the second option to ensure our photometry is based on the most up-to-date calibration reference data system (CRDS) files. Our final images were generated with jwst\_1126.pmap reference which introduced significant updates to the NIRCam flat fields and flux calibrations for both SW and LW filters. Using the latest CRDS reference files significantly improved the quality of these data products and yielded high-fidelity photometric catalogs (see \S~\ref{sec:phot} and \S~\ref{sec:CMDs} for photometry details).

\subsubsection{Photometry}\label{sec:phot}
To generate a robust catalog of stars to calibrate the TRGB across all six filters (3 SWs and 3 LWs) we ran point spread function (PSF)-fitting photometry simultaneously in all filters. We executed simultaneous PSF photometry with the software \Dolphot{}\footnote{\url{http://americano.dolphinsim.com/dolphot/} \\ \url{https://dolphot-JWST.readthedocs.io/en/latest/}}, a modified version of the WFPC2-specific \hst\ phot package that includes instrument-specific support for NIRCam modules \citep{Dolphin2002,Dolphin2016,Weisz2024}. 

\Dolphot{} requires a reference image to identify point sources on which to run photometry. We followed the recommendation from \citet{Weisz2024} and selected the mosaiced NIRCam F090W i2d.fits as the reference image for NIRCam photometry. The F090W filter has the highest angular resolution and therefore provides the best image for \Dolphot{} to perform optimally. 

We ran \Dolphot{} following a multi-step process.  First, we generated a source catalog by running \Dolphot{} on the F090W and F150W filters only. Second, we ran \Dolphot{} in warmstart mode on the SW and LW channels simultaneously. Note, for simplicity, we also ran \Dolphot{} separately for NIRCam modules A and B. Following the recommendation from the \Dolphot{} manual and \citet{Weisz2024}, we ran photometry on cal.fits images which are preferred over crf.fits images as additional image processing may produce sub-optimal photometry. Prior to running photometry we applied two pre-processing steps to the images. First, we ran the tool nircammask (included in \Dolphot{}) which finds and masks the bad pixels. Second, we ran the tool calcsky which creates a sky image and provides \Dolphot{} with an initial guess of the sky background for all pixels within a user specified annulus about a reference pixel. Finally we generate a parameter file, required for \Dolphot{} to run, which includes specific choices on how to fit the sources and sky background. We adopted the values reported in \citet{Weisz2024} where they executed an extensive set of photometry runs to optimize \Dolphot{} parameter values for both SW and LW detectors in NIRCam. We emphasize that our photometry is transformed to the updated \jwst\ NIRCam ``Vega" zeropoint calibrations that use Sirius as the color reference \citep[i.e. Vega-Sirius;][see \S\ref{sec:sh0es_comp} for the impact this choice has on comparisons to studies in the literature]{Rieke2022}.

\subsubsection{Artificial Star Tests}
The ``gold standard" technique for fully characterizing the photometric uncertainties and completeness function --the recovery fraction-- is with artificial star tests \cite[ASTs;][]{Stetson1987, Stetson1988}. ASTs consist of injecting fake stars \emph{with known properties} into every science image and re-running photometry on these fake stars in the exact way the photometry was initially run. There are three considerations we take into account when generating these fakes stars. First, we inject $\sim200\text{,}000$ sources into each image which provides a well characterized noise model and measure of completeness. Second, we adopt a uniform spatial distribution for the sources injected into each image since the stellar distributions in the target FOVs are approximately uniform.

Third, we distribute the fake stars in CMD space and pad both the color and the magnitude range of the observed photometry to ensure we recover an accurate completeness function for the data. We require that the ASTs cover the CMD space considering all filters, which dictates that we generate an input source list with magnitudes that follow realistic spectral energy distributions (SEDs) of stars.Paper I employed the program \Beast{} \citep{Gordon2016} to generate a multi-filter input list (see Paper I for details) with uniform CMD coverage. 
 
Here, we generate our AST input source lists using a flat initial mass function (IMF), a star formation history (SFH) and age-metallicity relation (AMR), and the PARSEC stellar library to generate synthetic photometry. We employ the package \texttt{MATCH} to generate our synthetic CMDs \citep{Dolphin2002, Dolphin2016}. Briefly, the CMDs are generated simultaneously across all 6 NIRCam filters. For each target we generate input colors and magnitudes to cover the full observed CMD space, optimized for the widest separation in filters (e.g., $F090W-F444W$). We then randomly assign spatial positions to each artificial star by drawing uniform xy coordinate pairs from the bounding box of the NIRCam LW modules (e.g., NIRCam A, NIRCam B). Our final input list of artificial stars cover the full FoV of the SW and LW footprints.

\subsubsection{Photometric Catalogs}
We produced high-fidelity stellar catalogs based on \Dolphot{} quality metrics. In particular we used the results from our ASTs to identify the optimal combination of quality cuts.  We focus on the quality metrics: signal-to-noise (SNR), crowding, sharpness, object type, and error flag. The object type is set to 1 following the \Dolphot{} manual since we set the photometry parameter Force1$=$1 which forces all identified objects into classes of 1 (bright star) or 2 (faint star). The error flag, which informs us how well a source was recovered in the image, is set to $\leq2$ in all filters following the recommendation from \citet{Weisz2024}. We set the SNR to $\geq5$ for all filters. We set crowding, a measure of how much brighter a source would be without the presence of nearby or overlapping sources, to $\leq0.15$ in the SW filters and $\leq0.2$ in the LW filters. Finally, we set the sharpness parameter, a measure of he angular extent of the source, compared to the PSF full width half maximum, to $\text{sharp}^2\leq0.01$ for SW and LW filters. 

\subsection{Converting to an Absolute Magnitude Scale}\label{sec:abs_mag}
In Paper I we used a combined stellar catalog based on the photometry from every target to characterize the TRGB across a wide range of colors. The combined catalog was generated by converting all magnitudes from apparent to absolute. To transform the magnitudes we use the uniformly measured distances to each field via the ACS $F814W$ TRGB. The exceptions are WLM and Sextans~A which we address later in this section. 

Briefly, the TRGB in $F814W$ has only a modest dependence on color and is well characterized \citep[e.g., ][]{Jang2017, Hoyt2021}. In Paper I we applied color-based corrections to the apparent magnitudes of the TRGB stars to rectify the TRGB at colors F606W-F814W$>1.5$~mag. We then employed a maximum likelihood (ML)-based approach to measure the magnitude of the TRGB \citep[see Paper I for details on the ML-method; ][]{Makarov2006}. After measuring the TRGB's apparent magnitude, we adopted $M^{\text{TRGB}}_{\text{F814W}}$ from \cite{Freedman2021} to calculate distance moduli for each target. We adopt the same distances measured in Paper I for the same targets in our calibration.

We adopt the distances to the two metal-poor dwarf galaxies WLM and Sextans~A from the literature. The apparent magnitude of the TRGB in WLM was measured using the same ML-based approached as we used in Paper I and the distance modulus was reported in \citet{Albers2019}. We adopt their TRGB measurement and updated the distance modulus to use the ACS $F814W$ zero-point from \citet{Freedman2021}.  Sextans~A has not been observed with the \hst\ ACS in $F814W.$ We instead adopt the TRGB-based distance to Sextans~A measured in \citep{Dolphin2003} using the \hst\ WFPC2 $F555W$ and $F814W$ filters. We summarize the adopted distances to WLM and Sextans~A in Table~\ref{tab:sexa_wlm_f814w_dist}.

\begin{table}[]
    \centering
    \caption{Summary of Distances Measured from \hst\ F814W LF}
    \begin{tabular}{|l|ccc|}
    \hline
         Galaxy & $m^{TRGB}_{F814W}$ & Distance Modulus &  Distance\\
         & (mag) & (mag) & (Mpc) \\
         \hline
         M81$^{a}$ &	23.74$^{+0.02}_{-0.04}$ & 27.79$^{+0.02}_{-0.04}$ & 3.61$^{+0.04}_{-0.07}$ \\
         NGC253$^{a}$ &	23.65$^{+0.01}_{-0.01}$ & 27.70$^{+0.01}_{-0.01}$ & 3.46$^{+0.02}_{-0.02}$ \\
         NGC300$^{a}$ &	22.39$^{+0.14}_{-0.08}$ & 26.44$^{+0.14}_{-0.08}$ & 1.94$^{+0.13}_{-0.08}$ \\
         NGC2403$^{a}$ &	23.38$^{+0.02}_{-0.01}$ & 27.43$^{+0.02}_{-0.01}$ & 3.06$^{+0.03}_{-0.02}$\\
        	 WLM$^{b}$ & 20.91$^{+0.02}_{-0.01}$ & 24.96$^{+0.03}_{-0.02}$ & 0.981$^{+0.01}_{-0.01}$\\
         Sextans A$^{c}$ & 21.76$\pm0.05$ & 25.69$\pm0.12$ & 1.45$\pm0.05$\\
         \hline
    \end{tabular}
    \tablecomments{{\bf$a$: \citet{Newman2024}; $b$: \citet{Albers2019}; $c$: \citet{Dolphin2003}}.}
    \label{tab:sexa_wlm_f814w_dist}
\end{table}

\section{\jwst\ NIRCam CMDs}\label{sec:CMDs}
\subsection{Comparing the NIRCam CMDs}
In this section we present and discuss the different CMD combinations based on our high-fidelity photometric catalogs. Figure~\ref{fig:all_cmds} shows 22 separate NIRCam CMDs from our data in a 5x6 grid. Each row corresponds to one filter on the ordinate axis and increases in wavelength from top to bottom (i.e., $F090W$ to $F444W$). Across a given row the two filters comprising a color increase in wavelength separation from left to right. For example, the top row ordinate is F090W magnitudes and the colors are ordered as: $F090W-F115W$, $F090W-F150W$, $F090W-F277W$, $F090W-F356W$, and $F090W-F444W$. The range of the x-axis is tailored to the color range of the stars.  

CMDs are composite datasets depending on the filters available for a given target (see Table~\ref{tab:SampleObservations} for details). We place all data  on an absolute scale (see \S~\ref{sec:abs_mag}). The breadth seen in color is due to the range of metallicity of the sample (which introduces the gaps in some CMDs where the metallicity range is not fully sampled by the targets). The prominent feature seen in each CMD is the RGB, identified by a dense column of sources extending at minimum $\sim2$ magnitudes from the bright to the faint end. While this RGB structure is similar across the CMDs, the distribution of brightest stars in the RGB -- the TRGB stars -- exhibit variations in the brightness as a function of their color. The TRGB in the $F090W$ (top row) is roughly flat (i.e., at a constant magnitude) in all CMDs. When we look at the TRGB stars in the redder filters the distribution is sloped to varying degrees (see \S~\ref{sec:calibration_method} for details). Note that the apparent ``gap" seen in a handful of these CMDs (e.g., $F090W$/$F277$ vs. $F090W-F277W$) is a result of having only a subset of the targets available in various filter combinations (i.e., the entire range of metallicity/age is not uniformly covered).

Above the RGB there is a well-defined AGB sequence with a diversity of structure in different CMDs. For example, in the CMD  \cmd{F090W}{F115W}{F090W} the AGB extends beyond the color range of the RGB, but is diffuse and extends only $\sim1$~mag above the RGB in $F090W$. When the same stars are shown in \cmd{F150W}{F444W}{F444W} they are offset to redder colors from the bulk RGB distribution and are confined to a narrow range of color. The AGB stars span a range of $\sim2$~mag above the RGB in $F444W$. 

AGB stars can impact the precision of TRGB measurements since they can overlap with RGB stars and therefore make identifying the TRGB more challenging. The significance of this can depend on (a) the total number of AGB stars (or the number of AGB stars relative to stars at the TRGB) present in the observed field, (b) the filters selected to measure the TRGB (e.g., blending of sources in the lower resolution LW NIRCam filters can increase scatter and blur the RGB/AGB sequences further), and (c) the chemical enrichment history of the galaxy. Recently \citet{Boyer2024}  presented a detailed study of thermally-pulsating AGB (TP-AGB) stars in four NIRCam filters ($F090W$, $F150W$, $F250M$, and $F430M$). They demonstrated that while TP-AGB stars are typically brighter than the TRGB, they can appear as RGB interlopers due to thermal pulses, variability and/or strong molecular absorption in the atmosphere \citep[see Figure~3;][]{Boyer2024}. Moreover,  the strength of the effect is wavelength dependent and increases at longer wavelengths (i.e., in their $F250M$ filter). In this work the potential impacts to our TRGB calibration precision due to AGB stars are not a concern. The relative number of AGB to RGB stars in our dataset is low and the TRGB is well-defined.

\begin{figure*}
    \centering
    \includegraphics[width=\textwidth]{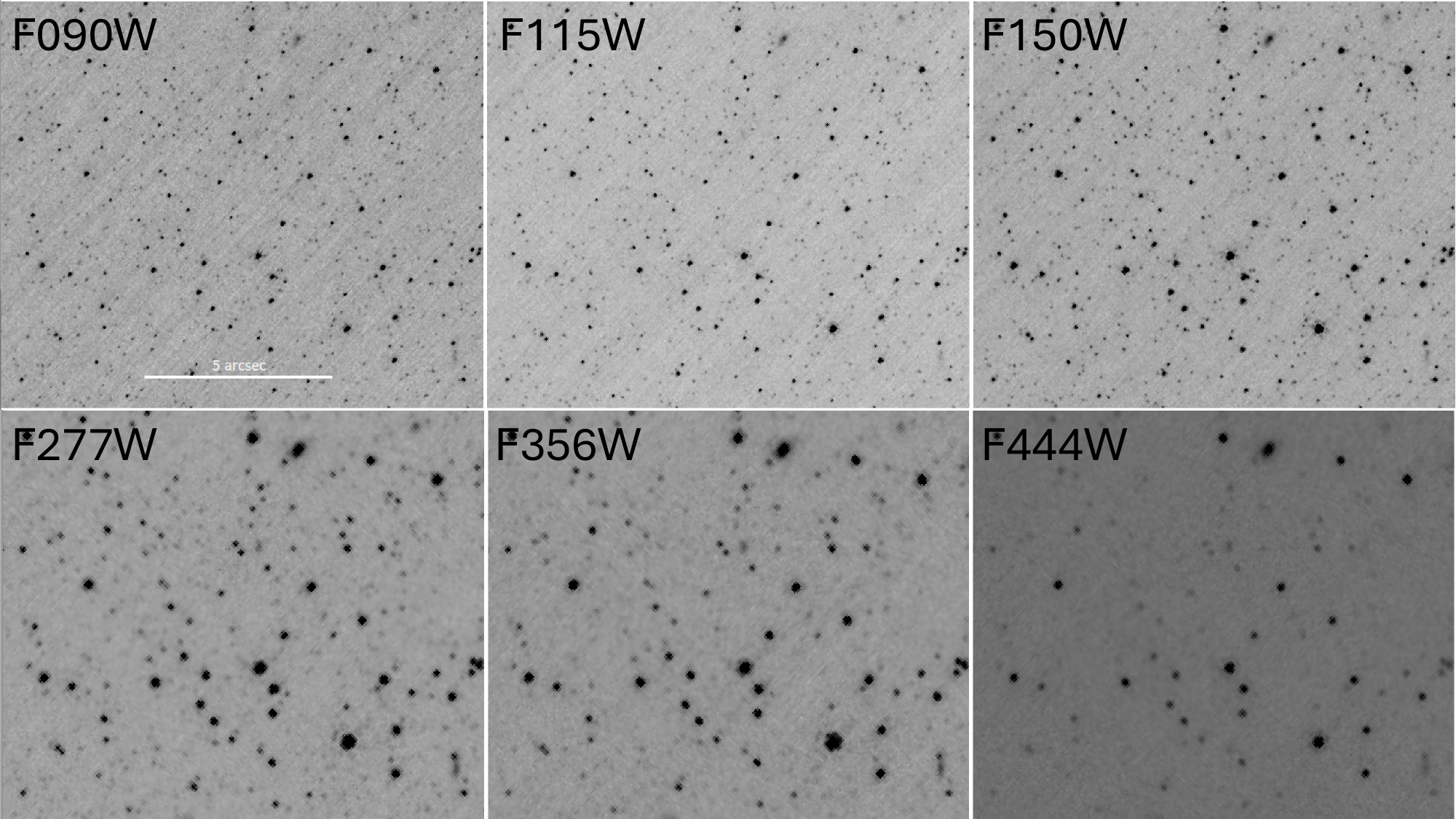}
    \caption{A demonstration of the changing image quality from the SW to LW filters. Each panel is a zoom-in region in the \jwst\ NIRCam module A footprint of NGC~253. The top row shows the SW filters and the bottom row shows the LW filters. Sources in $F090W$ that are well-resolved are fainter and blended in $F444W$ or have entirely disappeared from the image.}
    \label{fig:sw_to_lw_im}
\end{figure*}

\begin{figure*}[ht]
    \centering
    \includegraphics[width= \textwidth]{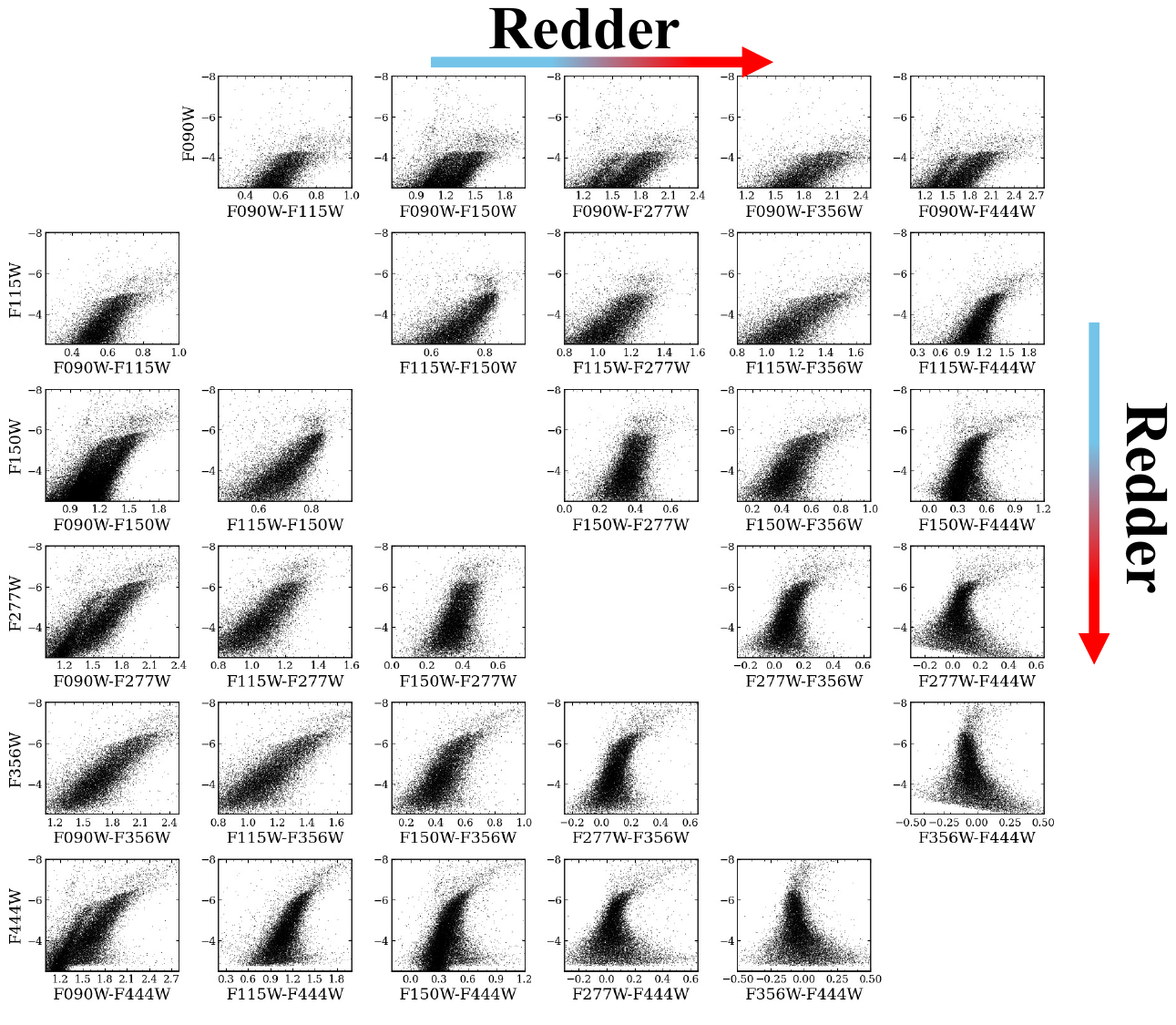}
    \caption{Composite Color-Magnitude Diagrams. Each panel combines the targets with observations available in a set of filters on an absolute magnitude scale (see Table~\ref{tab:SampleObservations} for details of observations). Top to bottom: $F090W$ to $F444W$ (the bluest to the reddest filter in these data). Left to right: narrowest to widest color separation. The flattest and faintest TRGB is observed in the top row ($F090W$) and the most narrow and brightest TRGB is observed in the bottom row ($F444W$). The properties of the AGB, RGB, and HeB stars change as a function of the color. Note that the apparent ``gap" seen in a handful of these CMDs is a result of having only a subset of the targets available in various filter combinations (i.e., the entire range of metallicity/age is not uniformly covered).}
    \label{fig:all_cmds}
\end{figure*}

The depth of the CMDs changes from SW to LW. The stars in the SW CMDs are even fainter in the LW filters, especially in $F444W$. In addition, the lower resolution in the LW relative to the SW filters introduces an increased susceptibility to blending of sources. These compounding effects are highlighted in Figure~\ref{fig:sw_to_lw_im} where we present zoom-in images of a small region in the footprint of NGC~253. In the top row we show the SW filters and in the bottom row we show the LW filters. It is evident that as one moves from the SW to LW end individual sources are fainter and blended or disappear completely. The results are fewer sources recovered in the LW filters, a wider spread in the recovered photometry at the faint end, and a completeness function at which the completeness limit sets in at brighter magnitudes as measured from ASTs.

Finally, we apply foreground-extinction-corrections to these data. We assume a Milky Way reddening slope of $R_V=3.1$ and determine values of $E(B-V)$ along the line-of-sight at each \jwst\ pointing. In Table~\ref{tab:extinction} we provide the all values used to determine the total extinction in a filter, $A_{\text{filter}}$ \citep{Schlegel1998, Schlafly2011}. 

\subsection{Exclusion of NGC~300 and M81 NIRCam B Photometry from the Calibration Data}\label{sec:excluded_data}
While the majority of the new observations yielded high-fidelity data with a sufficient number of stars near the TRGB, none of the NIRCam fields covering NGC~300 had sufficient star counts to contribute to the calibration sample. In Paper I, we found that while the observations of NGC~300 were not as well populated as the other targets there were still sufficiently high star counts to use in the \hst\ calibration. Unfortunately, our \jwst\ field contains even fewer stars than in the \hst\ footprint for NGC~300. We therefore chose to forgo using the NGC~300 data. In addition, we found that the M~81 photometry measured in NIRCam module B contained few RGB stars and primarily introduced contamination from non-RGB stars (e.g., BHeB stars). We only use photometry in NIRCam module A which overlaps with the M81~1 footprint in Paper I. 

\begin{table*}
    \caption{\jwst\ NIRCam Reddening Corrections}
    \label{tab:extinction}
    \centering
    \begin{tabular}{|lc|ccccc|}
    \hline
    Target & $A_{\text{F090W}}$ & $A_{\text{F115W}}$ & $A_{\text{F150W}}$& $A_{\text{F277W}}$& $A_{\text{F356W}}$& $A_{\text{F444W}}$\\
    & ($R_{\text{F090W}}=0.475$) & ($R_{\text{F115W}}=0.319$) & ($R_{\text{F150W}}=0.209$)& ($R_{\text{F277W}}=0.078$)& ($R_{\text{F356W}}=0.052$)& ($R_{\text{F444W}}=0.037$) \\
    \hline
    \hline
    M81  & 0.037 & 0.025 & 0.016 & 0.006 & 0.004 & 0.003 \\
    NGC~253  & 0.009 & 0.006 & 0.004 & 0.002 & 0.001 & 0.001 \\
    NGC~300  & 0.006 & 0.004 & 0.003 & 0.001 & 0.001 &  0.000 \\
    NGC~2403  & 0.019 & 0.013 & 0.008 & 0.003 & 0.002 & 0.002 \\
    Sextans A  & 0.021 & $\ldots$ & 0.009 & 0.003 & $\ldots$ & 0.002 \\
    WLM  & 0.018 & $\ldots$ & 0.008 &$\ldots$ &$\ldots$ & $\ldots$ \\
    \hline
    \end{tabular}
    \tablecomments{\raggedleft We adopt the extinction law slopes for each filter, $R_{\text{filter}}$, from PARSEC \citep{Cardelli1989,Bressan2012}\hfill.}
\end{table*}

\section{Characterizing the Mutliwavelength Behavior of the TRGB}\label{sec:calibration_method}

The brightness of the TRGB is a function of the color of the stars at the TRGB. This color dependence, which traces the age and/or metallicity of the stars, can manifest in the magnitude of the TRGB in two ways: (1) approximately constant at bluer colors and modestly decreasing at redder colors (e.g., in the I band); or (2) an increase from bluer to redder colors (e.g., at NIR wavelengths). For the TRGB to become a precise distance indicator in the \jwst\ NIR filters, its brightness as a function of color must be robustly characterized. In this section we present our methodology for measuring the slope (color dependence) and the zero point (ZP) of the TRGB.  

\subsection{Methodology: Fitting for the Slope and Zero Point of the TRGB}\label{sec:slope_fit_method}
To calibrate the TRGB across many different filter combinations, we adopted the approach detailed in Paper I in which we empirically calibrated the TRGB in each filter combination.  Here we describe our TRGB characterization model. 

We characterize the color dependence of the TRGB by assuming that it is described by a line with a single slope. Our slope finding method is similar to that presented in Paper I for the NIR \hst\ filters in their calibration. The functional form of the slope is captured in the following linear model:
\begin{equation}\label{eqn:rectify}
    m_{\text{NIR,rect}} = m_{\text{NIR}} - \beta\left[\left(\textrm{NIR}_{\textrm{blue}}-\textrm{NIR}_{\textrm{red}}\right)-\gamma\right]
\end{equation}
where $\textrm{NIR}_{\textrm{blue}}$ and $\textrm{NIR}_{\textrm{red}}$ represent the blue and red filters in a CMD, $m_{\text{NIR}}$ ($m_{\text{NIR,rect}}$) is the unrectified (rectified) apparent magnitudes of stars in either the $\textrm{NIR}_{\textrm{blue}}$ or $\textrm{NIR}_{\textrm{red}}$ filter, and $\beta$ and $\gamma$ are the slope and color pivot of the TRGB in a given filter, respectively. We emphasize that in our model the color pivot and the zero-point are directly related. In particular, the absolute magnitude of the TRGB at the color pivot is equivalent to the TRGB zero-point.

The process for measuring the slope in a given CMD is split up into three parts. First, we locate the starting point for $\beta$ and $\gamma$. Second, we iterate to optimize $\beta$. Third, we use Monte Carlo (MC) simulations to find our final value with uncertainties. We discuss each step in greater detail here.

First we determine $\gamma$ and visually identify a fiducial value for $\beta$ for each CMD combination. To measure $\gamma$ we take the mean color of the stars within $0.1-0.4$ mag of the TRGB depending on the change in brightness of the TRGB as a function of the color. We then measure $\beta$ by generating a line on the CMD and changing the parameters until the line sits appromixately along the TRGB. 

Second, we optimize the value of $\beta$ by iteratively changing the fiducial slope. We begin by rectifying the CMD with the initial $\beta$ value, then generate a LF from the rectified CMD magnitudes within the specified TRGB color range using a bin width of 0.01 mag. We then apply a Gaussian-weighted locally estimated scatterplot smoothing (GLOESS) method \citep{Persson2004,Monson2017,Wu2023} to smooth over the LF. GLOESS is a non-parametric weighted least squares algorithm designed to locally fit the data with weighting provided by a Gaussian profile of bandwidth $\tau$. We use a smoothing kernel width of $\tau=0.03$ (see \S~\ref{sec:err_budget} for uncertainties associate with the choice of kernel width). We then apply to the smoothed LF a Sobel filter kernel of width [-1,0,1] convolved with a Poissonian noise weighting scheme \citep[e.g.,][]{Madore2009, Hatt2017}.   

We then iterate on the $\beta$ values. In Paper I, we used a simple bounded local minimization routine and the Nelder-Mead formalism to determine an optimal value for $\beta$. Here we employ a more sophisticated method, \texttt{basinhopping} (within \texttt{Scipy}). This routine is a two-part method: a global stepping algorithm based on a Metropolis-Hastings resampling of the starting $\beta$ value and the local Nelder-Mead minimization algorithm at each step. We use this routine to explore a wider range of $\beta$ parameter space. Each local minimization optimizes $\beta$ by maximizing (the negative of the minimum) the Sobel response at the TRGB magnitude. We assume that the largest Sobel response corresponds to the slope best characterizing the TRGB color dependence. On each iteration we search for the Sobel peak response corresponding to the TRGB magnitude over a range of $\pm0.2$ magnitudes about the expected location of the TRGB. We choose 80 basin-hopping iterations with a step size of $\delta\beta=0.2$ (random displacement from the input $\beta$) and terminate the algorithm after the optimal slope value is the same on 5 separate iterations. 

Third, we measure our best-fit slopes, zero-points, and their associated statistical uncertainties through 1000 MC simulation trials. For each MC trial we repeat the second step in our calibration methodology for resampled colors and magnitudes of the stars. We simultaneously resample the magnitudes based on both the uncertainties measured for the $\mu_{\text{F814W}}$ measurements and the individual photometric uncertainties determined from the ASTs. The AST-based photometric uncertainties are measured by locating the 2000 recovered fake stars closest to each real star in a CMD (based on a Euclidean distance $d=\sqrt{\mathrm{magnitude}^2 + \mathrm{color}^2}$). We then calculate the standard deviation and median (i.e., the bias) of the difference in the 2000 ASTs recovered and input magnitudes (see Paper I for details). We measure best-fit values for zero-points and slopes at the $50$th-percentile of the MC output. We measure the positive and negative statistical uncertainties based on the $16$th- and $84$th-percentile (see \S~\ref{sec:results} for further discussion of the fits). 

\begin{figure} 
 \centering
\includegraphics[width=0.45\textwidth]{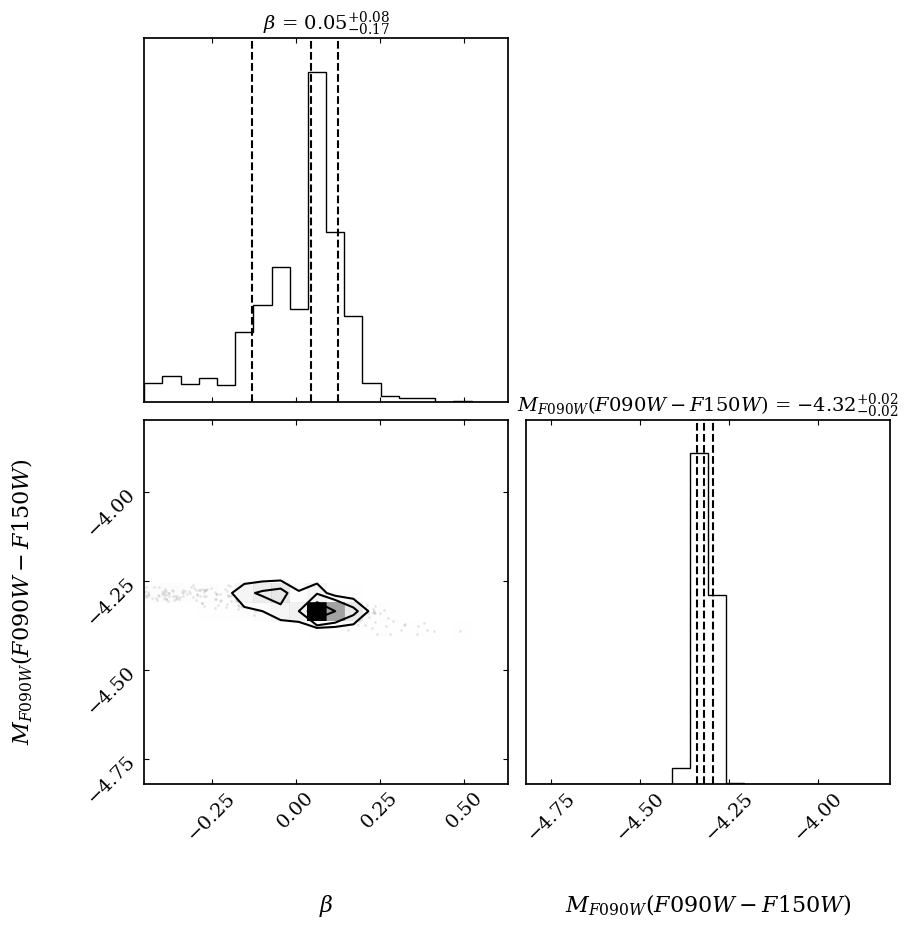}
\caption{A representative corner plot based on the results of the slope and zero-point MC simulations. The top left panel shows the 1d histogram for the slope fits. The bottom right panel shows the 1d histogram for the zero-point fits. The bottom left panel shows the 2d histogram for the simultaneous slope and zero-point MC simulation.}
\label{fig:corner_plot_f090w}
\end{figure}

In Figure~\ref{fig:corner_plot_f090w} we show a representative corner plot based on the MC trials for $F090W$ vs $F090W-F150W$. The bottom left panel shows the joint probability distribution for slope fits (top left) and zero-point fits (bottom right). We present all 18 corner plots in Appendix~\ref{sec:all_corner_plots}.

\section{Results}\label{sec:results}
As discussed and demonstrated, the TRGB luminosity changes as a function of the color of the stars. However, the trend with color is not the same across the many possible combinations of filters. In both the \hst\ F814W to the \jwst\ F090W the TRGB luminosity is close to constant with a decreasing slope toward redder colors\footnote{For a more detailed explanation see e.g., \citet{Jang2017}; \citet{Newman2024}}. In the NIR, the TRGB luminosity is characterized as increasing towards redder colors. The choice of filters in which to apply the TRGB method must be carefully considered and the advantages and disadvantages discussed. In this section we present 18 different combinations and provide detailed discussion on the use cases for each combination. We also provide a summary of the calibrations including the zero-points, slopes, color pivots, and their associated uncertainties in Table~\ref{tab:all_trgb_calibs}. 

\subsection{TRGB Calibration in $F090W$: The Workhorse Filter}
For precision distance work minimal corrections to standardize the TRGB luminosity are favorable. The TRGB in NIRCam $F090W$, much like in the \hst\  $F814W$ filter, is nearly constant across a range of colors. In addition, NIRCam $F090W$ has the highest angular resolution of the filters selected for our calibration and can therefore be used to minimize the impacts of photometric crowding and blending relative to other filters. 

\begin{figure*}[]
    \centering
    \includegraphics[width=0.645\textwidth]{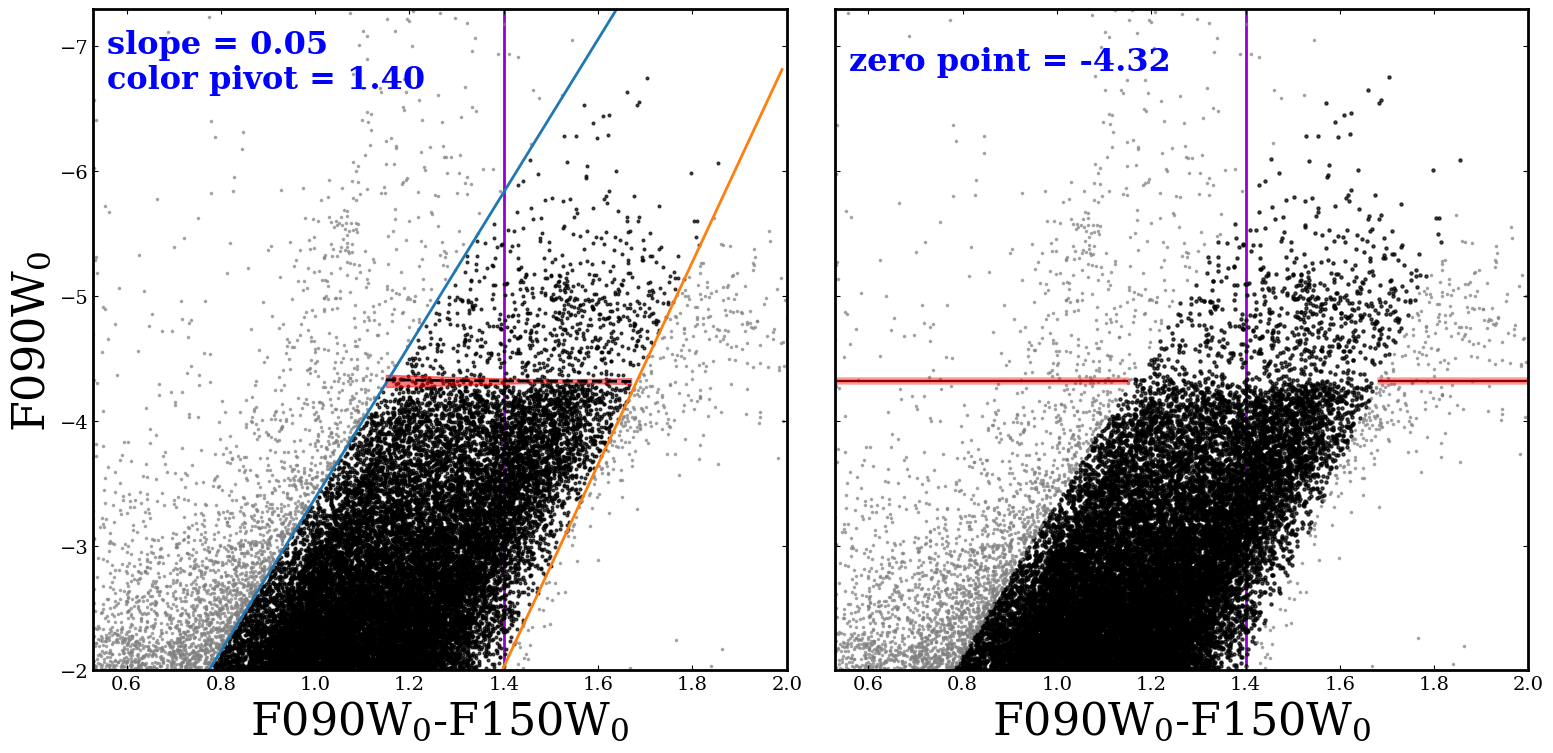}
    \includegraphics[width=0.645\textwidth]{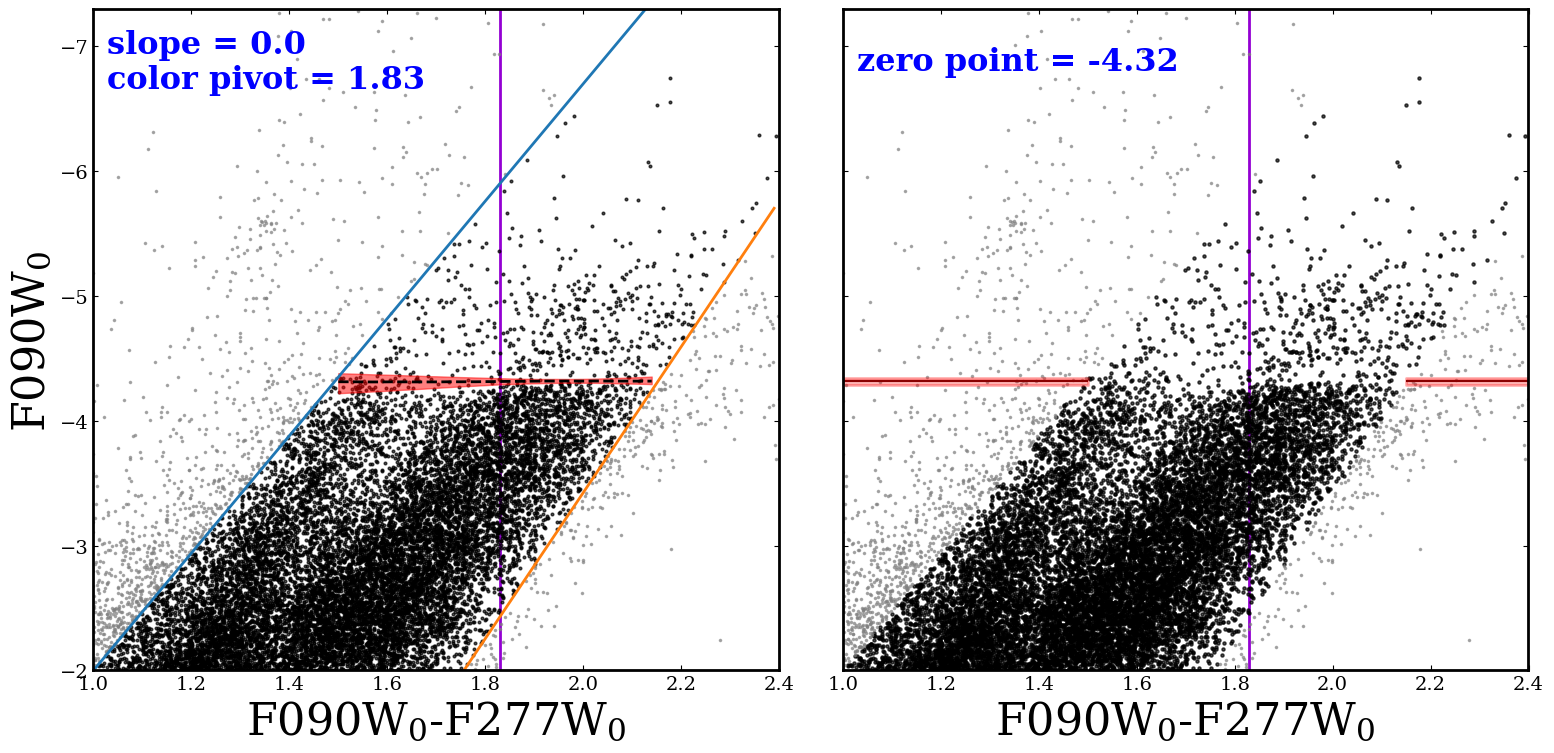}
    \includegraphics[width=0.645\textwidth]{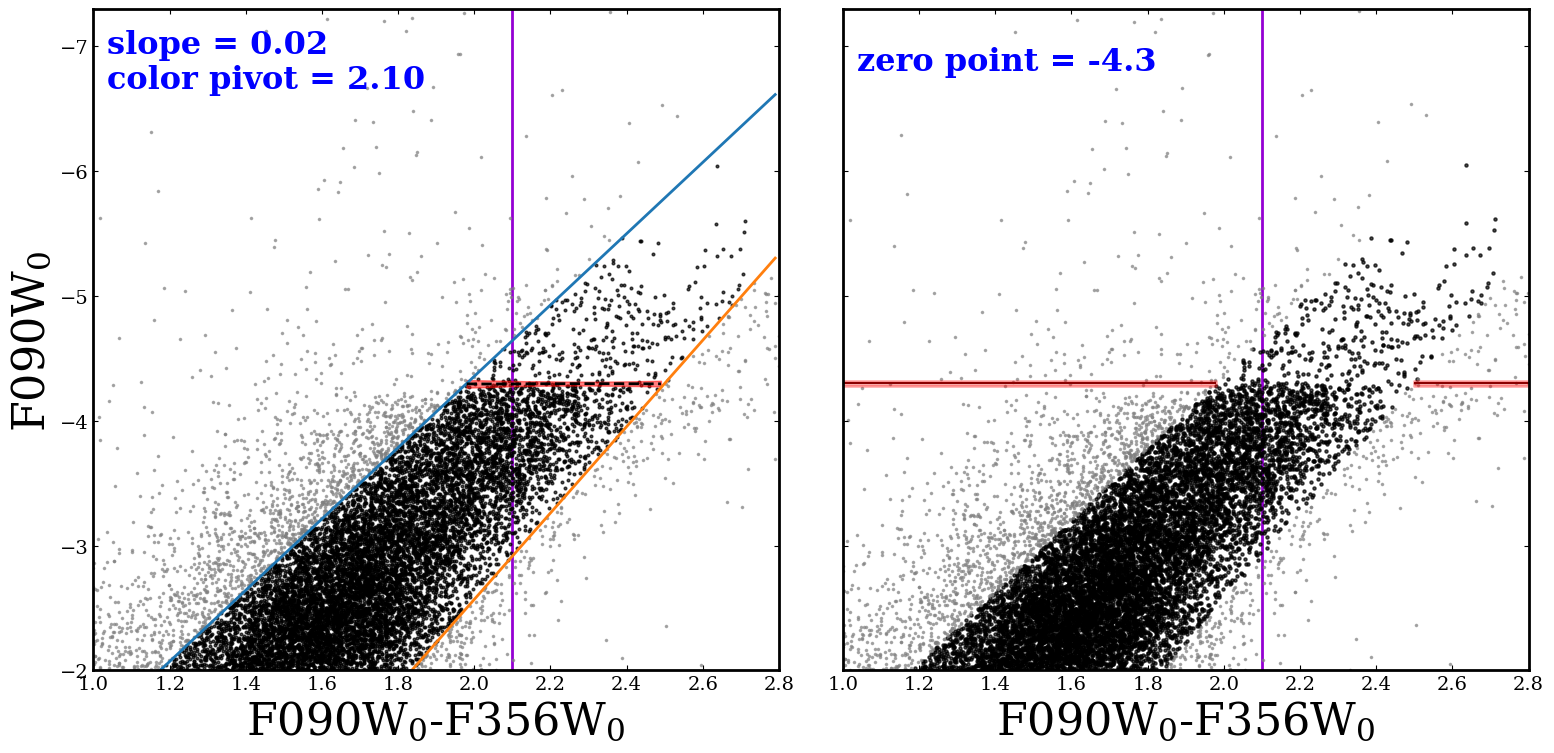}
    \includegraphics[width=0.645\textwidth]{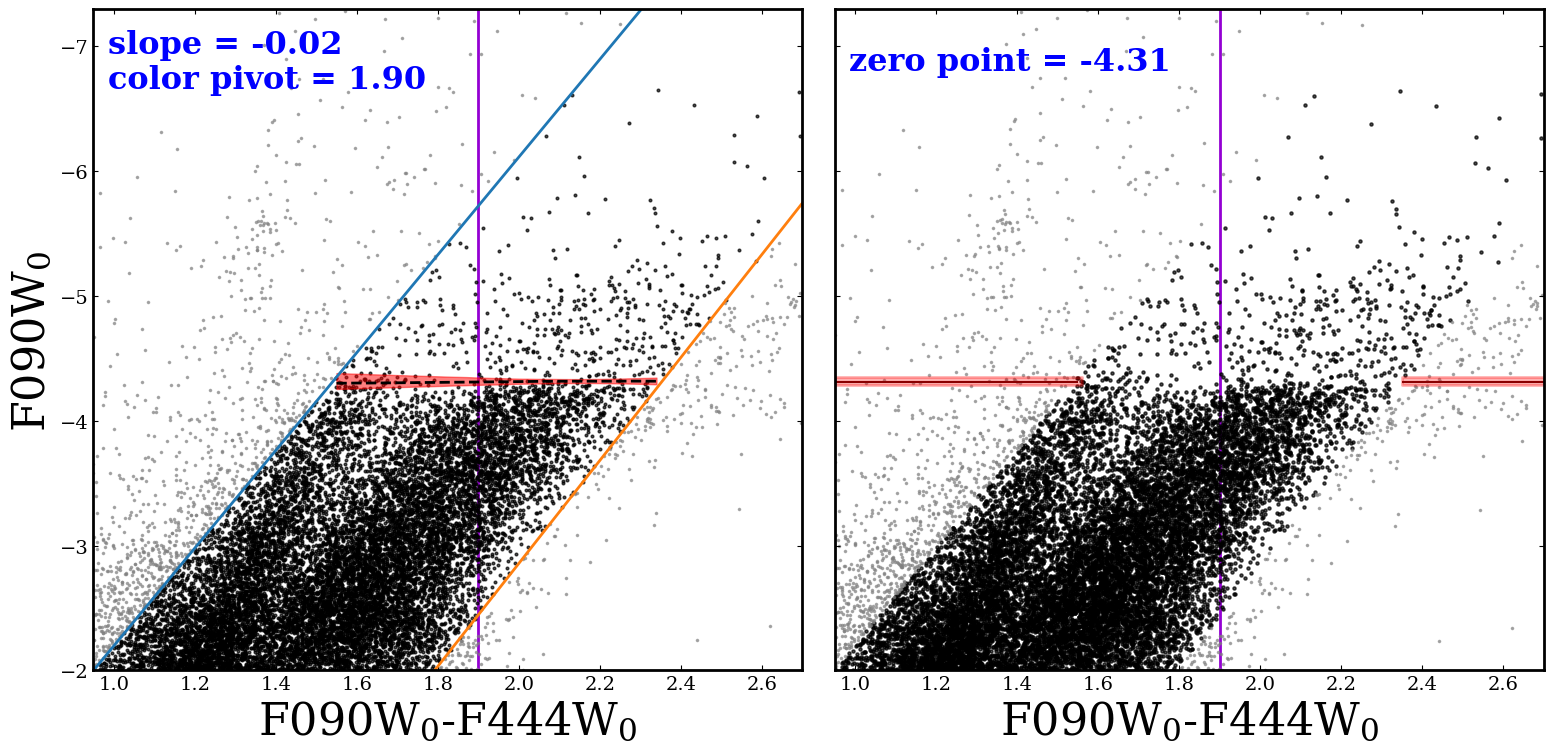}
    \caption{The NIRCam $F090W$ filter TRGB calibration solutions. We show in each of the four 2-panel figures the un-rectified (left) and rectified (right) CMDs. We calibrate the F090W TRGB for four different colors. From left to right, top to bottom are $F090W$ vs. $F090W-F150W$, $F090W-F277W$, $F090W-F356W$, and $F090W-F444W$. Overlayed in the un-rectified CMD panels are RGB selection criteria (blue and orange lines). The stars passing (excluded) from (by) our RGB selection are shown as solid black (solid grey) points. Best-fit slopes and zero-points are shown in the left and right panels, respectively. Uncertainties on the fits are shown as red shading. The color pivot point is shown as as a solid purple vertical line. We provide full calibration equations for each combination.}
    \label{fig:f090w}
    
\end{figure*}

When combined with NIRCam $F150W$ (which also has a high angular resolution the TRGB stars in our calibration extend over a wide color range $1.20\leq F090W-F150W\leq1.7$. In the top left panel of Figure~\ref{fig:f090w} we present the $F090W$ vs. $F090W-F150W$ calibration in absolute magnitudes. The left panel shows the original uncorrected magnitudes with the best-fit slope solution and MC-based uncertainties (dashed black line, shaded red region; see \S\ref{sec:slope_fit_method} for methodological details). The slope value is small and implies there is only a modest color dependence on the TRGB brightness. The right panel shows rectified $F090W$ magnitudes and the best-fit zero-point with uncertainties (solid black line, shaded red region). The full color-based correction equation for $F090W$ vs $F090W-F150W$ is:
\begin{equation}\label{eqn:f090w_f090w-f150w} 
 \resizebox{0.9\linewidth}{!}{$
\trgbcalfull{F090W}{-4.32}{0.05}{F090W}{F150W}{1.40}\\
$}
\end{equation}

We also provide three more calibrations of the $F090W$ TRGB for three additional colors: $F090W-F277W$, $F090W-F356W$, and $F090W-F444W$. The calibrations are shown in Figure~\ref{fig:f090w} clockwise from the top right and the corresponding color-based corrections are as follows:

\begin{equation}\label{eqn:f090w_f090w-f277w} 
 \resizebox{0.9\linewidth}{!}{$
\trgbcalfull{F090W}{-4.32}{0.0}{F090W}{F277W}{1.83}\\
$}
\end{equation}
\begin{equation}\label{eqn:f090w_f090w-f356w} 
 \resizebox{0.9\linewidth}{!}{$
\trgbcalfull{F090W}{-4.3}{0.02}{F090W}{F356W}{2.10}\\
$}
\end{equation}
\begin{equation}\label{eqn:f090w_f090w-f444w} 
 \resizebox{0.9\linewidth}{!}{$
\trgbcalfull{F090W}{-4.31}{-0.02}{F090W}{F444W}{1.90}\\
$}
\end{equation}

While the $F090W$ zero-points agree within their uncertainties, there  are differences in the best-fit $F090W$ zero-points. The $F090W$+$F150W$ calibration (Eqn~\ref{eqn:f090w_f090w-f150w}) is based on the data from all targets included in the analysis. The other three $F090W$ calibrations (Equations~\ref{eqn:f090w_f090w-f277w} to~\ref{eqn:f090w_f090w-f444w}) are derived from a subset of the targets all which exclude the metal-poor target WLM. As a result the color pivot shifts to redder colors where the TRGB luminosity is expected to be fainter and therefore the zero-points for F090W are slightly different depending on the filter with which it is paired. We omit the uncertainties in Equations~\ref{eqn:f090w_f090w-f150w} to \ref{eqn:f090w_f090w-f444w} for clarity but report them in Table~\ref{tab:all_trgb_calibs}.

We also note how the best-fit slopes vary with the filter pairs. All combinations demonstrate a remarkable flatness (e.g., minimal color dependence), and moreover are all statistically consistent with having no slope (i.e., $\beta=0$). 
\subsection{The TRGB in the $F115W$ Filter}
While the TRGB in $F090W$ demonstrates its strengths in the ease of standardization via minimal color-based magnitude corrections and reduced impacts due to crowding, the TRGB is the faintest in $F090W$ compared with other NIRCam filter considered here and is less than what can be achieved when observing at longer wavelengths. The $F115W$ filter stands out as one of the more robust filters for applications of the TRGB-based distance method. Like $F090W,$ $F115W$ has a high angular resolution. However, the TRGB absolute magnitude is $\sim0.7$ magnitudes brighter than in the $F090W$ filter. In addition, the $F115W$ filter's PSF sampling and spatial resolution sits between that of F090W and $F150W$ and therefore enables precision distance measurements at greater distances without substantial losses compared to $F090W$ image, and therefore photometric, quality. 

\begin{figure*}[]
    \centering
    \includegraphics[width=0.8\textwidth]{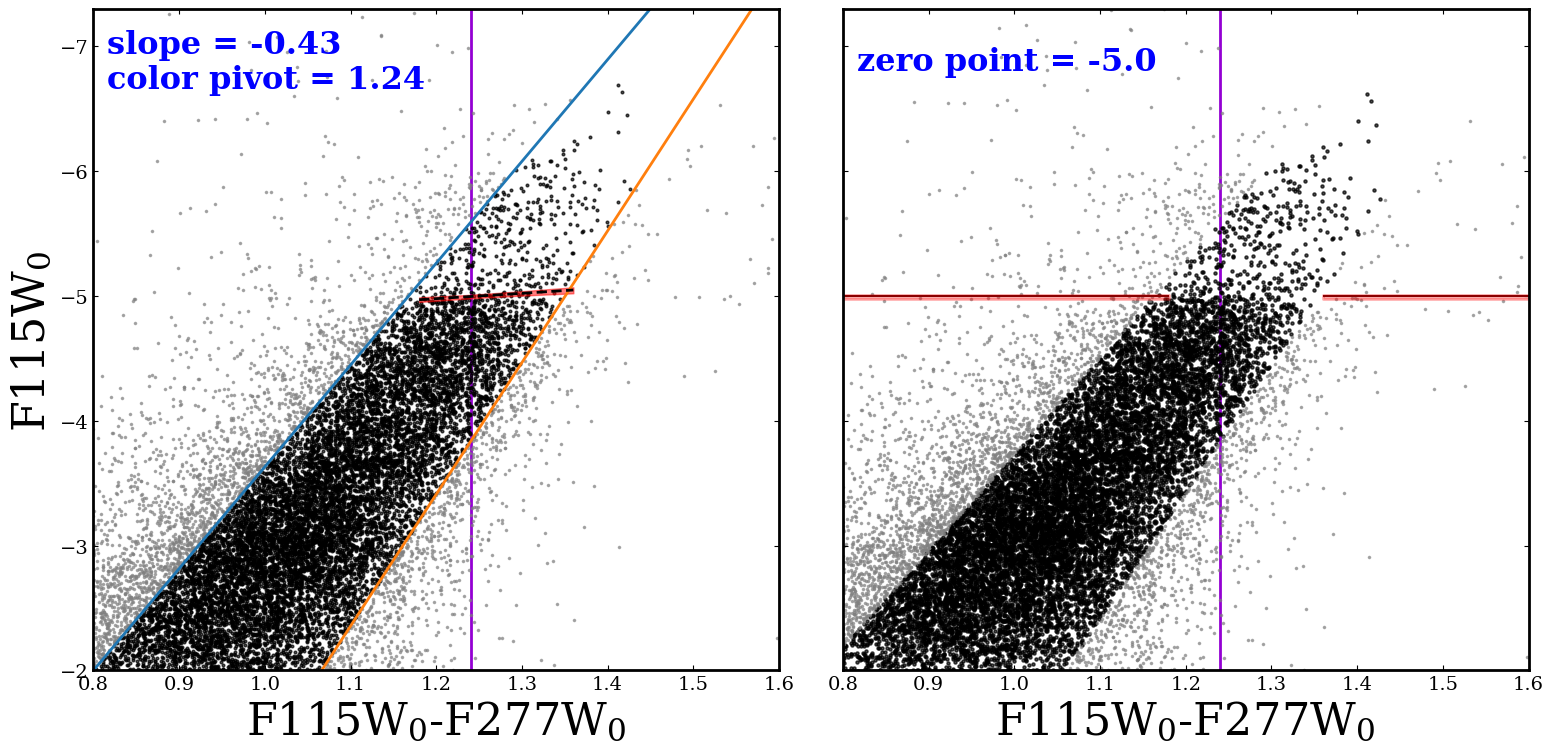}
    \includegraphics[width=0.8\textwidth]{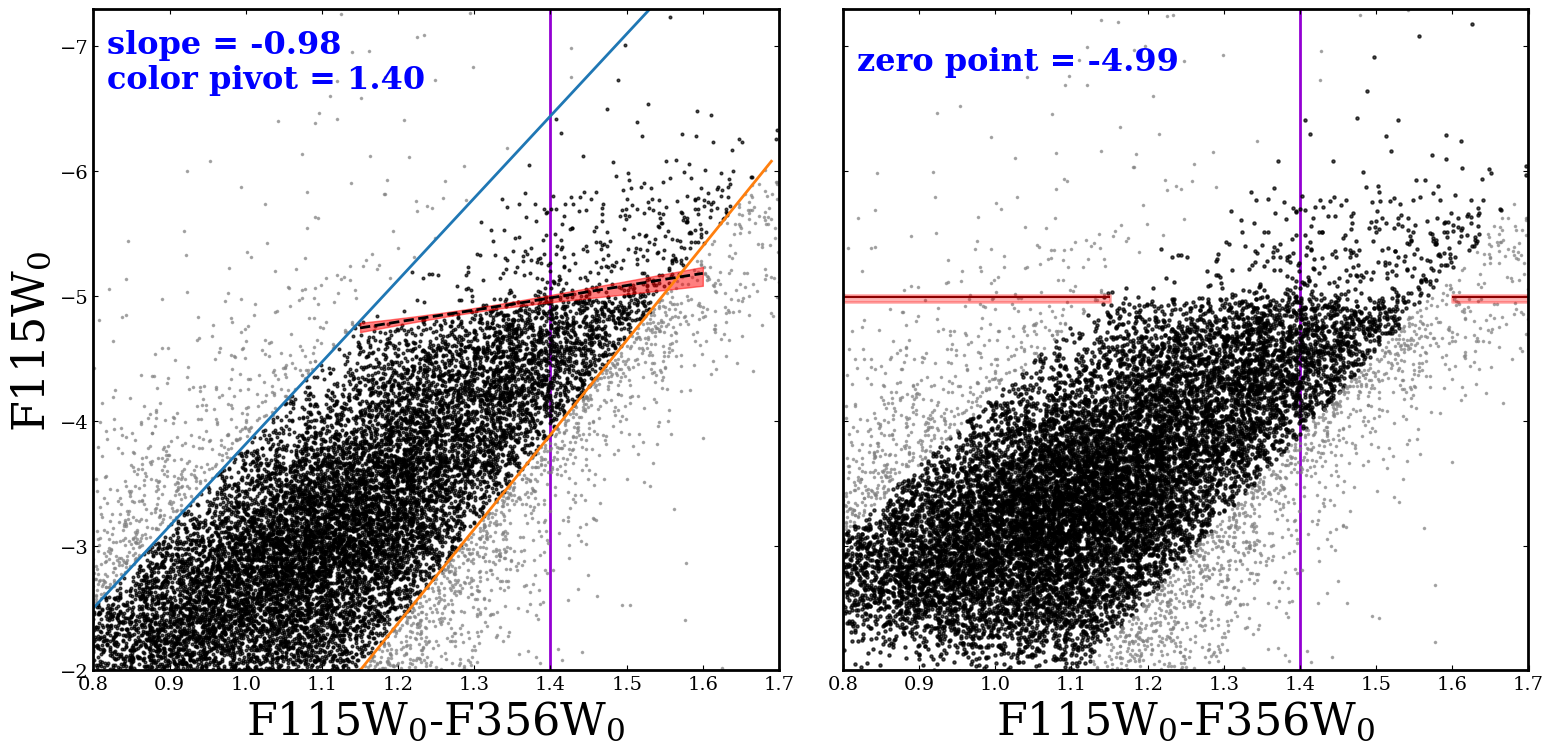}
    \includegraphics[width=0.8\textwidth]{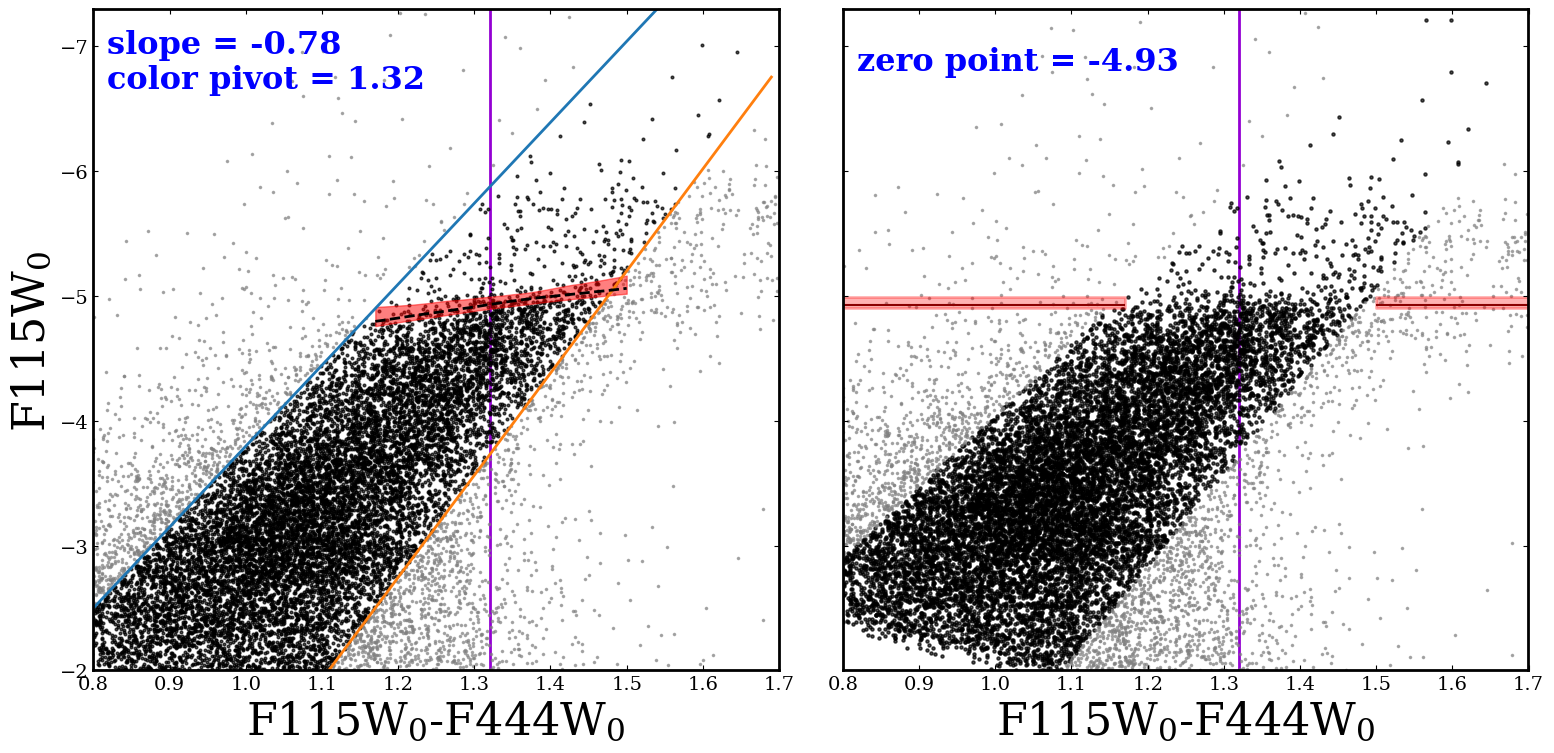}
    \caption{The NIRCam $F115W$ filter TRGB calibration solutions. The three panels are ordered from left to right by increasing wavelength separation in the color baseline: $F115W-F277,$ $F115W-F356W,$ and $F115W-F444W.$ See Figure~\ref{fig:f090w} for descriptions of the sub-panels.}
    \label{fig:f115w}
\end{figure*}

In Figure~\ref{fig:f115w} we present three $F115W$ TRGB calibrations. The top left, top right, and bottom panels show the $F115W-F277W$, $F115W-F356W$, and $F115W-F444W$, respectively. Each panel follows the same format as described in Figure~\ref{fig:f090w}. The color-based TRGB corrections are:
\begin{equation}\label{eqn:f115w_f115w-f277w} 
 \resizebox{0.88\linewidth}{!}{$
\trgbcalfull{F115W}{-5.0}{-0.43}{F115W}{F277W}{1.24}\\
$}
\end{equation}
\begin{equation}\label{eqn:f115w_f115w-f356w} 
 \resizebox{0.88\linewidth}{!}{$
\trgbcalfull{F115W}{-4.99}{-0.98}{F115W}{F356W}{1.40}\\
$}
\end{equation}
\begin{equation}\label{eqn:f115w_f115w-f444w} 
 \resizebox{0.88\linewidth}{!}{$
\trgbcalfull{F115W}{-4.93}{-0.78}{F115W}{F444W}{1.32}\\
$}
\end{equation}
The difference in the color pivot points between Equations~\ref{eqn:f115w_f115w-f277w} to \ref{eqn:f115w_f115w-f444w} is due to the shifts in the colors of stars. Each filter samples the SED of a star in a specific wavelength range and the flux of the star within the filter changes.

\subsection{The TRGB in the $F150W$ Filter}
\begin{figure*}
\centering
\includegraphics[width=0.8\textwidth]{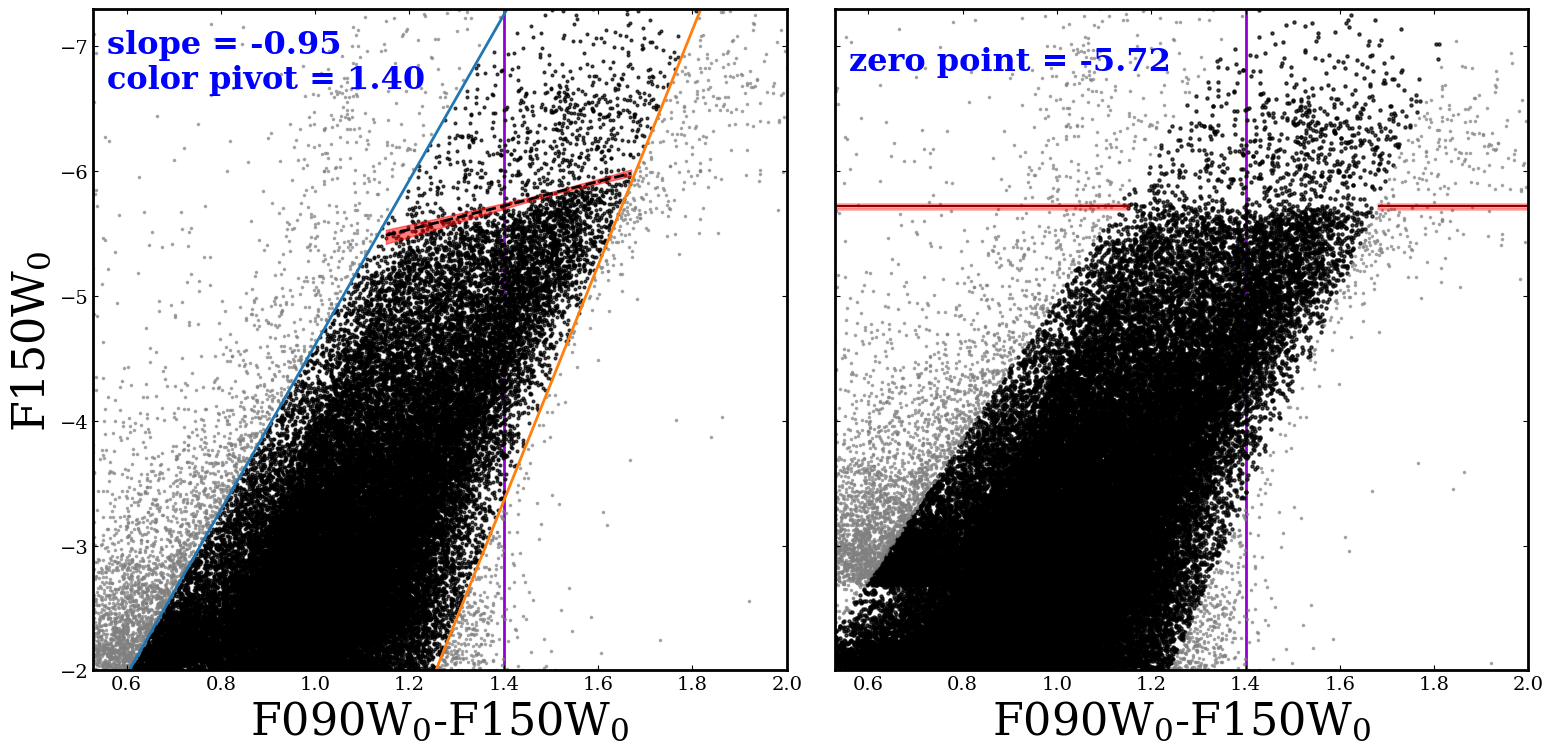}
\includegraphics[width=0.8\textwidth]{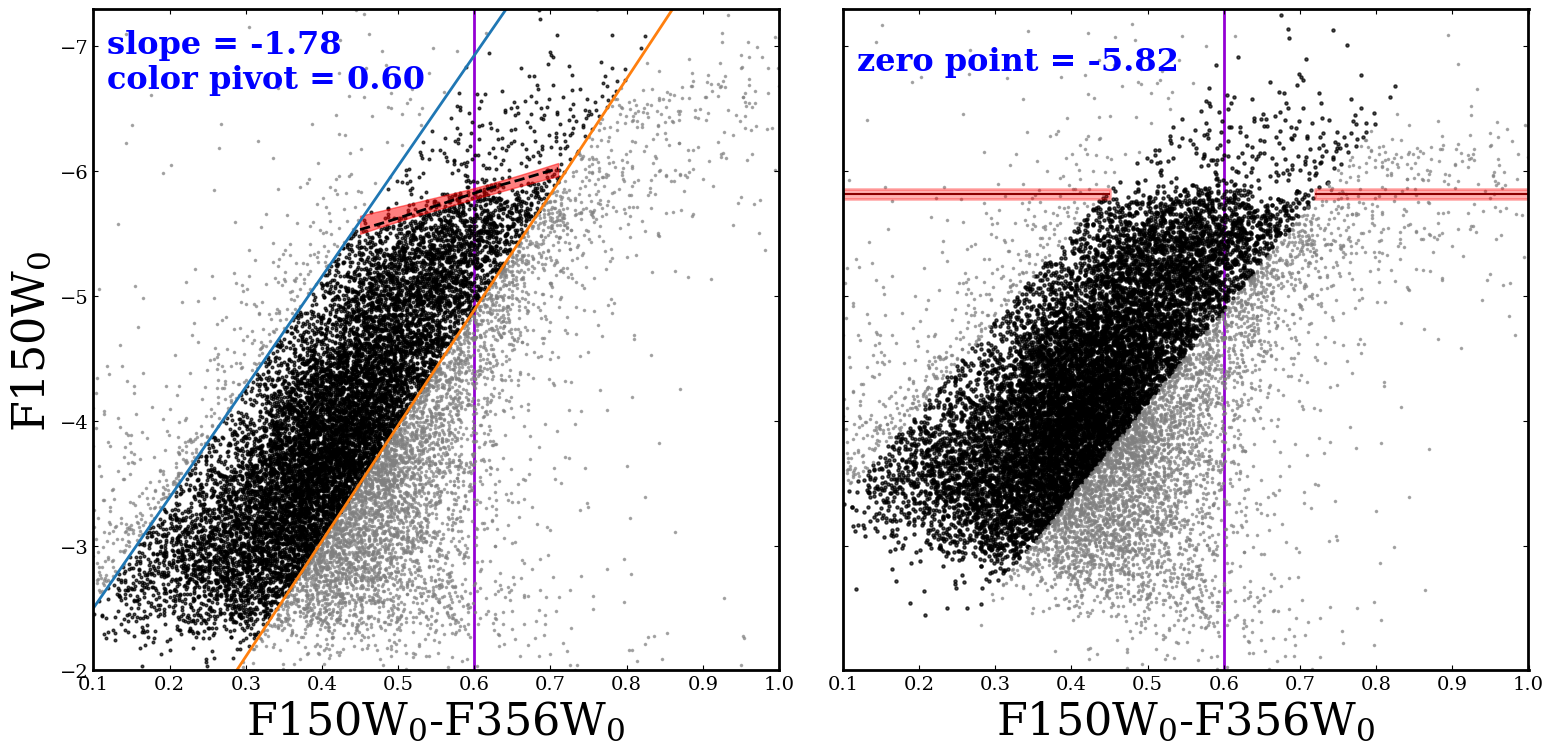}
\includegraphics[width=0.8\textwidth]{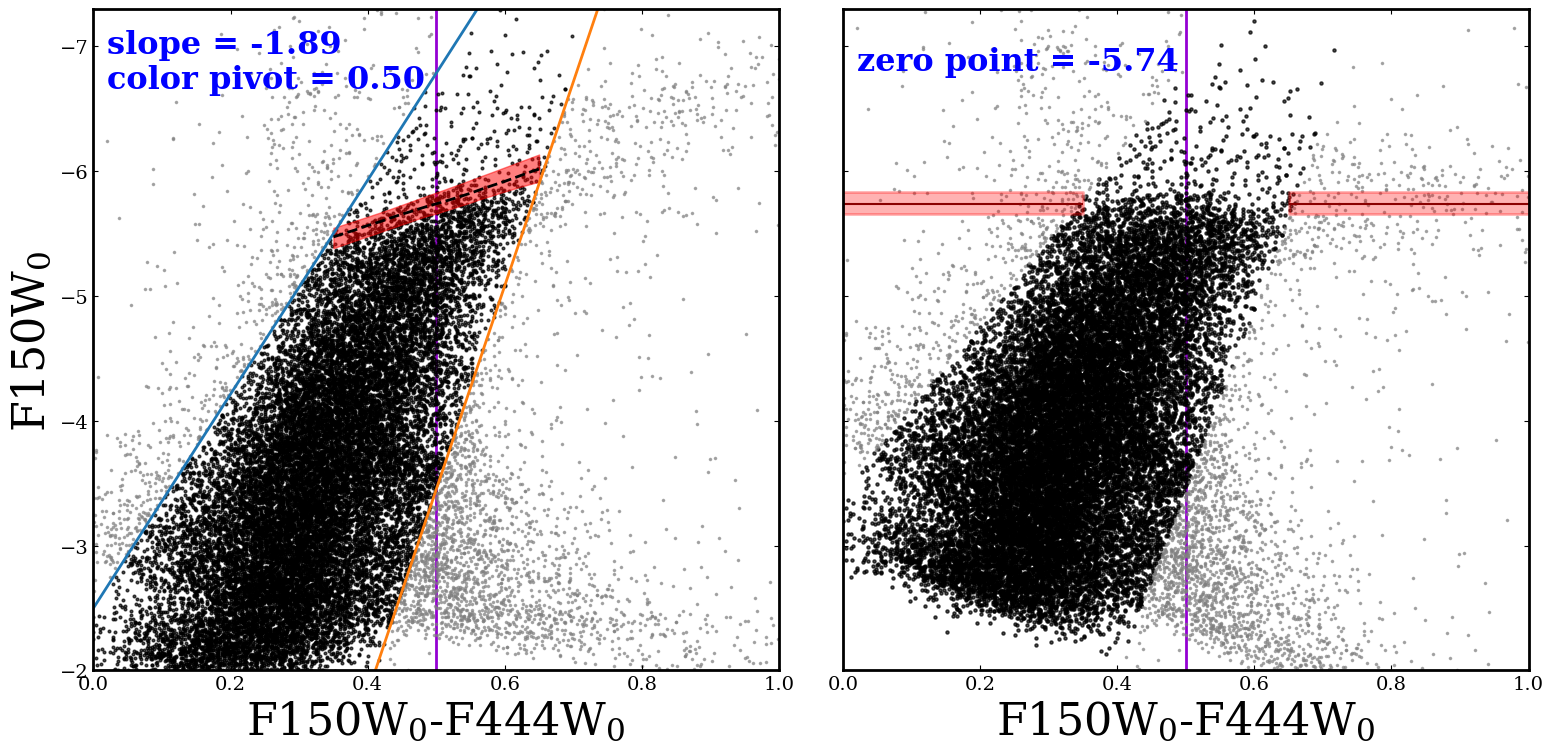}
\caption{The NIRCam $F150W$ filter TRGB calibration solutions. See the caption in Figure~\ref{fig:f115w} for a description of the ordering and content in the panels.}
\label{fig:f150w}
\end{figure*}

 At still redder wavelengths, the TRGB appears brighter and its distance reach is extended further over what is possible with the $F115W$ filter. The panels of Figure~\ref{fig:f150w} are ordered similarly to those in Figure~\ref{fig:f115w}. The un-rectified $F150W$ magnitudes demonstrate that, similar to the $F115W$ filter, the TRGB luminosity is not as constant as a function of color, but increases linearly at redder colors. Applying the slope finding algorithm discussed in \S~\ref{sec:calibration_method} we derive the $F150W$ TRGB calibration:
\begin{equation}\label{eqn:f150w_f090w-f150w} 
 \resizebox{0.88\linewidth}{!}{$
\trgbcalfull{F150W}{-5.72}{-0.95}{F090W}{F150W}{1.40}\\
$}
\end{equation}
\begin{equation}\label{eqn:f150w_f150w-f356w} 
 \resizebox{0.88\linewidth}{!}{$
\trgbcalfull{F150W}{-5.82}{-1.78}{F150W}{F356W}{0.60}\\
$}
\end{equation}
\begin{equation}\label{eqn:f150w_f150w-f444w} 
 \resizebox{0.88\linewidth}{!}{$
\trgbcalfull{F150W}{-5.74}{-1.89}{F150W}{F444W}{0.50}\\
$}
\end{equation}

\subsection{The TGRB in NIRCam LW Filters}
The \jwst\ NIRCam LW filters are less practical for precision TRGB distance measurements due to a lower angular resolution relative to the SW filters. These two compounding effects cause the imaging and photometry to be more susceptible to crowding and blending of point sources. However, the RGB stars in LW filters appear up to $2$~mag brighter than in the $F090W$. Therefore LW filters can be used to detect TRGB stars well above the completeness limit of the data at appreciable gains in observational efficiency and at greater distances. The trade-off in precision and the ability to detect the TRGB at large distances must be considered on the basis of the study.

\begin{figure*}
\centering
\includegraphics[width=0.8\textwidth]{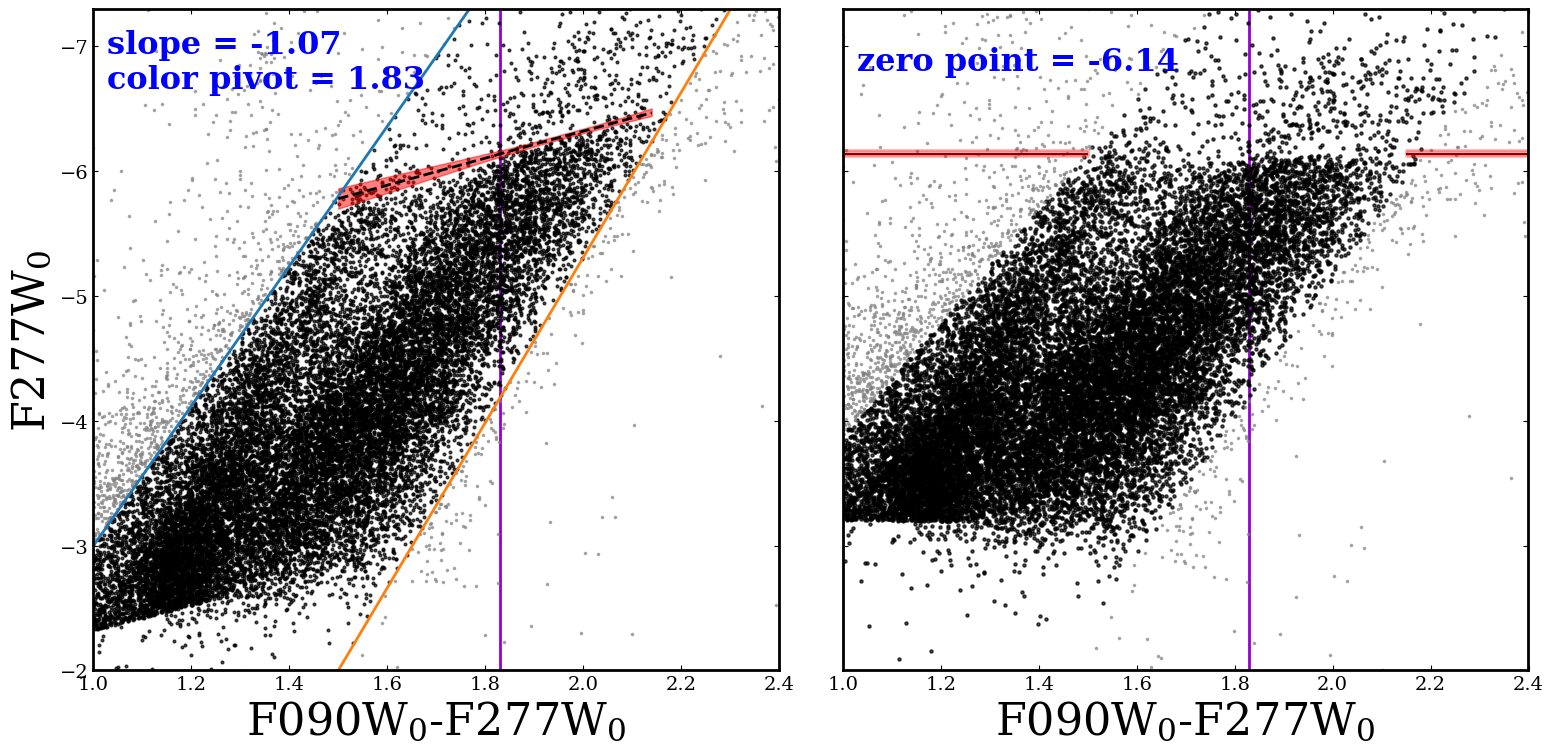}
\includegraphics[width=0.8\textwidth]{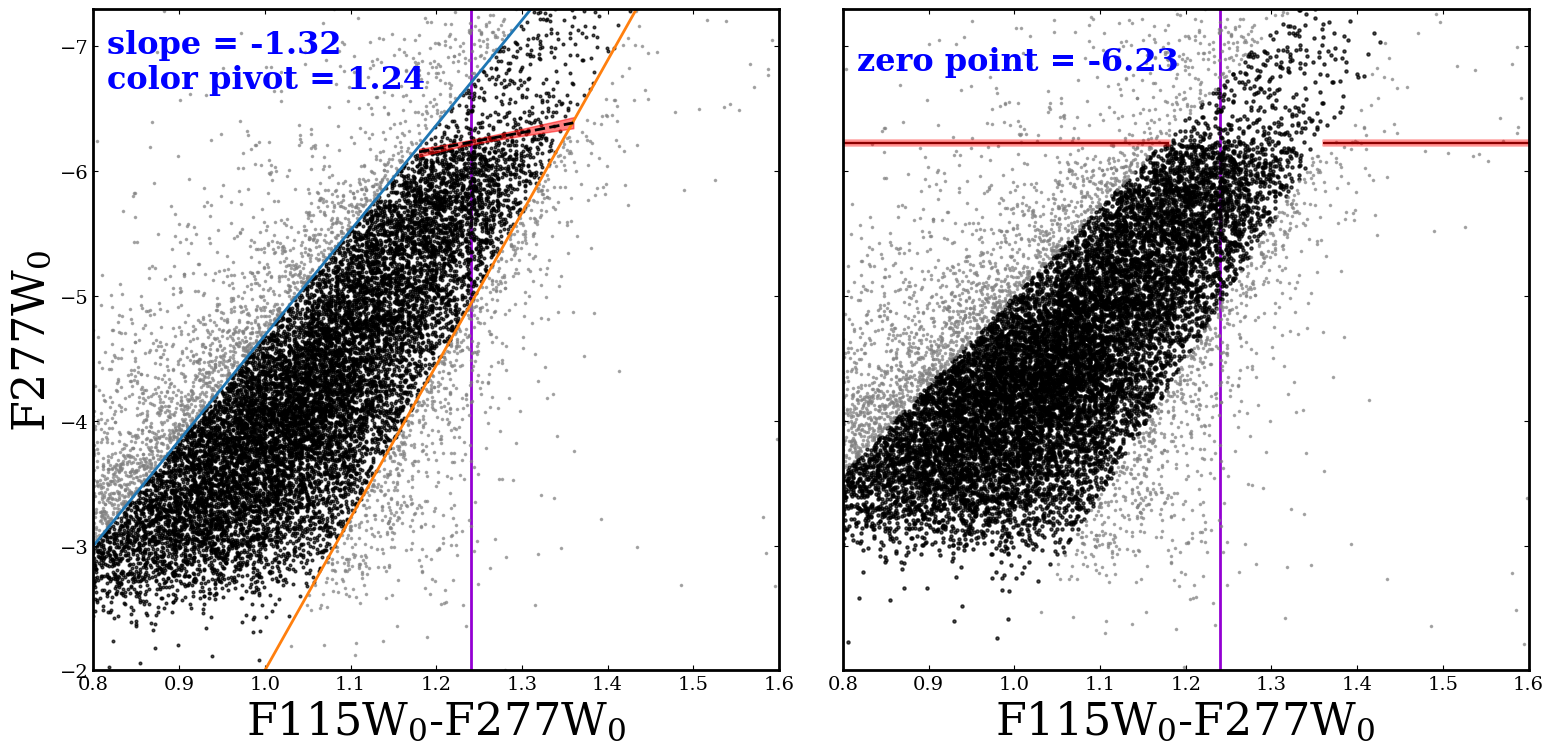}
\caption{The NIRCam LW filter TRGB calibration solutions for $F277W$. See the caption in Figure~\ref{fig:f090w} for a description of the ordering and content in the panels.}
\label{fig:f277w}
\end{figure*}

\begin{figure*}
\centering
\includegraphics[width=0.8\textwidth]{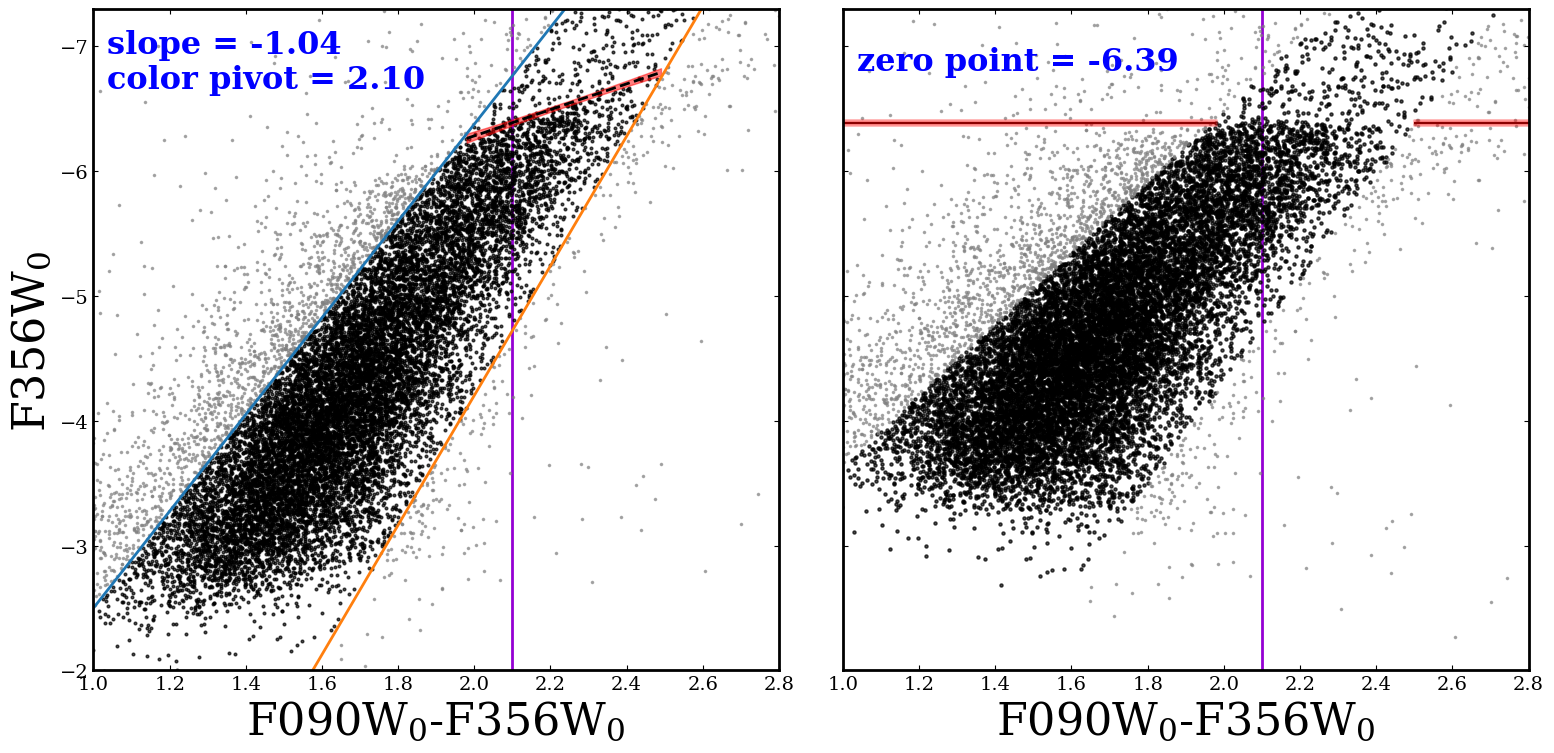}
\includegraphics[width=0.8\textwidth]{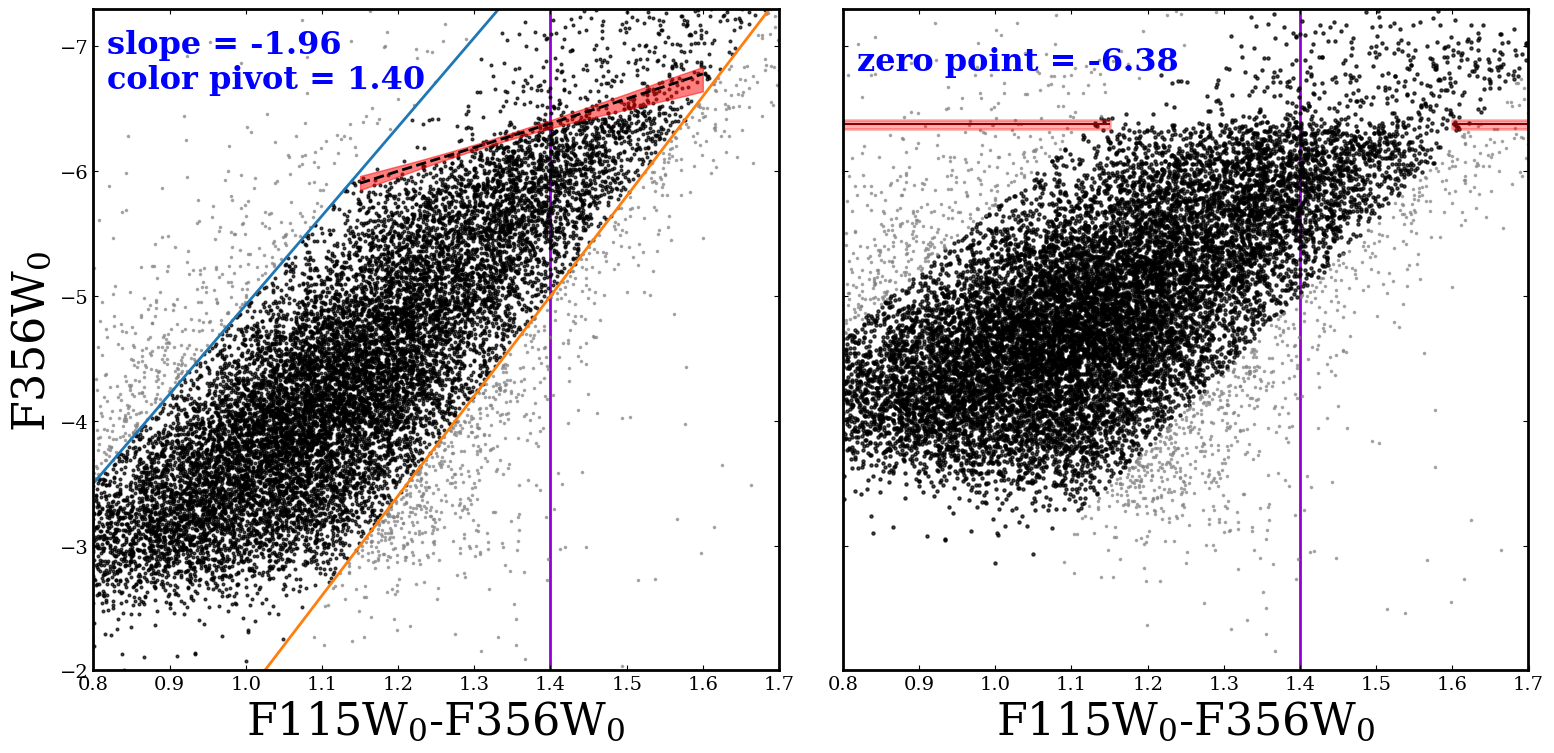}
\includegraphics[width=0.8\textwidth]{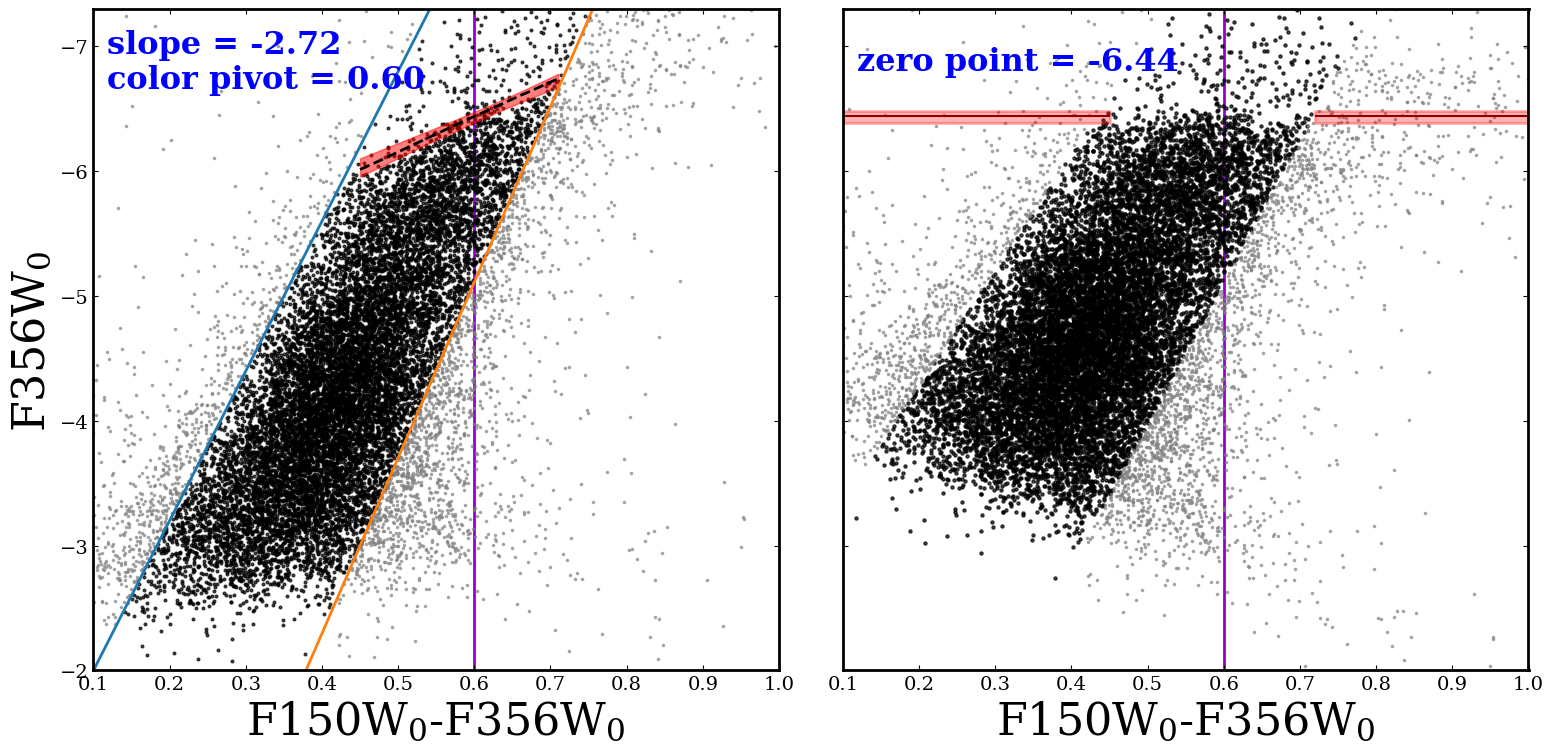}
\caption{The NIRCam LW filter TRGB calibration solutions for $F356W$. See the caption in Figure~\ref{fig:f090w} for a description of the ordering and content in the panels.}
\label{fig:f356w}
\end{figure*}

\begin{figure*}
\centering
\includegraphics[width=0.8\textwidth]{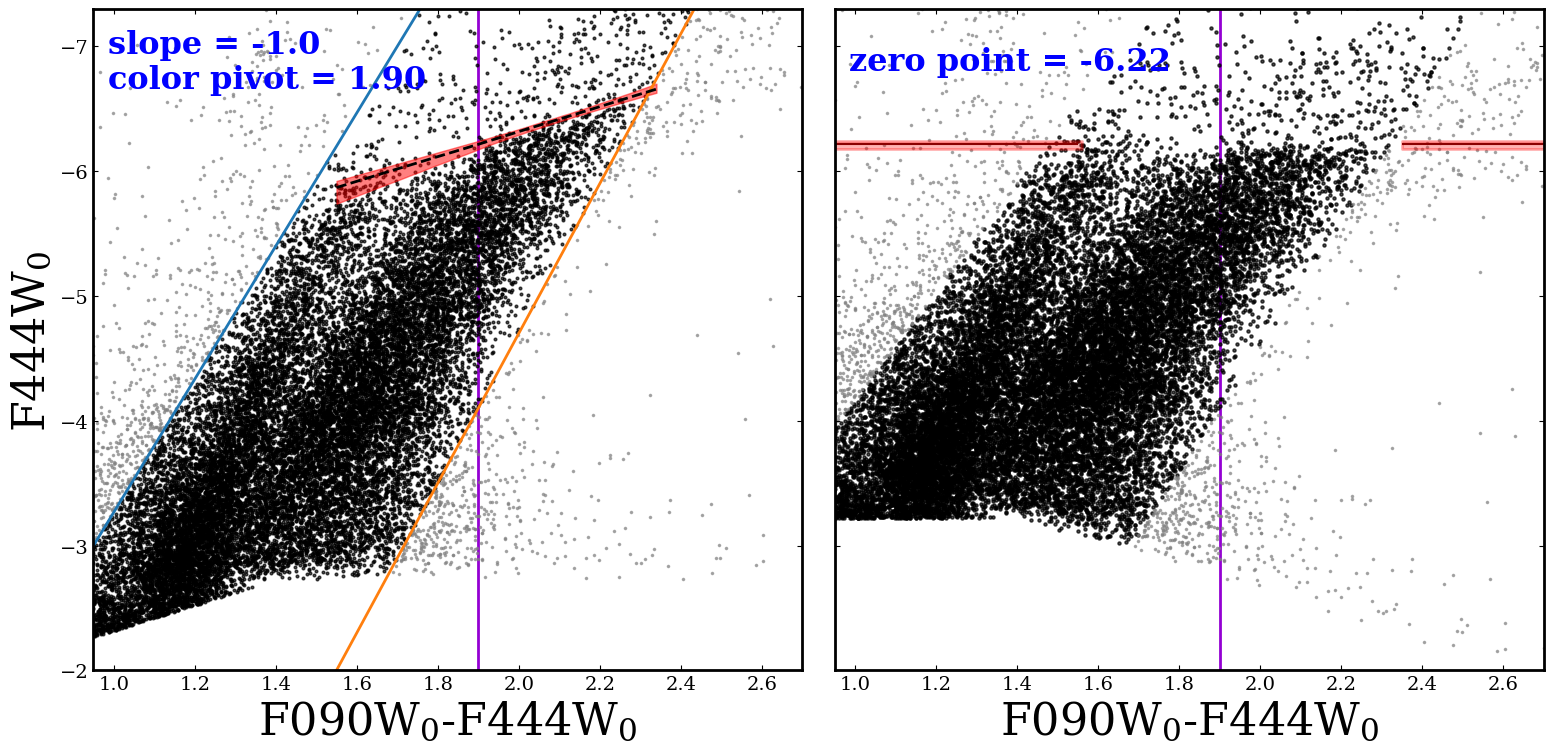}
\includegraphics[width=0.8\textwidth]{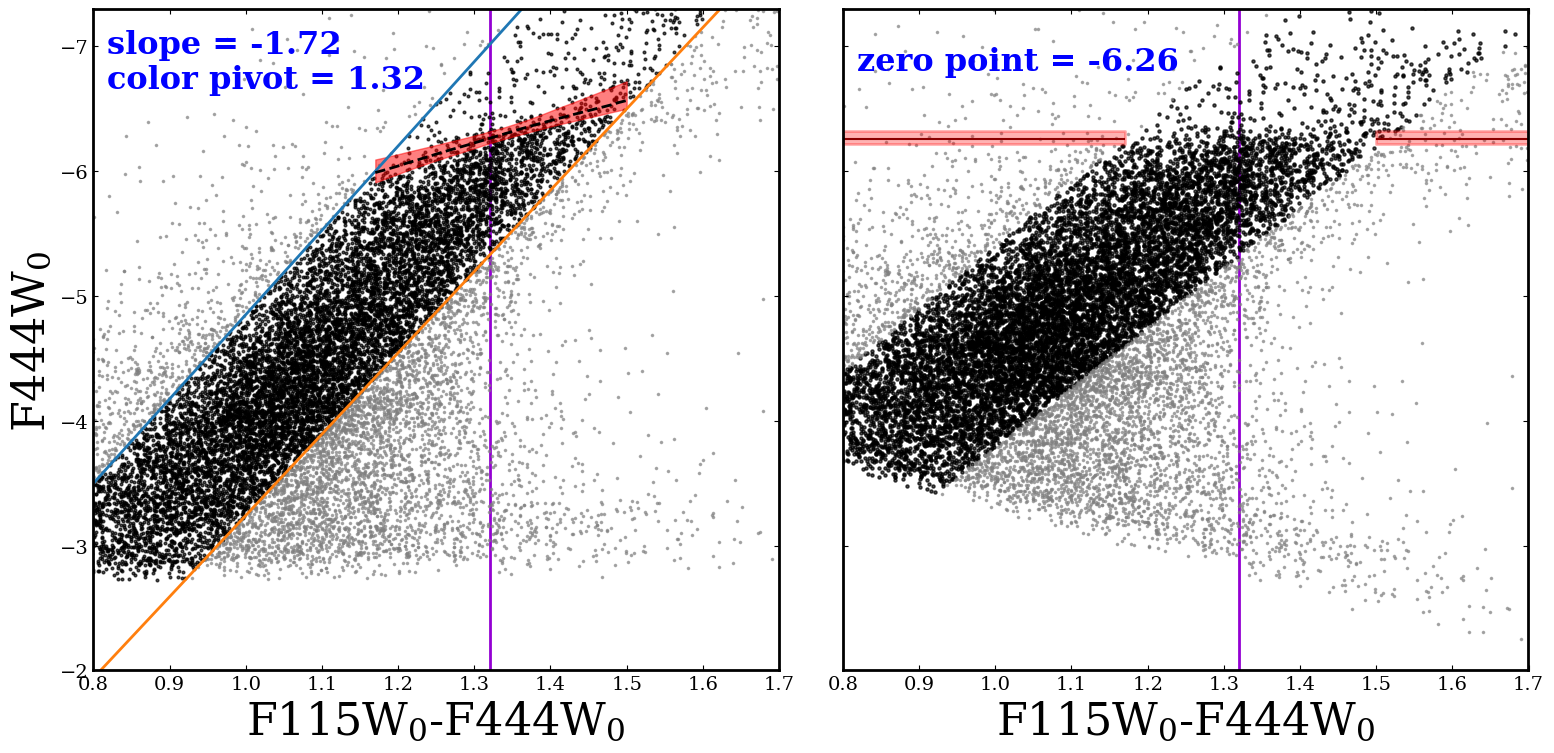}
\includegraphics[width=0.8\textwidth]{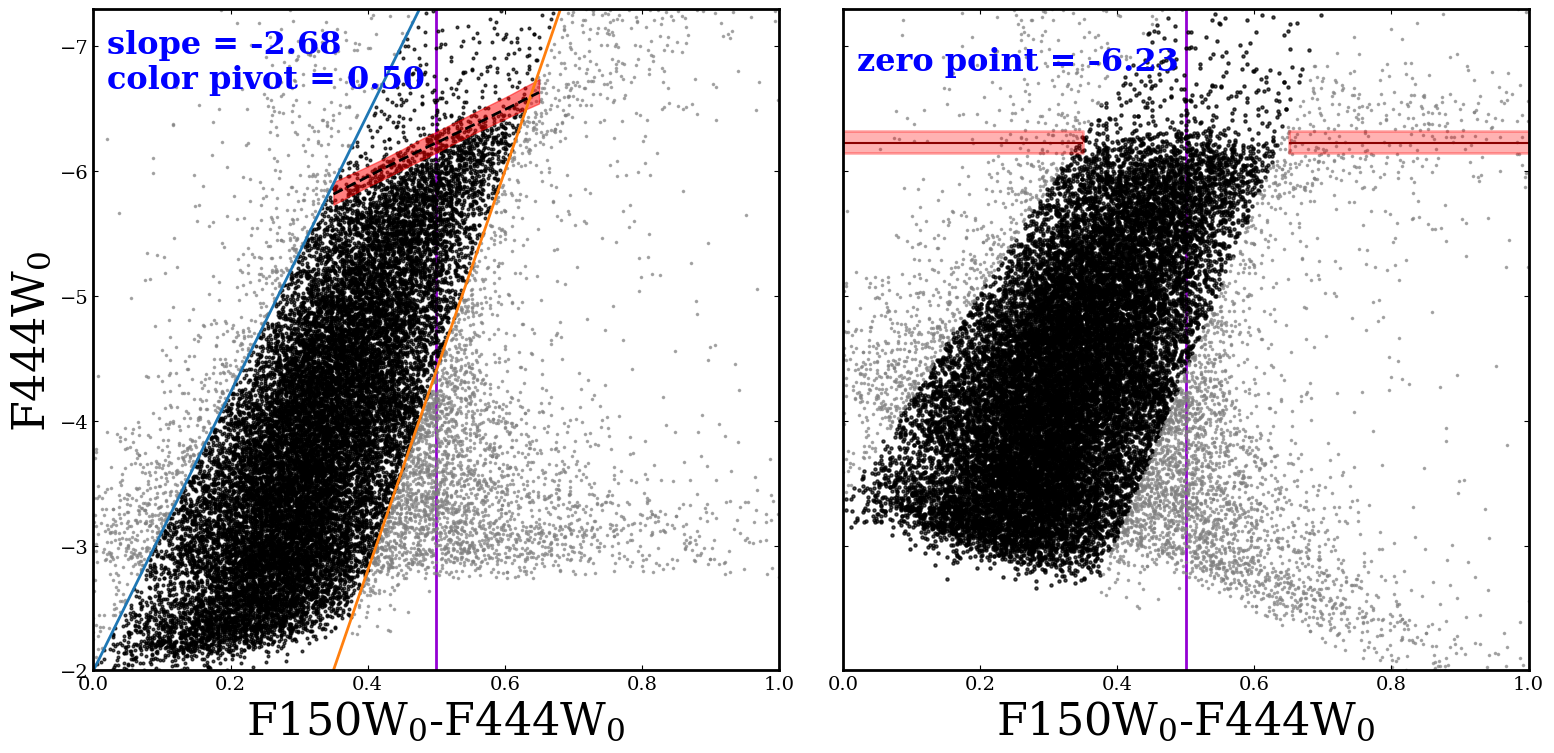}
\caption{The NIRCam LW filter TRGB calibration solutions for $F444W$. See the caption in Figure~\ref{fig:f090w} for a description of the ordering and content in the panels.}
\label{fig:f444w}
\end{figure*}

We provide calibrations for each of the three LW filters $F277W,$ $F356W,$ and $F444W$ only when paired with one of the available SW filters. We do not calibrate the TRGB in pairs of LW filters based on the lower angular resolution, and a narrow range in the color of the stars in a CMD leading to little separation in the RGB and AGB stellar populations (see the lower right panels of Figure~\ref{fig:all_cmds}). In Figures~\ref{fig:f277w} to \ref{fig:f444w} we present 8 total calibrations. The corresponding color-based rectification equations are provided:
\begin{equation}\label{eqn:f277w_f090w-f277w} 
 \resizebox{0.88\linewidth}{!}{$
\trgbcalfull{F277W}{-6.14}{-1.07}{F090W}{F277W}{1.83}\\
$}
\end{equation}
\begin{equation}\label{eqn:f277w_f115w-f277w} 
 \resizebox{0.88\linewidth}{!}{$
\trgbcalfull{F277W}{-6.23}{-1.32}{F115W}{F277W}{1.24}\\
$}
\end{equation}
\begin{equation}\label{eqn:f356w_f090w-f356w} 
 \resizebox{0.88\linewidth}{!}{$
\trgbcalfull{F356W}{-6.39}{-1.04}{F090W}{F356W}{2.10}\\
$}
\end{equation}
\begin{equation}\label{eqn:f356w_f115w-f356w} 
 \resizebox{0.88\linewidth}{!}{$
\trgbcalfull{F356W}{-6.38}{-1.96}{F115W}{F356W}{1.40}\\
$}
\end{equation}
\begin{equation}\label{eqn:f356w_f150w-f356w} 
 \resizebox{0.88\linewidth}{!}{$
\trgbcalfull{F356W}{-6.44}{-2.72}{F150W}{F356W}{0.60}\\
$}
\end{equation}
\begin{equation}\label{eqn:f444w_f090w-f444w} 
 \resizebox{0.88\linewidth}{!}{$
\trgbcalfull{F444W}{-6.22}{-1.0}{F090W}{F444W}{1.90}\\
$}
\end{equation}
\begin{equation}\label{eqn:f444w_f115w-f444w} 
 \resizebox{0.88\linewidth}{!}{$
\trgbcalfull{F444W}{-6.26}{-1.72}{F115W}{F444W}{1.32}\\
$}
\end{equation}
\begin{equation}\label{eqn:f444w_f150w-f444w} 
 \resizebox{0.88\linewidth}{!}{$
\trgbcalfull{F444W}{-6.23}{-2.68}{F150W}{F444W}{0.50}\\
$}
\end{equation}

\begin{table*}[]
    \caption{JWST NIRCam TRGB Calibrations}
    \centering
    \begin{tabular}{|cccccc|}
    \hline
    Mag Filter & Color Filters & Zero-Point & Slope & Color Pivot  & Color Range\\
    & & (mag) & (mag color$^{-1}$) & (mag)  & (mag, mag)\\
    \hline\hline
$F090W$ & $F090W-F150W$ & -4.32$^{+0.02}_{-0.02}$ & 0.05$^{+0.08}_{-0.18}$  & 1.40 & 1.15 - 1.68\\
 & $F090W-F277W$ & -4.32$^{+0.03}_{-0.02}$ & 0.00$^{+0.13}_{-0.19}$  & 1.83 & 1.50 - 2.15\\
 & $F090W-F356W$ & -4.30$^{+0.02}_{-0.02}$ & 0.02$^{+0.04}_{-0.11}$  & 2.10 & 1.98 - 2.50\\
 & $F090W-F444W$ & -4.31$^{+0.02}_{-0.04}$ & -0.02$^{+0.12}_{-0.09}$  & 1.90 & 1.55 - 2.35\\
$F115W$ & $F115W-F277W$ & -5.00$^{+0.02}_{-0.01}$ & -0.43$^{+0.17}_{-0.13}$  & 1.24 & 1.18 - 1.36\\
 & $F115W-F356W$ & -4.99$^{+0.04}_{-0.02}$ & -0.98$^{+0.31}_{-0.17}$  & 1.40 & 1.15 - 1.60\\
 & $F115W-F444W$ & -4.93$^{+0.03}_{-0.06}$ & -0.78$^{+0.30}_{-0.45}$  & 1.32 & 1.17 - 1.50\\
$F150W$ & $F090W-F150W$ & -5.72$^{+0.02}_{-0.02}$ & -0.95$^{+0.10}_{-0.19}$  & 1.40 & 1.15 - 1.68\\
 & $F150W-F356W$ & -5.82$^{+0.04}_{-0.04}$ & -1.78$^{+0.30}_{-0.32}$  & 0.60 & 0.45 - 0.72\\
 & $F150W-F444W$ & -5.74$^{+0.08}_{-0.09}$ & -1.89$^{+0.45}_{-0.37}$  & 0.50 & 0.35 - 0.65\\
$F277W$ & $F090W-F277W$ & -6.14$^{+0.02}_{-0.03}$ & -1.07$^{+0.16}_{-0.18}$  & 1.83 & 1.50 - 2.15\\
 & $F115W-F277W$ & -6.23$^{+0.02}_{-0.02}$ & -1.32$^{+0.31}_{-0.35}$  & 1.24 & 1.18 - 1.36\\
$F356W$ & $F090W-F356W$ & -6.39$^{+0.02}_{-0.02}$ & -1.04$^{+0.10}_{-0.11}$  & 2.10 & 1.98 - 2.50\\
 & $F115W-F356W$ & -6.38$^{+0.04}_{-0.03}$ & -1.96$^{+0.45}_{-0.22}$  & 1.40 & 1.15 - 1.60\\
 & $F150W-F356W$ & -6.44$^{+0.05}_{-0.04}$ & -2.72$^{+0.31}_{-0.36}$  & 0.60 & 0.45 - 0.72\\
$F444W$ & $F090W-F444W$ & -6.22$^{+0.04}_{-0.02}$ & -1.00$^{+0.10}_{-0.22}$  & 1.90 & 1.55 - 2.35\\
 & $F115W-F444W$ & -6.26$^{+0.04}_{-0.06}$ & -1.72$^{+0.40}_{-0.68}$  & 1.32 & 1.17 - 1.50\\
 & $F150W-F444W$ & -6.23$^{+0.08}_{-0.09}$ & -2.68$^{+0.36}_{-0.51}$  & 0.50 & 0.35 - 0.65\\
    \hline
    \end{tabular}
    \label{tab:all_trgb_calibs}
\end{table*}
\subsection{Verifying the Precision of \jwst\ TRGB  Calibrations}\label{sec:dm_verify}
\begin{figure*}[!tbh]
    \centering    \includegraphics[width=0.425\textwidth]{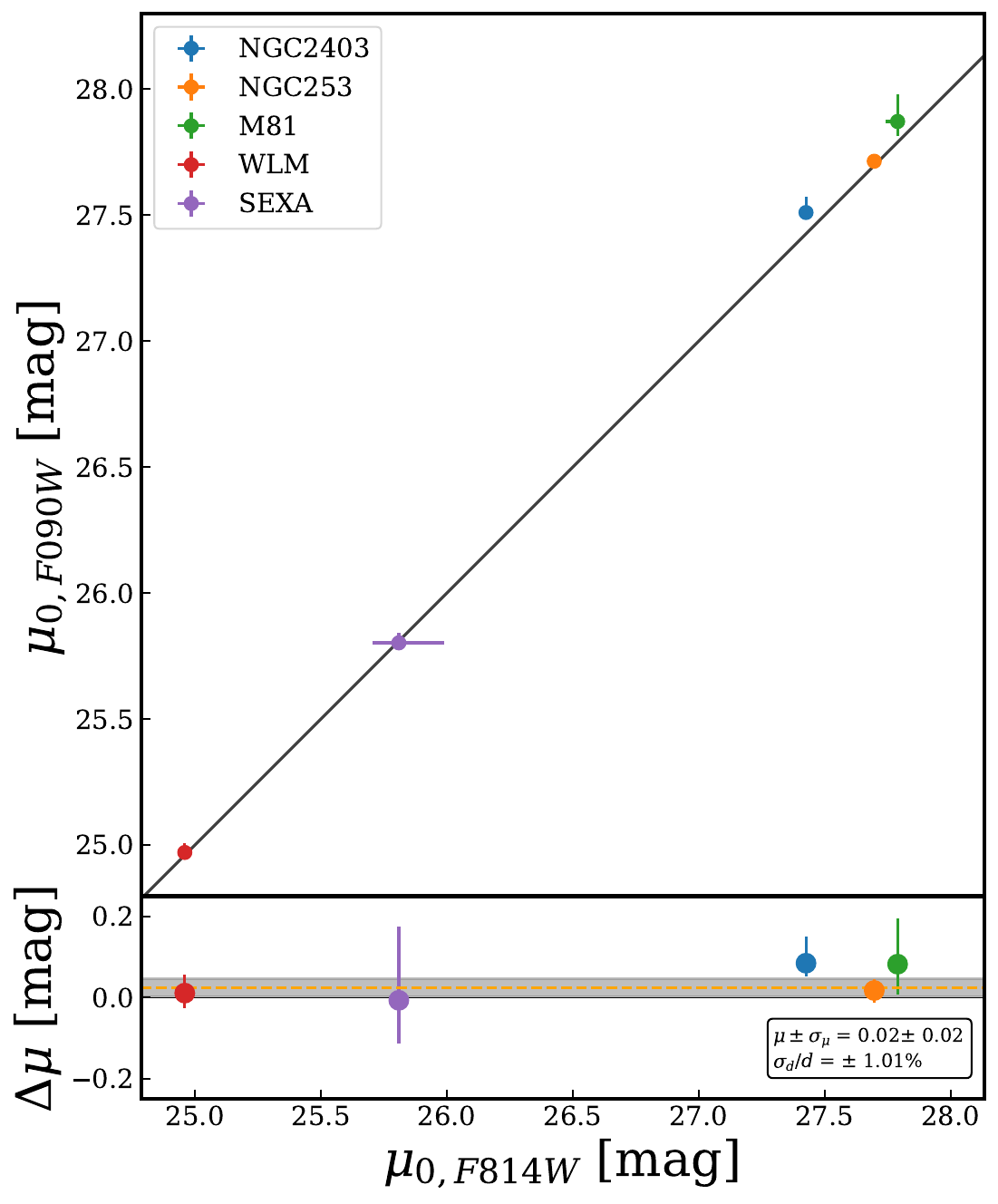}
     \includegraphics[width=0.51\textwidth]{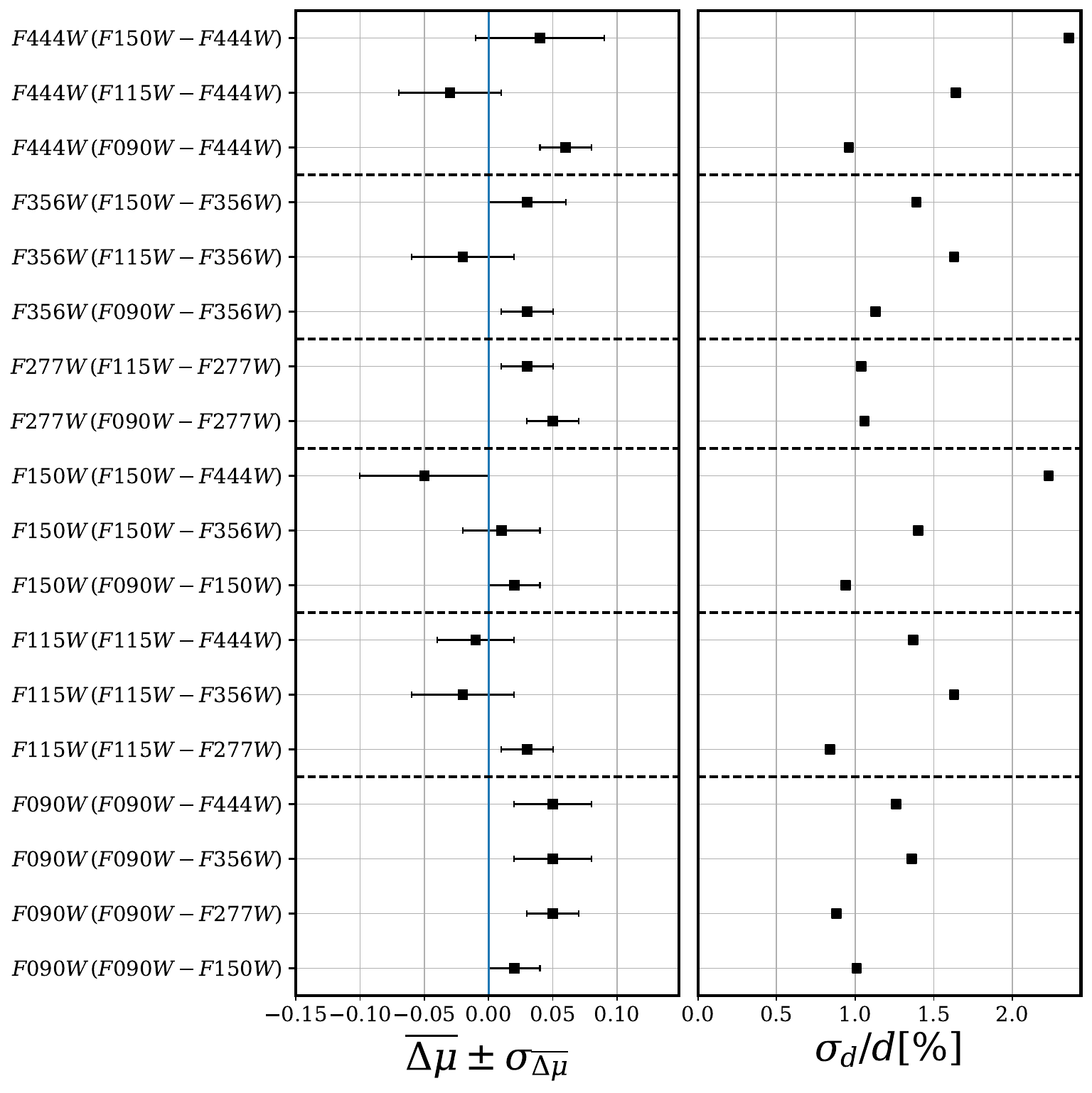}
    \caption{\textit{Left}: Comparison between the foreground-extinction-corrected distance moduli measured for each target in $F090W$ vs $F090W-F150W$ (Eqn~\ref{eqn:f090w_f090w-f150w}) the \hst\ F814W distance moduli which anchor the distance scale. Top panel: $F814W$ distance moduli ($\mu_{0,F814W}$) plotted against $F090W$ distance moduli ($\mu_{0,F090W}$). Each color-coded point represents one galaxy: M~81 (green), NGC~253 (orange), NGC~2403 (blue), Sextans A (purple), and WLM (red). Bottom panel: The residuals in distance modulus $\delta\mu=\mu_{0,F090W}-\mu_{0,F814W}$ versus $\mu_{0,F814W}$. The null residual is shown as a black solid line. Points above the line are measured as fainter and therefore more distance in $F090W$. The orange dashed line and gray shaded region are the mean and standard deviation of the mean weighted by the uncertainties on the data. The residual plot demonstrates that the distance moduli are in agreement within the uncertainties based on both the weighted mean and standard deviation of the weighted mean of the residuals ($\overline{\Delta\mu}=0.02\pm0.02$). We note that the mean is consistent with an offset of $\overline{\Delta\mu}=0$~mag. As a fractional uncertainty in distance, this represents a precision of $\sim1\%$. \textit{Middle}: The fits to the distance modulus residuals $\overline{\Delta\mu}$ for all 18 TRGB calibrations. Each black square corresponds to the calibration show on the y-axis. Error bars represent the $1\sigma$ standard deviation of the mean weighted by the uncertainties in distance modulus for each galaxy. Dashed horizontal lines separate the TRGB calibrations by the luminosity function filter. The solid vertical blue line marks $\overline{\Delta\mu}=0$. \textit{Right}: The fractional uncertainty in distance for each TRGB calibration. The y-axis and horiztonal dashed lines are the same as in the middle panel. Note that the fractional uncertainties increase from blue to red filters (up along the y-axis).}
    \label{fig:f090w_dm_comp}
\end{figure*}

To validate the precision of our \jwst\ TRGB calibrations we take the same approach as Paper I. We re-measure the distances to each field in their individual CMDs and compare them to the F814W distances used to anchor the calibrations. In the left panel of Figure~\ref{fig:f090w_dm_comp} we compare the anchoring distance moduli measured in \hst\ $F814W$ to the distance moduli measured from individual targets. We demonstrate the high precision in our $F090W$ vs.\ $F090W-F150W$ calibration. In the top panel we show the distance moduli comparison. Three of five points intersect the 1-to-1 line (black solid line) within their uncertainties. M~81 and NGC~2403 are modestly offset from the 1-to-1 line. In the bottom panel we show the distance moduli residuals, $\mu_{F090W}-\mu_{F814W}$, and the mean and standard deviation of the mean weighted by the uncertainties on the data (orange dashed line and grey shaded region, respectively). The mean of the residual is consistent with having a residual of $\overline{\Delta\mu}=0$.

In the middle panel of Figure~\ref{fig:f090w_dm_comp} we show $\overline{\Delta\mu}\pm\sigma_{\overline{\Delta\mu}}$ for all 18 TRGB calibrations. Black dashed horizontal lines separate the TRGB magnitude filters. The blue solid vertical line marks $\overline{\Delta\mu}=0$~mag. We note that for $F090W$ there is a modest bias in the residual toward positive values of $\overline{\Delta\mu}$. In the right panel of Figure~\ref{fig:f090w_dm_comp} we present the fractional uncertainty in distance $\sigma_d/d$ as percentages for the 18 calibrations. We calculate $\sigma_d/d$ as:

\begin{equation}
    \sigma_d/d = \frac{\sigma_{\overline{\Delta\mu}}0.461*10^{\overline{\Delta\mu}/5+1}}{10^{\overline{\Delta\mu}/5+1}}*100
\end{equation}
Note that the fractional uncertainties tend toward greater values at the reddest filter $F444W$. 
We provide the tabulated results of the same type of analysis for the remaining calibrations in Table~\ref{tab:dm_resid_res}.

\begin{table}[!h]
\caption{Summary of Distance Modulus Comparison and Fits to Residuals}
\centering
	\begin{tabular}{|cccc|}
	\hline
	 Mag Filter & Color & $\overline{\Delta\mu}\pm\sigma_{\overline{\Delta\mu}}$ & $\sigma_d/d$ \\
	  &  & (mag) & (\%) \\
	\hline
	\hline
	$F090W$ & $F090W-F150W$ & $0.02\pm0.02$ & 1.01 \\
		& $F090W-F277W$ & $0.05\pm0.02$ & 0.88\\
		& $F090W-F356W$ & $0.05\pm0.03$ & 1.36\\
		& $F090W-F444W$ & $0.05\pm0.03$ & 1.26\\
	\hline
	$F115W$ & $F115W-F277W$ &$0.03\pm0.02$ & 0.84\\
		& $F115W-F356W$ & $-0.02\pm0.04$ & 1.63\\
		& $F115W-F444W$ & $-0.01\pm0.03$ & 1.37\\
	\hline
	$F150W$ & $F090W-F150W$ & $0.02\pm0.02$ & 0.94\\
		& $F150W-F356W$ &$0.01\pm0.03$ & 1.40\\
		& $F150W-F444W$ &$-0.05\pm0.05$ & 2.23\\
	\hline
	$F277W$ & $F090W-F277W$ & $0.05\pm0.02$ & 1.06\\
		& $F115W-F277W$ &$0.03\pm0.02$ & 1.04\\
	\hline
	$F356W$ & $F090W-F356W$ & $0.03\pm0.02$ &1.13 \\
		& $F115W-F356W$ &$-0.02\pm0.04$ & 1.63\\
		& $F150W-F356W$ &$0.03\pm0.03$ & 1.39\\
	\hline
	$F444W$ & $F090W-F444W$ & $0.06\pm0.02$ &0.96 \\
		& $F115W-F444W$ & $-0.03\pm0.04$ & 1.64\\
		& $F150W-F444W$ & $0.04\pm0.05$ & 2.36\\
	\hline
	\end{tabular}
\label{tab:dm_resid_res}
\end{table}

\subsection{Error Budget for the TRGB Calibration}\label{sec:err_budget}
The statistical uncertainties include the $\pm0.015$~mag uncertainty from the \hst\ ACS $F814W$ zero-point \citep{Freedman2021} and the MC simulation \jwst\ TRGB zero-point uncertainties. The systematic uncertainties include contributions from the foreground extinction measured in $F814$, the sensitivity of the measured \jwst\ TRGB zero-points to variations in the GLOESS smoothing scale $\tau$, the color-based luminosity correction applied to the $F814$ TRGB measurements, and the $\pm0.035$~mag uncertainty on the $F814W$ zero-point. In Table~\ref{tab:error_budget} we provide a breakdown of the contributions to our total error budget, both statistical and systematic. 
\begin{table*}[!]
\centering
    \footnotesize
    \caption{Summary of the Error Budget for the NIR TRGB Calibration}
    \label{tab:error_budget}
    \begin{tabular}{lll}
    \hline
    \hline
    Source & Stat. (mag) & Sys. (mag) \\
    \hline
    Foreground Extinction & $\ldots$ & $10\% \text{max}(A_{\text{F814W}})=0.01$ \\
    Smoothing Scale & $\ldots$ & 0.02\\
    QT Color Correction\footnote{Only 2 of the 12 targets contribute to this systematic.} & $\ldots$ & 0.01\\
    $M_{\text{F814W}}^{TRGB}$ & 0.015 & 0.035\\
    NIR zero point & see the zero-point column in Table~\ref{tab:all_trgb_calibs} & $\ldots$\\
    \hline
    \multicolumn{3}{c}{\bf{Totals}}\\
    \hline
    & $\ldots$ & $0.043$\\
    \hline
    \end{tabular}
    \tablecomments{The statistical error column does not list individual uncertainties for each TRGB calibration. Instead we refer the reader to the \textit{Zero-Point} column in Table~\ref{tab:all_trgb_calibs} for individual uncertainties. We add the zero-point statistical uncertainties in quadrature with the statistical uncertainties listed in this table. }
\end{table*}

\section{Comparison to the SH0ES $F090W$ TRGB Calibration}\label{sec:sh0es_comp}
The first absolute calibration of the TRGB in any \jwst\ filter was reported in \citet{Anand2024}.  They used NIRCam $F090W$ and $F150W$ observations of the galaxy megamaser host galaxy NGC~4258 to measure the slope and zero-point of the $F090W$ TRGB. NCG~4258 is one of the few extragalactic systems used to geometrically anchor resolved stellar population-based distance indicators \citep[e.g., the TRGB, Cepheids, J-AGB;][]{Humphreys2013}. Additionally, to demonstrate the applicability of the F090W filter for \jwst\  TRGB measurements they applied their calibration to two type Ia supernova host galaxies at distances of $\sim20$~Mpc. The final value for the absolute calibration of the \jwst\  $F090W$ TRGB is determined from the geometric distance to NGC~4258 and a weighted foreground-extinction-corrected average of the apparent TRGB magnitude in NGC~4258 of $m^{F090W}_{\text{TRGB}}=25.035\pm0.02$ mag: $M^{F090W}_{\text{TRGB}}=-4.362\pm0.033\text{ (stat) }\pm0.045\text{ (sys) }$mag. 

In comparing our $F090W\text{ vs. }F090W-F150W$ TRGB absolute calibration $M^{F090W}_{\text{TRGB}}=-4.32\pm0.03$ mag with the result from \citet{Anand2024} we find that our calibration for the F090W TRGB zero-point and the calibration reported in \citet{Anand2024} are in agreement within the uncertainties. Moreover,  \citet{Anand2024} present their calibration in the \jwst\ NIRCam Vega-Vega system. The Sirius-Vega and Vega-Vega calibrations can differ up to 0.04 mag. After applying a conservative 0.04 mag correction, the  \citet{Anand2024} TRGB zero-point is $M^{F090W}_{\text{TRGB}}=-4.322$ mag, demonstrating a remarkable agreement between the two calibrations. We note that our calibration methodologies and data are different (e.g., we have a color term in our calibration). Thus it is interesting that the results are so similar.

\section{Conclusion}\label{sec:conclusion}
We have calibrated the \jwst\ TRGB in 6 NIRCam filters and in 18 CMD combinations by characterizing the color-dependence of the TRGB luminosity function. Our calibration is based on new and archival observations of galaxies that span a range of age, mass, and metallicity. We found that the TRGB zero-points span the range from $-4.3$ mag in $F090W$ to $-6.4$ mag in $F444W$ or $~\sim2$ magnitudes. 

Next, we discussed the appropriate use cases for TRGB measurements using $\jwst$ SW and LW filters. We found that for the most precise and robust TRGB measurements the combinations of $F090W$ and $F150W$, or $F115W$ and $F277W$ are ideal. If lower precision is tolerable or shorter integration times are desired for one's science goal we recommend pairing $F115W$ or $F150W$ with $F356W$. We do not recommend $F444W$ for any precision TRGB measurements. The $F444W$ filter has the lowest angular resolution out of all the filters in our calibration, and sources in $F444W$ are the faintest as demonstrated in Figure~\ref{fig:sw_to_lw_im}. Our recommendations are based on the distance modulus comparison tests in \S~\ref{sec:dm_verify}. 

Finally, we compared our $F090W$ vs.\ $F090W-F150W$ (Eqn~\ref{eqn:f090w_f090w-f150w}) calibration to the only other calibration in the literature \citep{Anand2024}. We found that the TRGB zero-points were consistent within the uncertainties.

We provide these new TRGB calibrations to the community as a first step in measuring robust distances to galaxies with \jwst.\ In the future, as more datasets become available, \jwst\ TRGB calibrations will be updated. Additional observations of galaxies spanning a similar metallicity range to those presented in this study can help to better constrain the color dependence of the TRGB luminosity. New observations targeting the outer stellar fields of more metal rich/poor RGB stars will provide the data to measure any potential non-linearities in the color-dependence of the TRGB \citep{McQuinn2019}.  

\facility{\jwst\ (NIRCam)}
\software{Astropy \citep{AstropyCollaboration2022}, \Dolphot{} \citep{Dolphin2002,Dolphin2016}, corner \citep{corner}, Matplotlib \citep{Hunter:2007}, NumPy \citep{harris2020array}, SciPy \citep{2020SciPy-NMeth}, UltraNest \citep{Buchner2021}, \jwst\ Pipeline \citep{Bushouse2023}.}

\begin{acknowledgments}
We thank the referee for their valuable feedback on this work.This work is based on observations with the NASA/ESA/CSA James Webb Space Telescope obtained at the Space Telescope Science Institute, which is operated by the Association of Universities for Research in Astronomy, Incorporated, under NASA contract NAS5-03127. Support for Program number JWST-GO-01638 was provided through a grant from the STScI under NASA contract NAS5-03127.

This research was supported by the Munich Institute for Astro-, Particle and BioPhysics (MIAPbP), which is funded by the Deutsche Forschungsgemeinschaft (DFG, German Research Foundation) under Germany's Excellence Strategy – EXC-2094 – 390783311. This research has made use of the NASA/IPAC Extragalactic Database (NED), which is funded by the National Aeronautics and Space Administration and operated by the California Institute of Technology.
\end{acknowledgments}
\appendix
\section{Corner plots from MC Simulations}\label{sec:all_corner_plots}
Our MC simulation method simultaneously fits for the slope and zero-point of the TRGB. In Figure~\ref{fig:corner_plots} we present the full suite of corner plots generated from the TRGB calibration MC outputs. Each corner plot shows in the top panel the 1D histogram for the slope ($\beta$) fit, in the right panel the 1D histogram for the zero-point fit, and in the bottom left panel the 2D joint probability for the MC simulation. 
\section{Overview of All \jwst\ TRGB Zero-Points}
\begin{figure*}[!t]
    \centering
    \includegraphics[width=\textwidth]{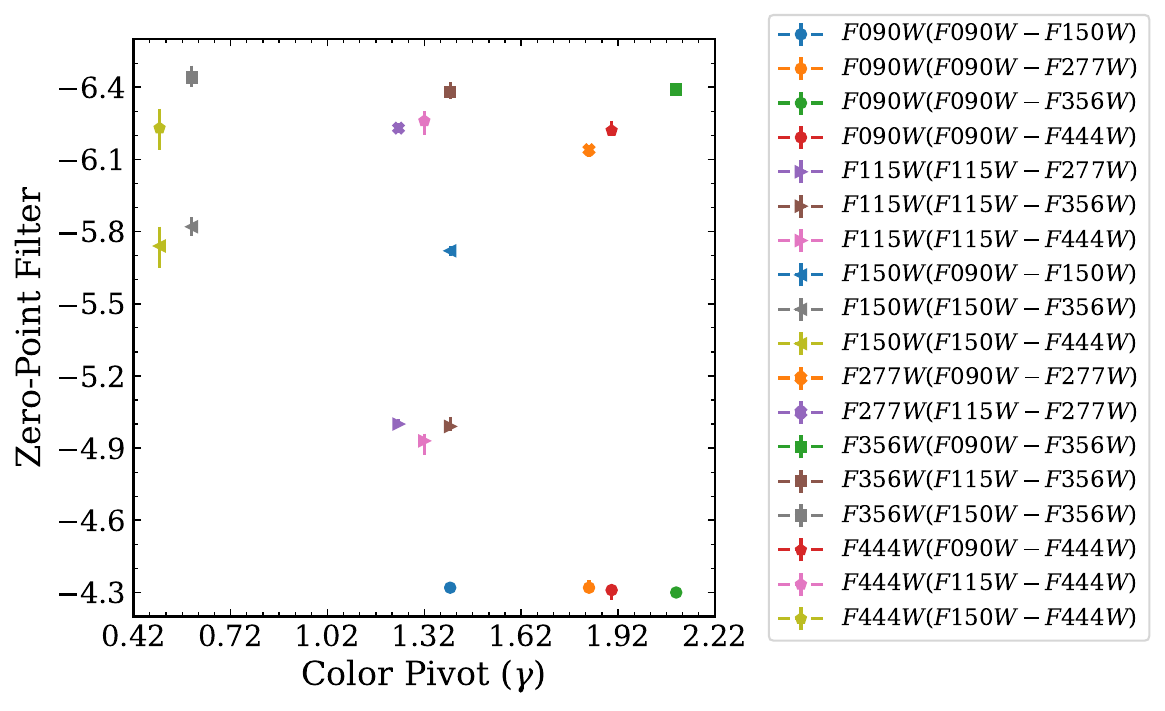}
    \caption{The \jwst\ TRGB zero-point overview. Shapes represent magnitude filters as follows: $F090W$ (filled circle); $F115W$ (right arrow); $F150W$ (left arrow); $F277W$ (filled ``X"); $F356W$ (square); and $F444W$ (pentagon). Shape colors represent color filter combinations as follows:  $F090W-F150W$ (blue);  $F090W-F277W$ (orange);  $F090W-F356W$ (green);  $F090W-F444W$ (red); $F115W-F277W$ (purple); $F115W-F356W$ (brown); $F115W-F444W$ (pink); $F150W-F356W$ (gray); and $F150W-F444W$ (yellow).}
    \label{fig:trgb_zp_overview}
\end{figure*}

\begin{figure*} 
 \centering
\includegraphics[width=0.24\textwidth]{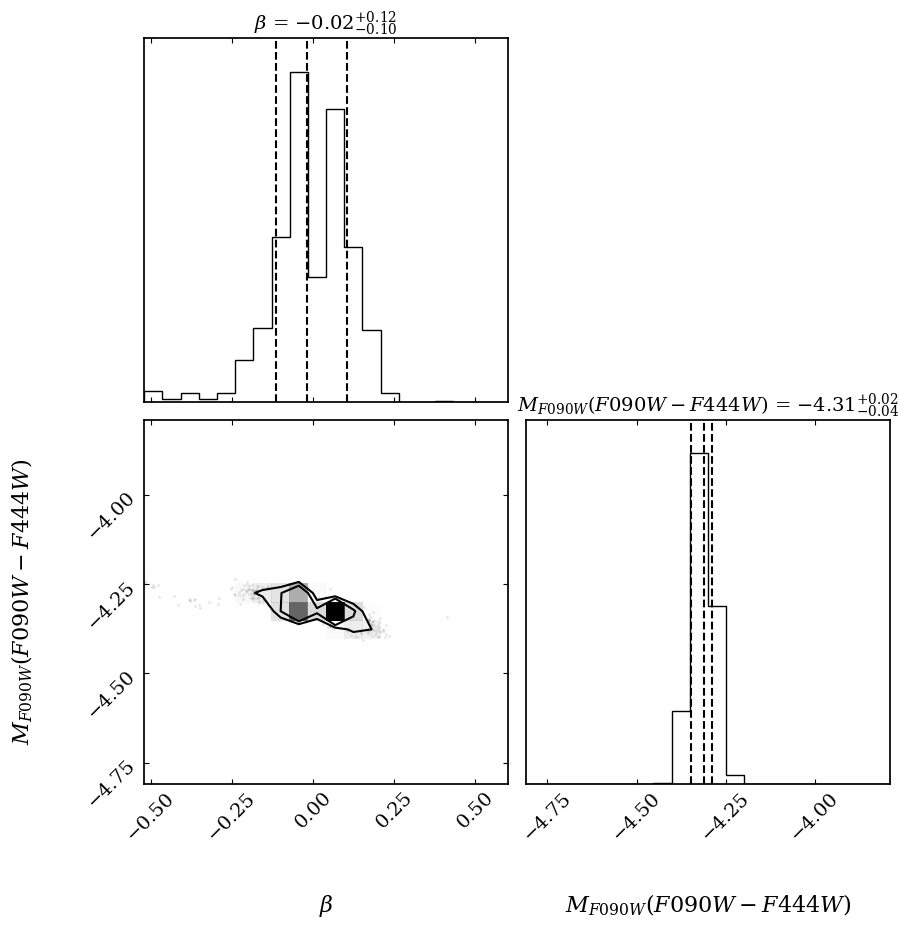}
\includegraphics[width=0.24\textwidth]{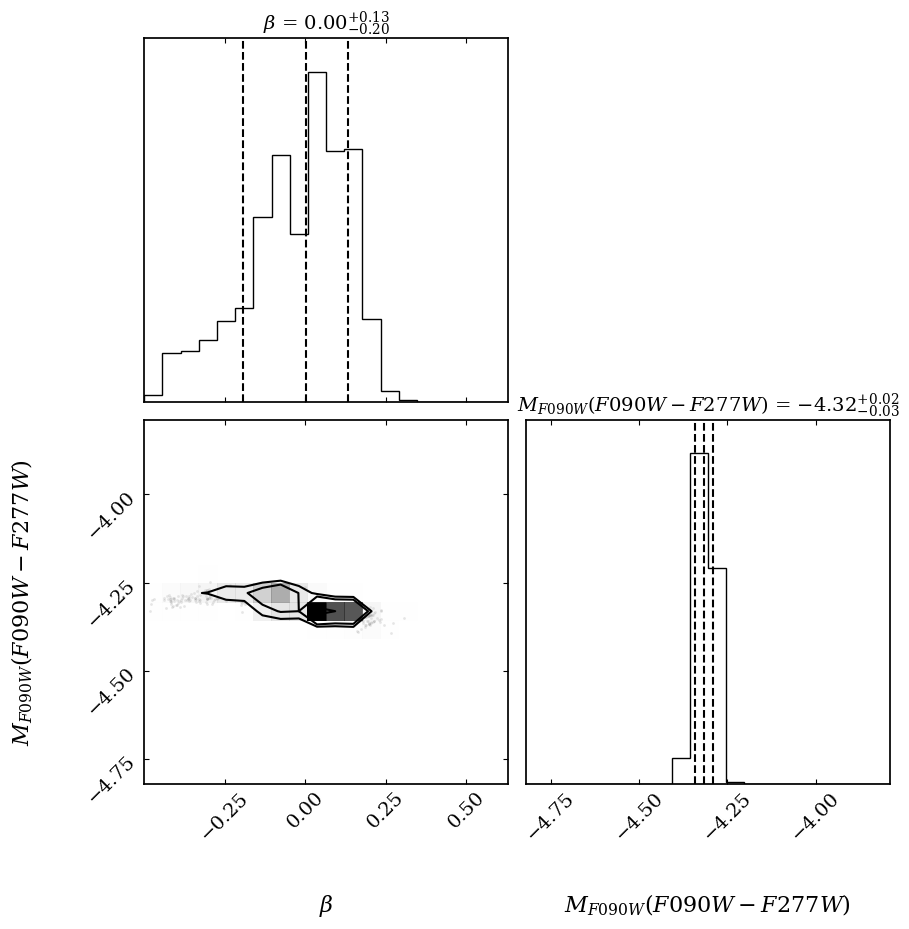}
\includegraphics[width=0.24\textwidth]{F090W_vs_F090W-F150W_linear_mc_output_beta0.05_bnds_basinhop_gamma1.40_tau0.030_cmin1.15_cmax1.68_corner_plot.png}
\includegraphics[width=0.24\textwidth]{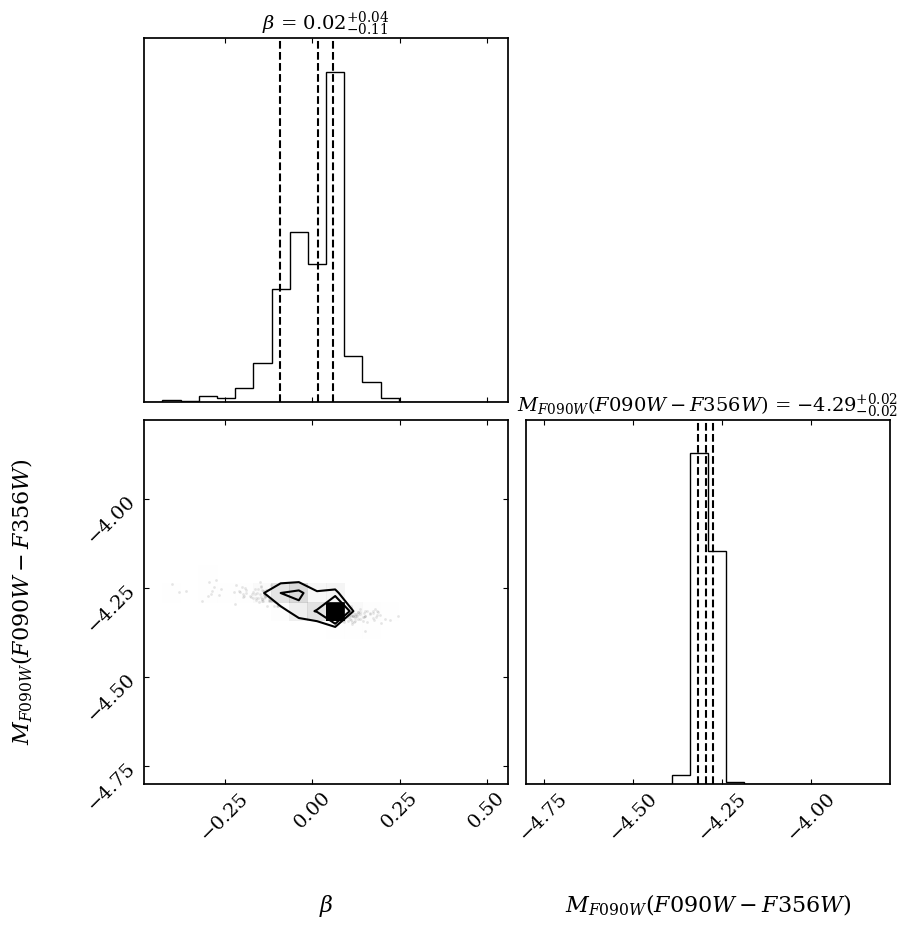}
\includegraphics[width=0.24\textwidth]{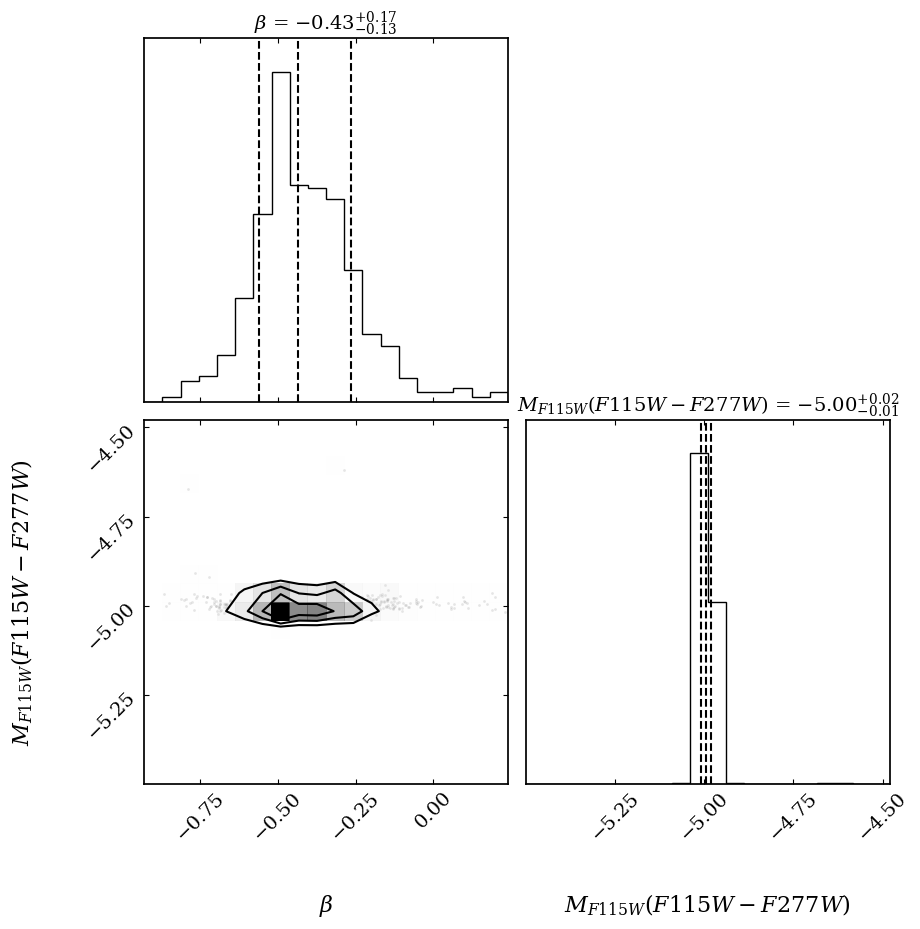}
\includegraphics[width=0.24\textwidth]{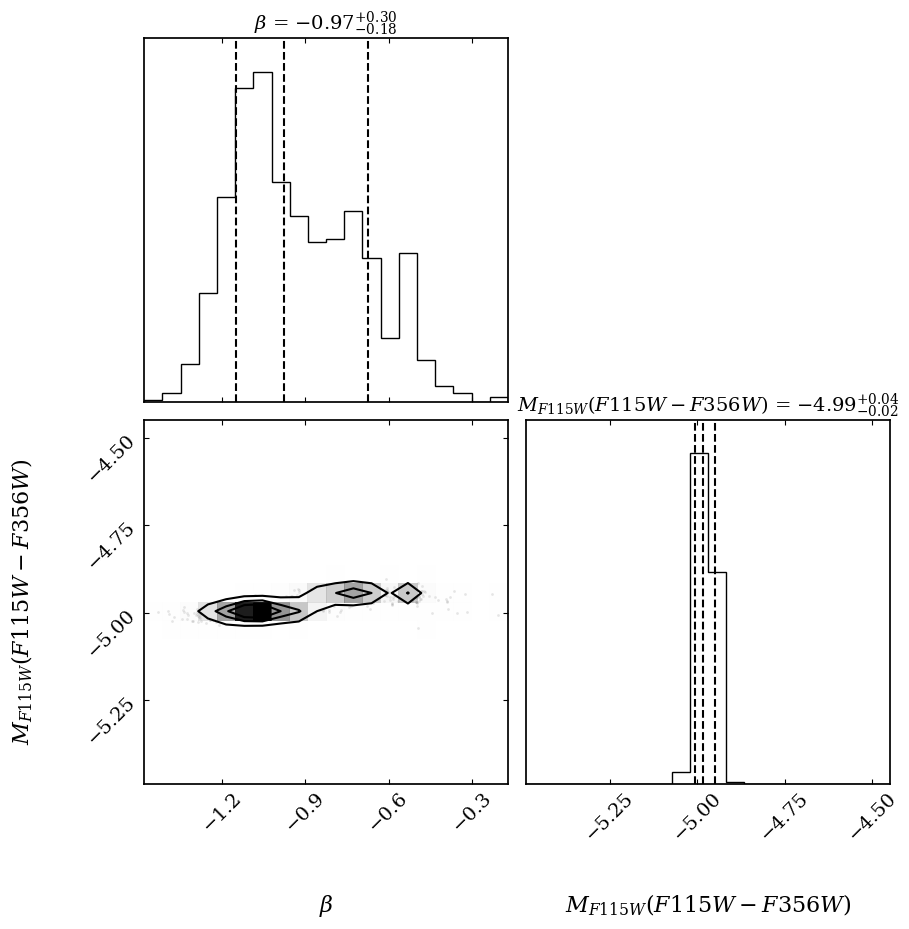}
\includegraphics[width=0.24\textwidth]{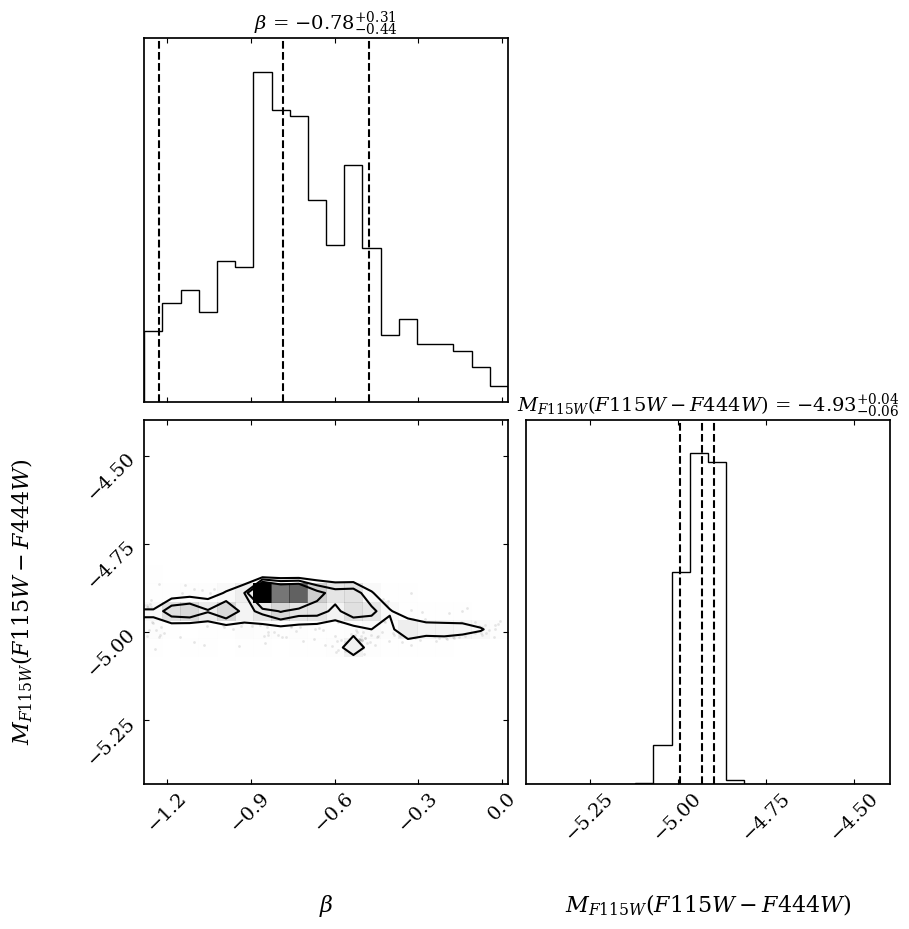}
\includegraphics[width=0.24\textwidth]{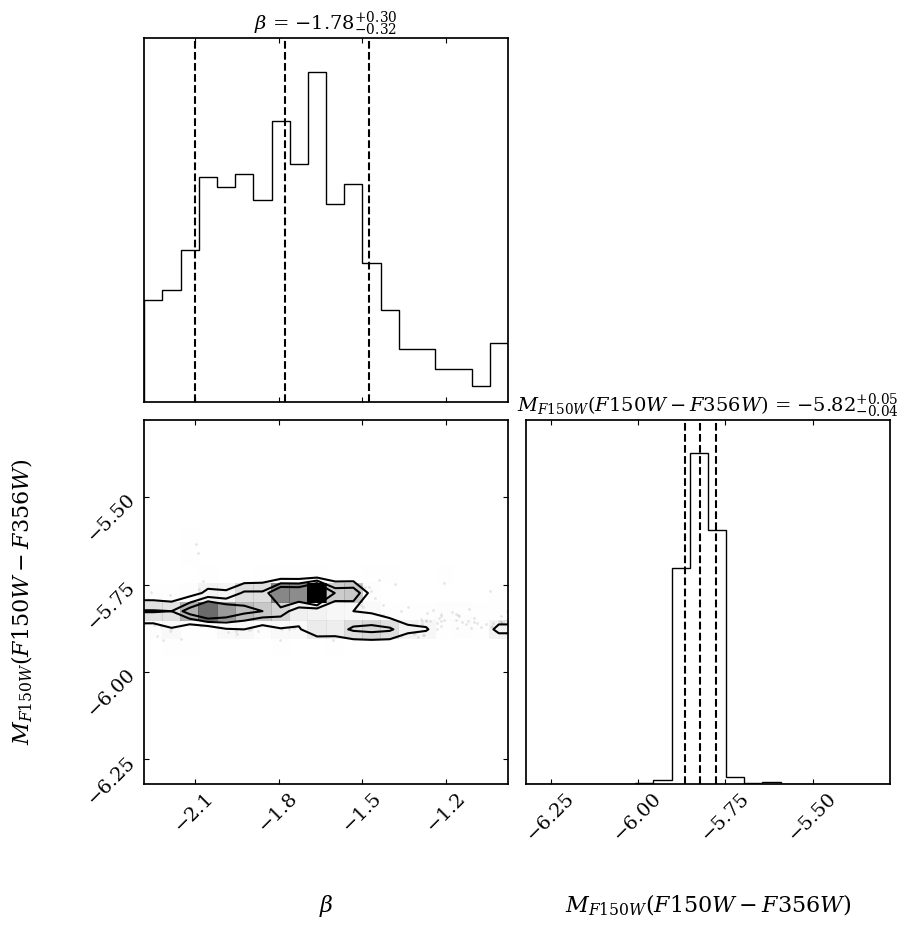}
\includegraphics[width=0.24\textwidth]{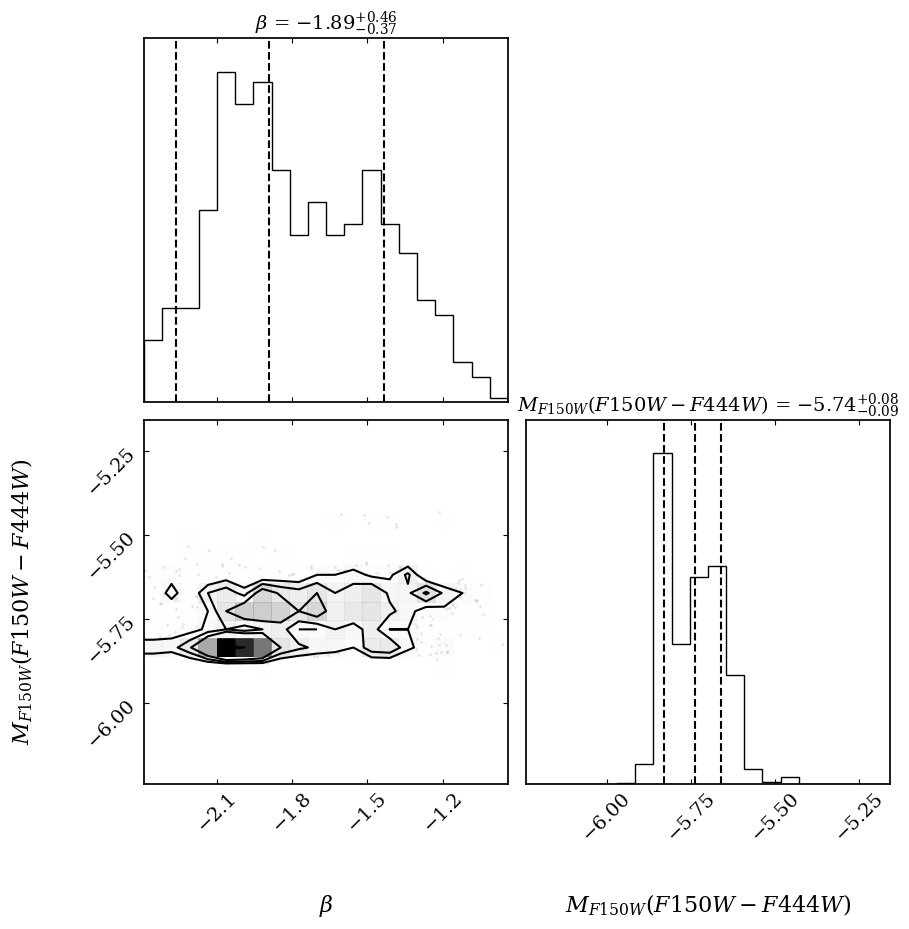}
\includegraphics[width=0.24\textwidth]{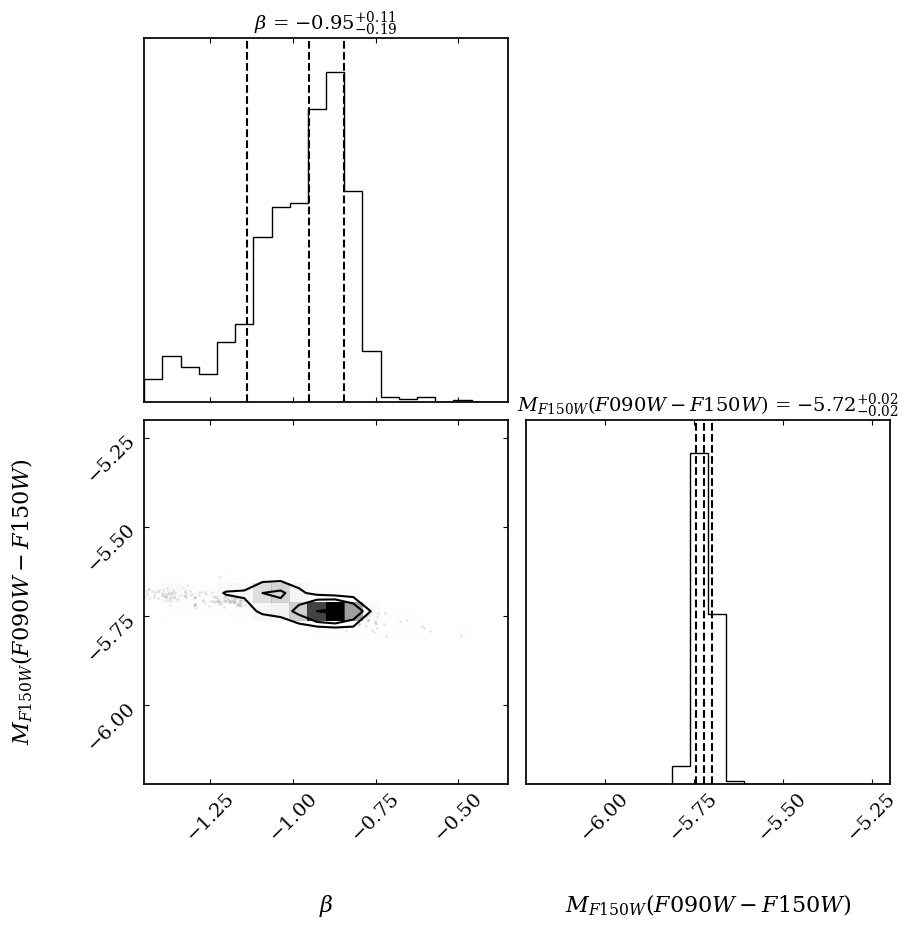}
\includegraphics[width=0.24\textwidth]{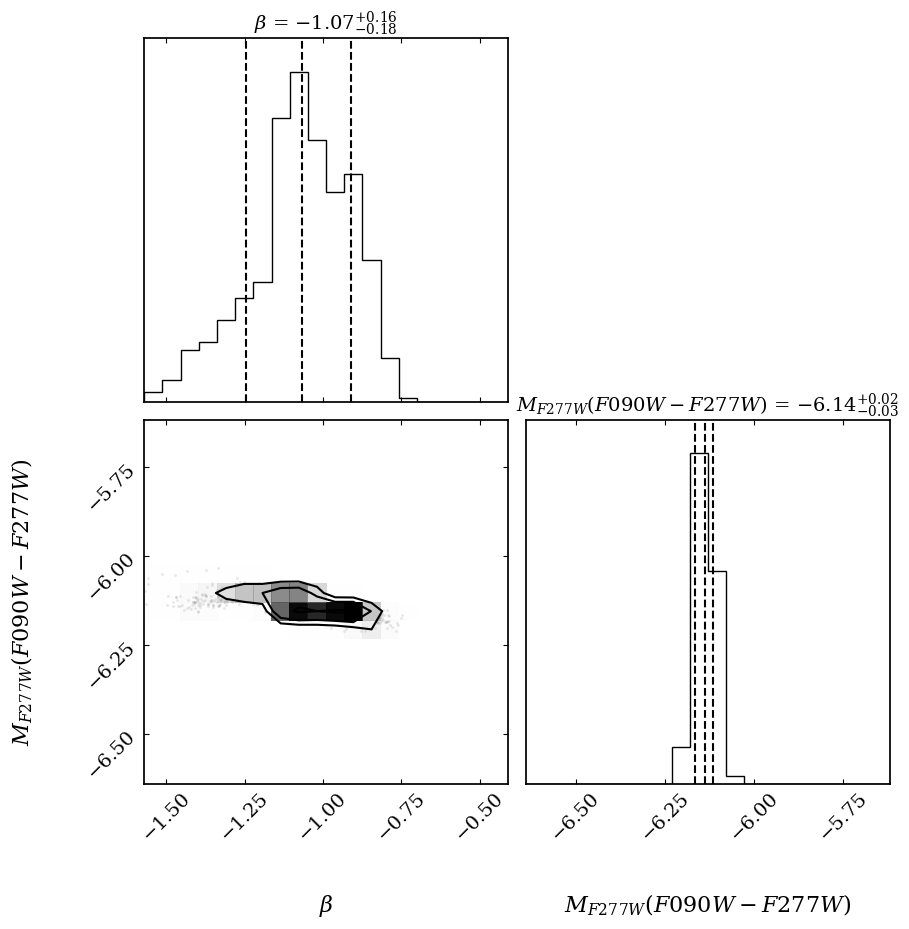}
\includegraphics[width=0.24\textwidth]{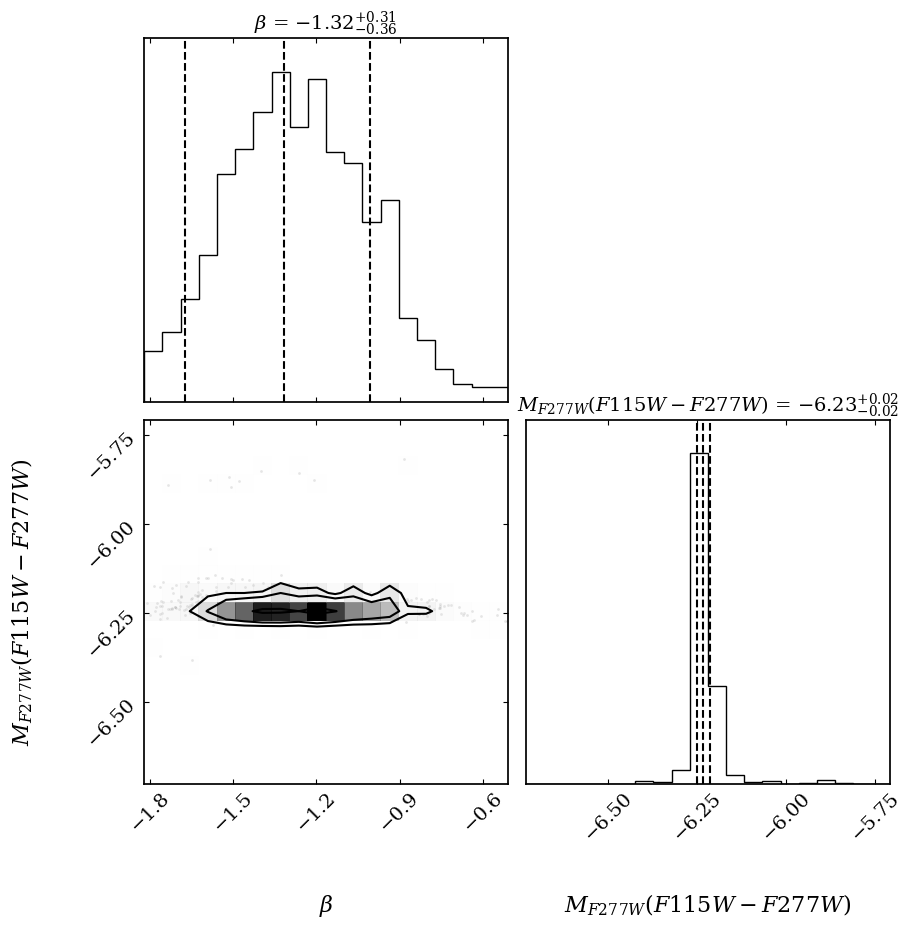}
\includegraphics[width=0.24\textwidth]{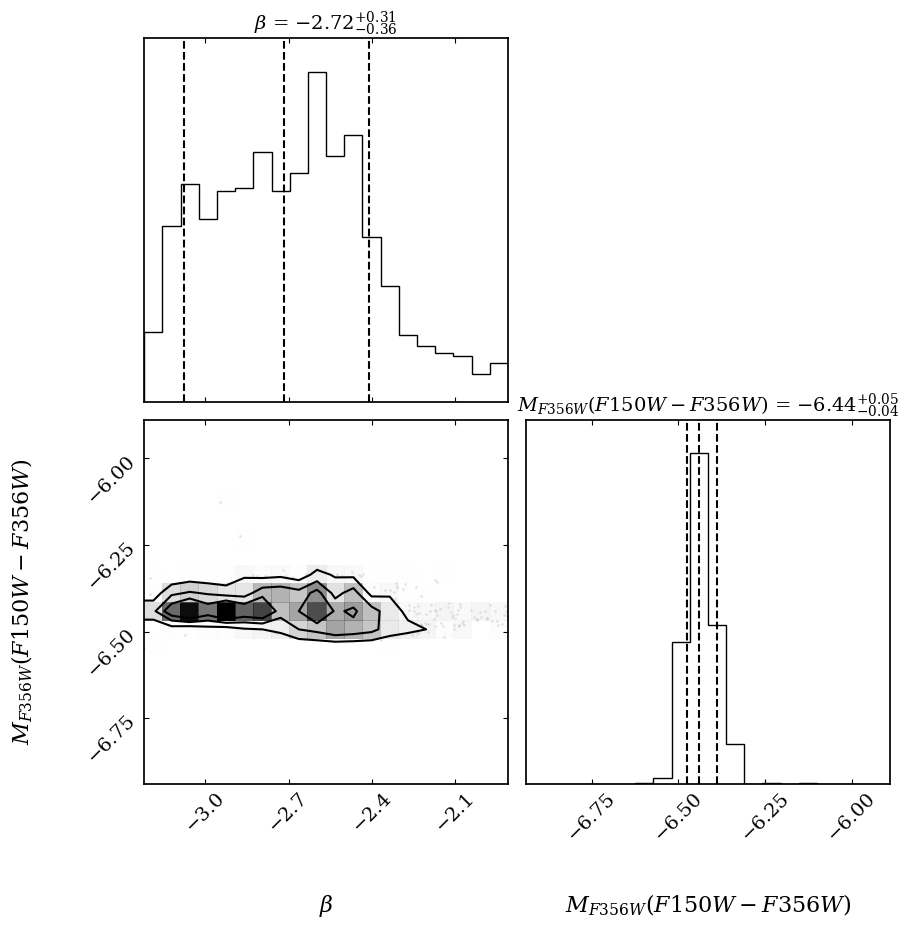}
\includegraphics[width=0.24\textwidth]{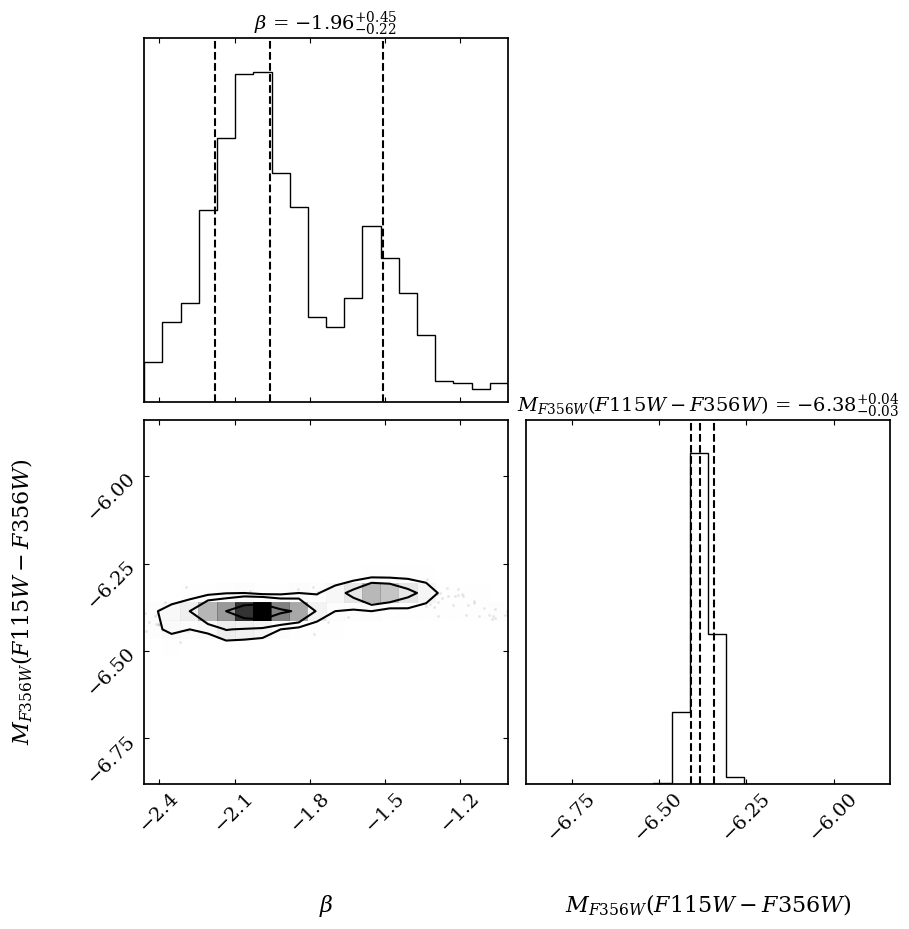}
\includegraphics[width=0.24\textwidth]{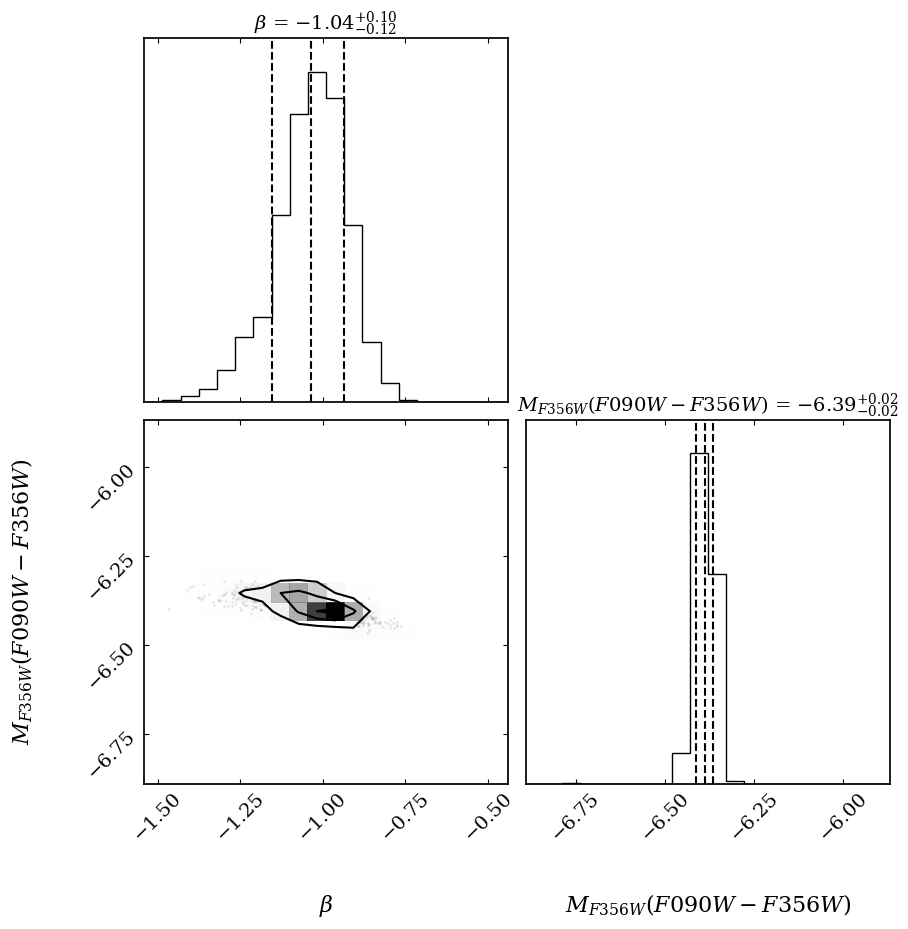}
\includegraphics[width=0.24\textwidth]{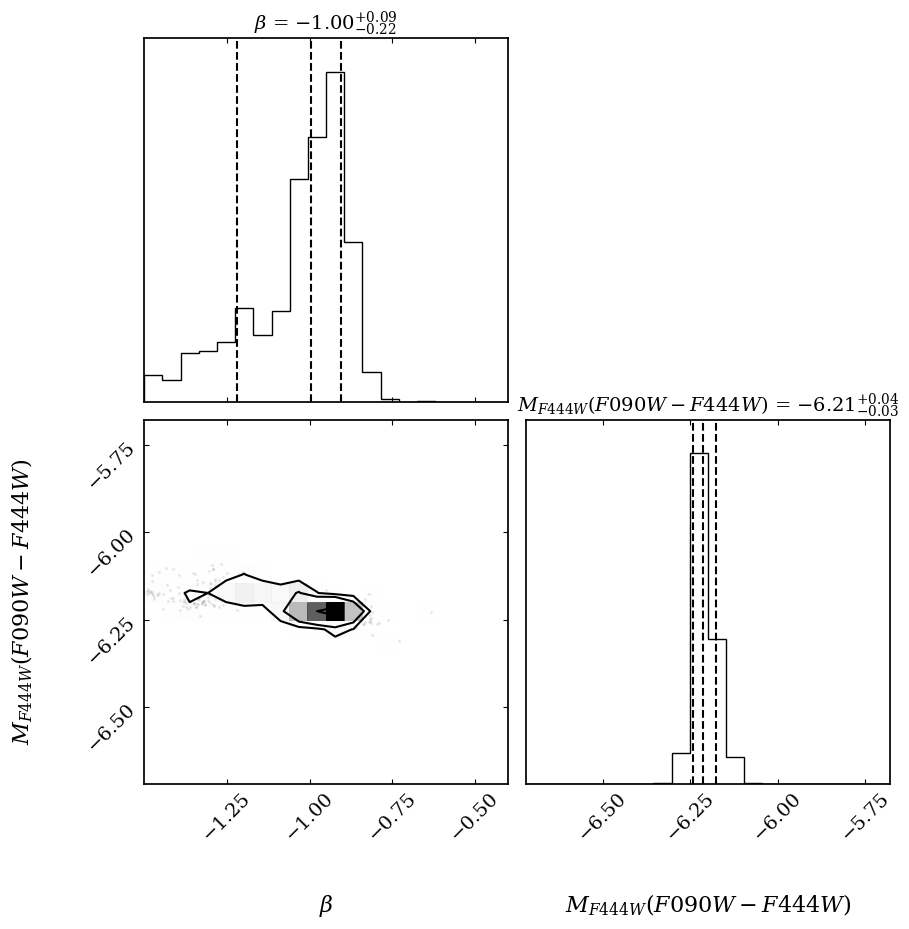}
\includegraphics[width=0.24\textwidth]{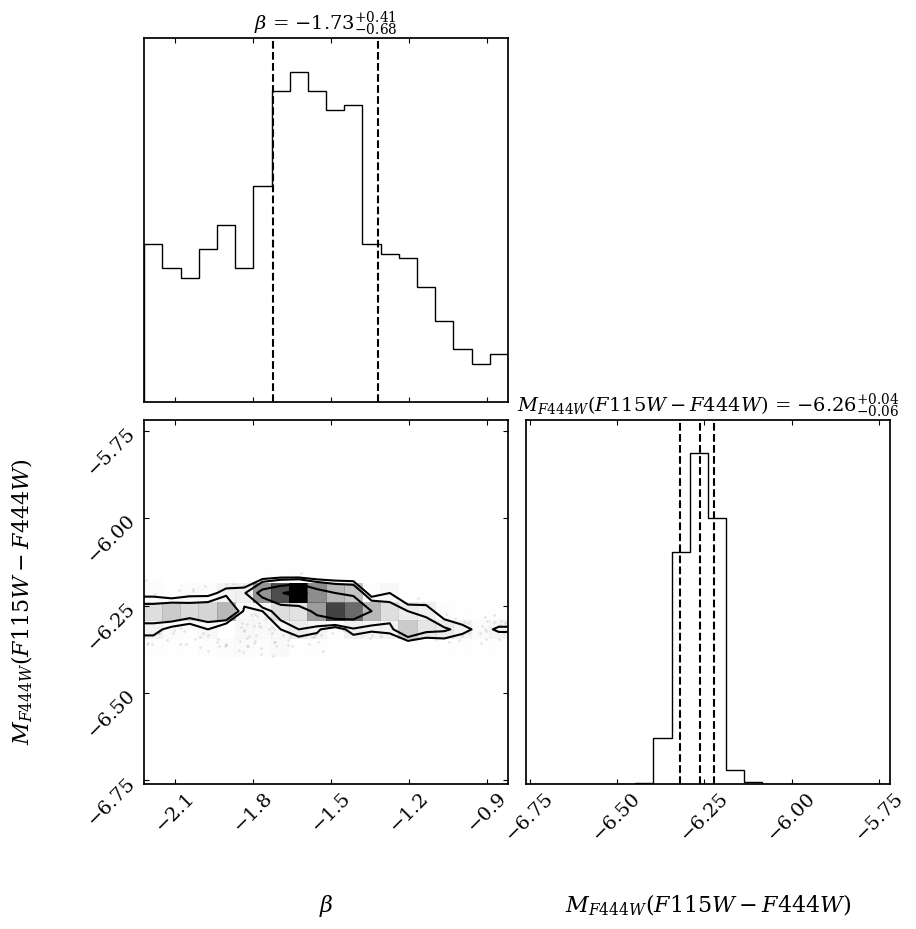}
\includegraphics[width=0.24\textwidth]{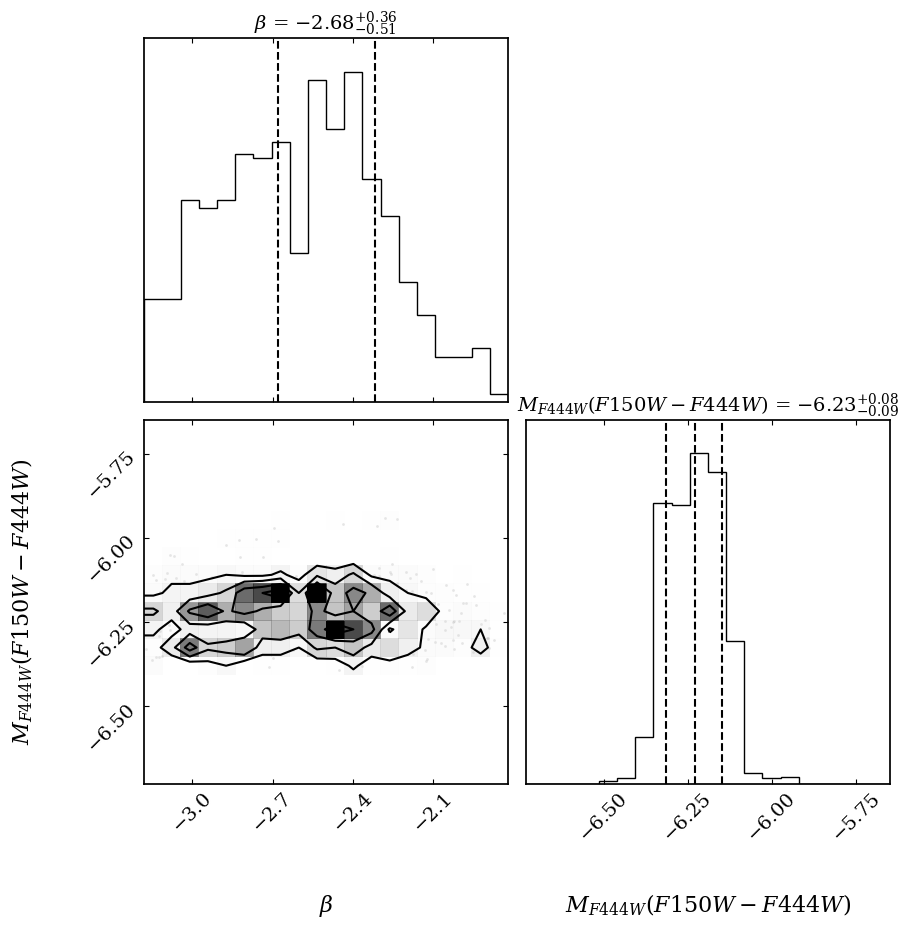}
\caption{Corner plots for all \jwst\ TRGB calibrations. }
\label{fig:corner_plots}
\end{figure*}

In Figure~\ref{fig:trgb_zp_overview} we present all \jwst\ TRGB zero-point fits from Table~\ref{tab:all_trgb_calibs}. Zero-points are plotted as a function of the color pivot point ($\gamma$). The shape of the points corresponds to the magnitude filters and the color of the points corresponds to the color baseline. 


\section{CMDs of Individual targets compared to the global TRGB slope fit}
\begin{figure*}
\centering
	\includegraphics[width=\textwidth]{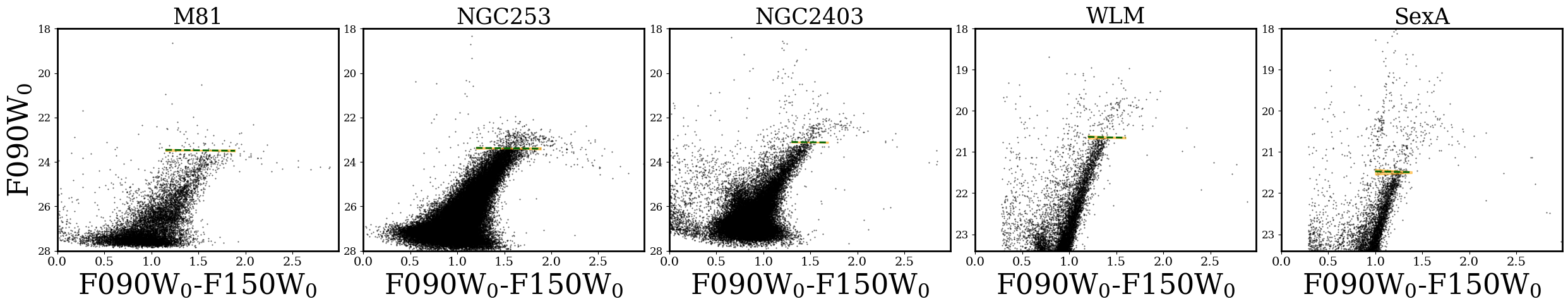}
	\includegraphics[width=\textwidth]{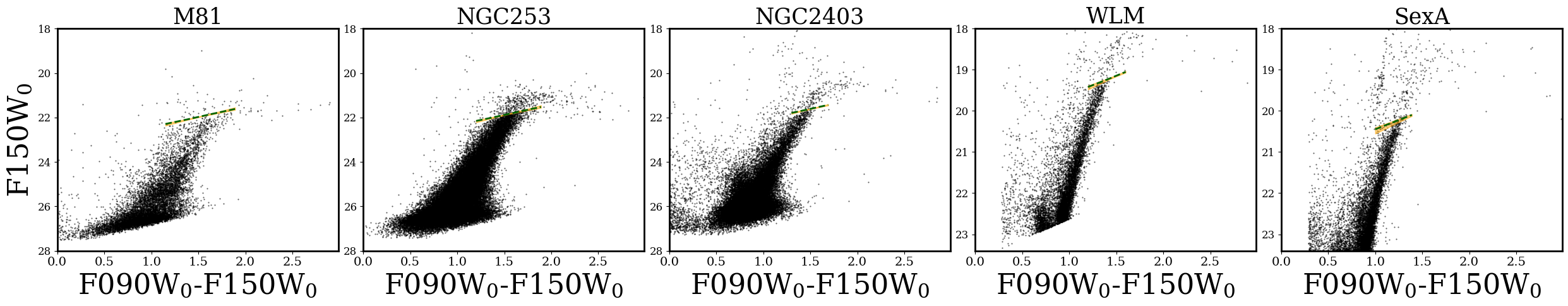}
	\caption{Color-Magnitude Diagrams for selected target galaxies compared to the global TRGB best-fit slope. The targets shown are those used in the corresponding TRGB calibration for the $F090W$ and $F150W$ filters and include: M~81 (detector A), NGC~253, NGC~2403, WLM, and Sextans A.  Top: $F090W$ vs. $F090W-F150W$. Bottom: $F150W$ vs. $F090W-F150W$. Best-fit slopes (dashed green) and uncertainties (orange shaded) are overlayed in each panel.}
	\label{fig:ind_cmd_slope_f090wf150w}
\end{figure*}

\begin{figure*}
\centering
	\includegraphics[width=\textwidth]{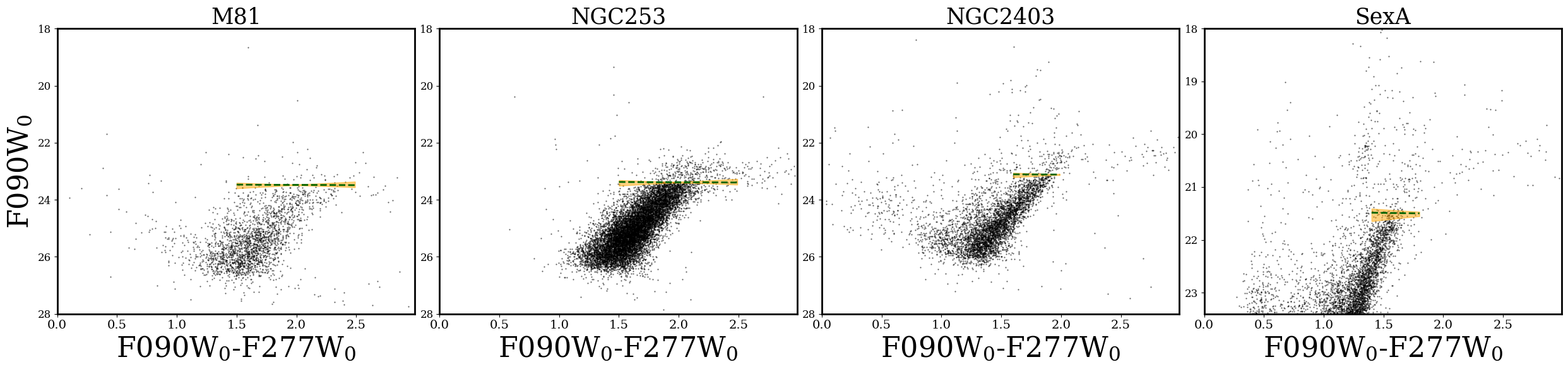}
	\includegraphics[width=\textwidth]{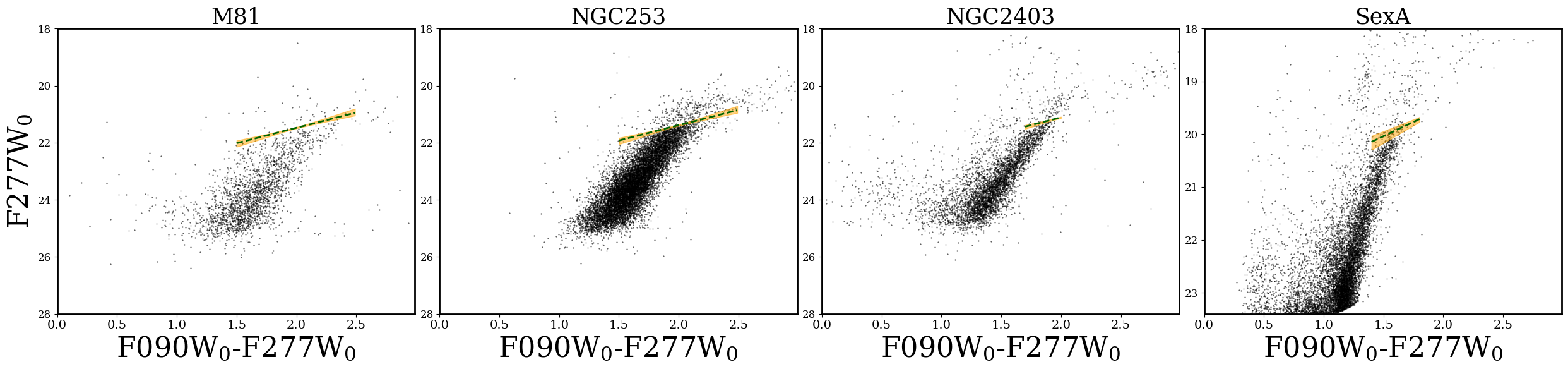}
	\caption{Similar to Figure~\ref{fig:ind_cmd_slope_f090wf150w} but shown now for $F090W$ and $F277W$.}
	\label{fig:ind_cmd_slope_f090wf277w}
\end{figure*}

In Figures~\ref{fig:ind_cmd_slope_f090wf150w} and ~\ref{fig:ind_cmd_slope_f090wf277w} we present CMDs of our sample broken down by target. Each panel shows the CMD of a single target and the best-fit \emph{global} TRGB slope (see \S\ref{sec:results}; see also Table~\ref{tab:all_trgb_calibs}). As expected based on the calibration precision achieved in \S~\ref{sec:dm_verify}, we find that the global TRGB slope for the selected CMD combinations agrees visually with the expected location of the TRGB in individual target CMDs.

\renewcommand\bibname{{References}}
\bibliography{main.bib}
\end{document}